\documentclass[apj]{emulateapj}

\usepackage{url}

\usepackage{apjfonts}
\usepackage{natbib}
 
\usepackage{epsfig}
\usepackage{amssymb}
\usepackage{natbib}
\usepackage{aas_macros}

\usepackage{comment}

\slugcomment{Accepted for publication in ApjS}

\shorttitle{morphologies in CANDELS}
\shortauthors{Huertas-Company et al.}
\usepackage{url}
\usepackage{amsmath}

\begin{document}

\voffset=-0.50in


\title{A catalog of \emph{visual-like} morphologies in the 5 CANDELS fields using deep-learning}


\author{\textsc{M. Huertas-Company\altaffilmark{1}, R. Gravet\altaffilmark{1}, G. Cabrera-Vives\altaffilmark{2,3}, P.G.P\'erez-Gonz\'alez\altaffilmark{4}, J.~S. Kartaltepe\altaffilmark{5}, G. Barro\altaffilmark{6}, M. Bernardi\altaffilmark{7}, S. Mei\altaffilmark{1}, F. Shankar\altaffilmark{8}, P. Dimauro\altaffilmark{1}, E.F. Bell\altaffilmark{9}, D. Kocevski\altaffilmark{10}, D.~C. Koo\altaffilmark{6}, S.~M. Faber\altaffilmark{6}, D.~H. Mcintosh\altaffilmark{11}}}

\affil{\scriptsize $^1$ GEPI, Observatoire de Paris, CNRS, Universit\'e Paris  Diderot, 61, Avenue de l'Observatoire 75014, Paris  France}
\affil{\scriptsize $^2$ Center for Mathematical Modeling and Department of Computer Science, University of Chile, Santiago, Chile.}
\affil{\scriptsize $^3$AURA Observatory in Chile, La Serena, Chile}
\affil{\scriptsize $^4$ Departamento de Astrof\'isica, Facultad de CC. F\'isicas, Universidad Complutense de Madrid, E-28040 Madrid, Spain}
\affil{\scriptsize $^5$School of Physics and Astronomy, Rochester Institute of Technology, 84 Lomb Memorial Drive, Rochester, NY 14623, USA}
\affil{\scriptsize $^6$ UCO/Lick Observatory, Department of Astronomy and Astrophysics, University of California, Santa Cruz, CA 95064}
\affil{\scriptsize $^{7}$ Department of Physics and Astronomy, University of Pennsylvania, Philadelphia, PA 19104, USA}
\affil{\scriptsize $^{8}$ Department of Physics and Astronomy, University of Southampton, Southampton SO17 1BJ, UK}
\affil{\scriptsize $^{9}$ Department  of  Astronomy,   University  of  Michigan,   500Church St., Ann Arbor, MI 48109}
\affil{\scriptsize $^{10}$Department of Physics and Astronomy,  University of Kentucky, Lexington, KY 40506, USA}
\affil{\scriptsize $^{11}$Department of Physics \& Astronomy, University of Missouri-Kansas City, 5110 Rockhill Rd., Kansas City, MO 64110, USA}



\begin{abstract}
We present a catalog of \emph{visual like} H-band morphologies of $\sim50.000$ galaxies ($H_{f160w}<24.5$) in the 5 CANDELS fields (GOODS-N, GOODS-S, UDS, EGS and COSMOS). Morphologies are estimated with Convolutional Neural Networks (ConvNets). The median redshift of the sample is $<z>\sim1.25$. The algorithm is trained on GOODS-S for which visual classifications are publicly available and then applied to the other 4 fields. Following the CANDELS main morphology classification scheme, our model retrieves the probabilities for each galaxy of  having a spheroid, a disk, presenting an irregularity, being compact or point source and being unclassifiable.  ConvNets are able to predict the fractions of votes given a galaxy image with zero bias and $\sim10\%$ scatter. The fraction of miss-classifications is less than $1\%$. Our classification scheme represents a major improvement with respect to \emph{CAS (Concentration-Asymmetry-Smoothness)-based} methods, which hit a $20-30\%$ contamination limit at high z. The catalog is released with the present paper via the Rainbow database~(\url{http://rainbowx.fis.ucm.es/Rainbow\_navigator\_public/}).

\end{abstract}





\section{Introduction}

Since the pioneer works in the first half of the XXth century by E. Hubble, galaxies have been classified according to their visual aspect (see e.g. \citealp{1926ApJ....64..321H,1936rene.book.....H}). This very first optical classification revealed that galaxies in the local Universe are broadly bimodal, with or without a stellar disk (\emph{Hubble Fork}). Understanding the physical processes that lead to such a bimodality - i.e. how bulges and disks form and evolve - is one of the major challenges in the field of galaxy evolution and the main goal of deep field surveys. Classification of galaxies at different cosmic epochs is therefore a key step towards understanding how the progenitors of today's Hubble Fork were shaped. The main difficulty is that is hampered by the impressive amount of data which are and will be available from large galaxy surveys.

A question naturally arises: can human classifiers be replaced by automatic techniques? There have been some efforts led by different groups towards that direction consisting on using existing visual morphologies on a smaller dataset to train automated machine learning algorithms~(e.g.~\citealp{2004MNRAS.348.1038B,2008A&A...478..971H, 2014MNRAS.443.3528S}). The basic idea behind these approaches is to find a set of parameters that correlate with the visual morphology of a galaxy and define the space of parameters that best characterize a given morphological type. (e.g \citealp{1996ApJS..107....1A,2000ApJ...529..886C, 2008ApJ...672..177L}). In astronomy, the parameters defining morphology traditionally include concentrations, asymmetries, clumpiness (or smoothness), gini coefficient, moments of light etc.\\

In the last years, we proposed a generalization of this approach with the development of galSVM \citep{2008A&A...478..971H, 2009A&A...497..743H, 2011A&A...525A.157H}, which enables an n-dimension classification with optimal non-linear boundaries in the parameter space as well as a quantification of errors following a probabilistic approach (see also~\citealp{2015arXiv150401751P,2007ApJS..172..406S}). These \emph{CAS (Concentration-Asymmetry-Smoothness)-based} methods have been proven to be relatively useful but are also affected by several limitations. The values of the parameters strongly depend on the data quality and redshift and they only provide rough morphological classifications in 2 or 3 classes. The most evident shortcoming with such techniques is that the fraction of miss-classifications is high especially at high redshifts ($\sim20-30\%$, \citealp{2014arXiv1406.1175H}). The latter is possibly the main reason why their popularity among the astronomical community is still quite low (see review by \citealp{2010IJMPD..19.1049B}). \\

The problem might reside in the parameters that people traditionally adopt. Concentrations, asymmetries etc and by extension principal components are useful because they reduce the complexity of the problem by globally describing a galaxy with just a few parameters. However, this approach at the same time, neglects an enormous amount of information contained in the pixels themselves. As a consequence CAS-based methods might not be suited to actually represent the capability of the human brain to capture the full, complex distribution of light. Using all the pixels as parameter space is now possible with the advent of powerful computing resources such as Graphic Processor Units (GPUs). At the same time, there exist very powerful machine learning algorithms suited to mimic the human perception (such as \emph{deep learning}) which are able to learn the best set of parameters for a given problem.This new approach has been first used in astronomy at low redshift earlier this year, in the framework of an online competition led by the Galaxy Zoo team (see~\S~\ref{sec:class} for more details) yielding to very promising results (\citealp{D14}, hereafter D15). \\

In this paper, we extend this new methodology to high redshift by classifying $\sim50.000$ galaxies with median redshift $<z>\sim1.25$ in the CANDELS fields where detailed visual classifications are available for a subsample of $\sim8.000$ objects \citep{2014arXiv1401.2455K}. We show that the use of deep learning yields to an almost free-of-contaminations classification that closely mimics the human perception. The resulting catalog on the 5 CANDELS fields (GOODS-S, GOODS-N, UDS, EGS  and COSMOS) is released with the present work. \\

The paper is structured as follows. In section~\ref{sec:dataset} we describe the dataset. In section~\ref{sec:class} we describe the method and how the CANDELS data are pre-processed before feeding the algorithm. In sections~\ref{sec:acc} and~\ref{sec:all_CANDELS} we discuss the performance and accuracy of the resulting classification and in section~\ref{sec:catalog} we describe the properties of the catalog which is released. We conclude with a summary of the main results (section~\ref{sec:summary}).

\section{Dataset}
\label{sec:dataset}
 Our starting-point catalogs are the CANDELS public photometric catalogs for UDS \citep{2013ApJS..206...10G} and GOODS-S \citep{2013ApJS..207...24G}. Preliminary CANDELS catalogs were used for COSMOS, EGS and GOODS-N (private communication). We select all galaxies in the F160W filters with F160W$<$24.5~mag (AB system) which is the magnitude limit imposed by Karltatepe et al. (2014) to perform reliable visual morphological classifications. Since our goal is to provide a morphological classification as close as possible to the visual one, we restrict our selection to the same criteria in all considered fields. \\

\begin{figure*}
\begin{center}
$\begin{array}{c c}
\includegraphics[width=0.40\textwidth]{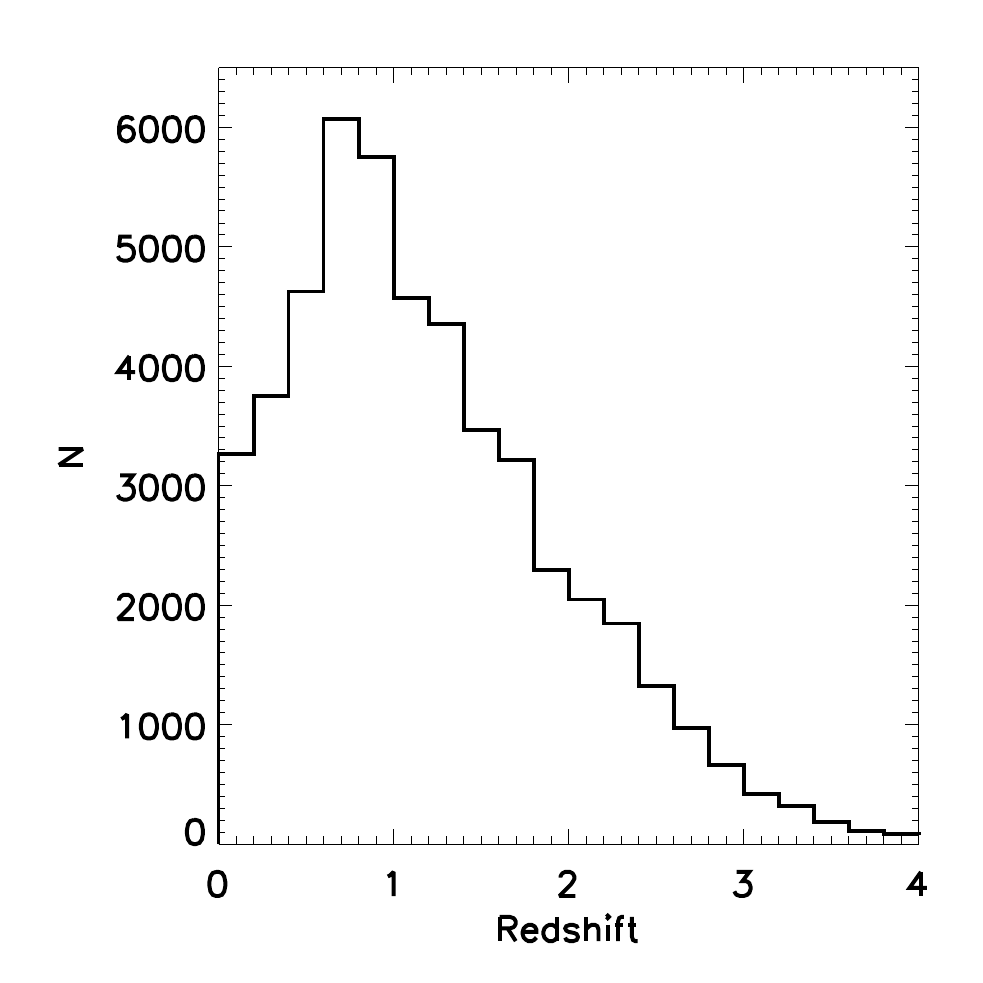} & \includegraphics[width=0.40\textwidth]{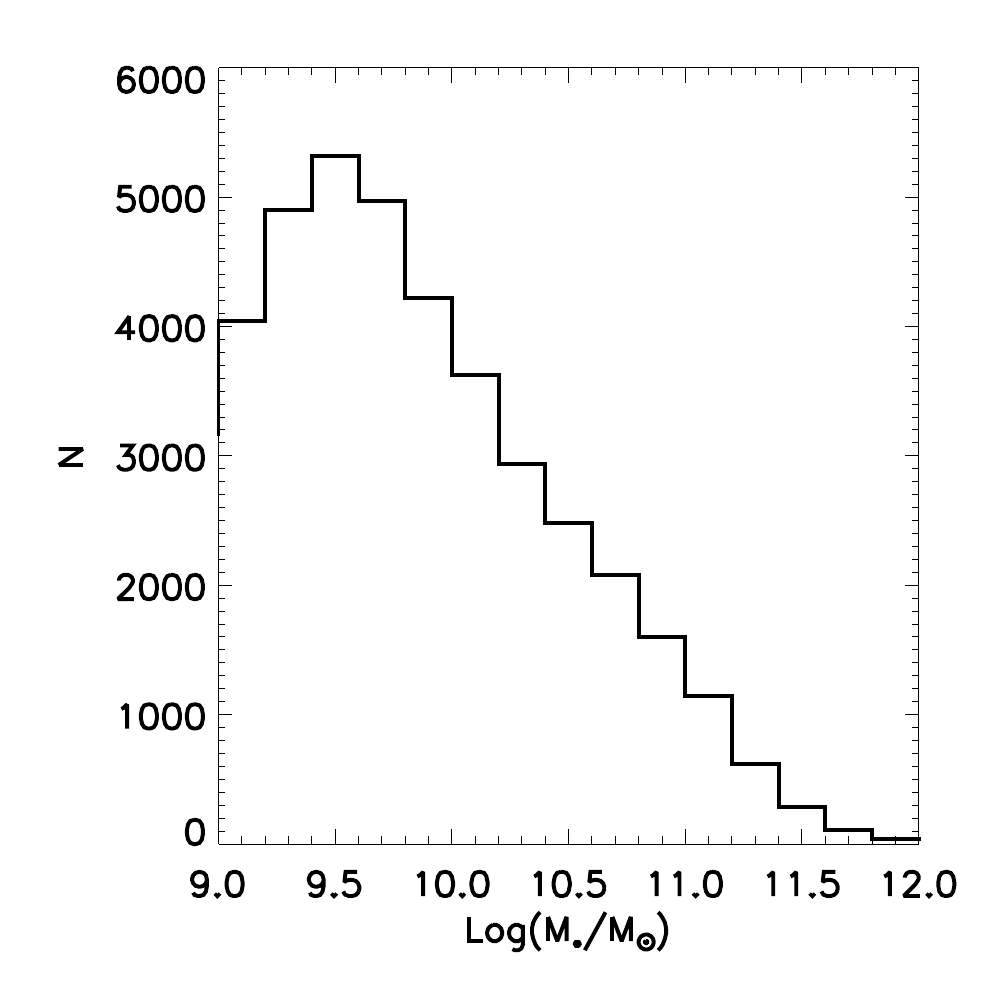}\\
\end{array}$
\caption{Redshift (left) and stellar mass (right) distributions of the selected sample for morphological classifications. The dataset contains more than 20.000 galaxies at $z>1$ where the CANDELS fields probe the optical rest-frame morphologies.} 
\label{fig:sample_props}
\end{center}
\end{figure*}

The resulting sample consists of 50.000 galaxies, which increases by a factor of 5 the visual catalog published in CANDELS up to date. About $50\%$ of the sources are between $1<z<3$ (fig.~\ref{fig:sample_props}), where the CANDELS filters probe optical rest-frame morphologies. As extensively discussed in \cite{2014arXiv1401.2455K}, the sample is $\sim80\%$ complete down to $log(M_*/M_\odot)\sim10$ (see their figure 1). 

\section{CANDELS morphological classification with deep learning}
\label{sec:class}
\subsection{Convolutional Neural Network (ConvNet) configuration}

In this work we mimic the human perception with \emph{deep
  learning} using
convolutional neural networks (ConvNets). Although it is clearly beyond the
scope of the present paper to give a complete description of how
convolutional neural networks work, we provide a brief introduction below.  We refer the interested reader to D15 for more details.

Deep learning is a methodology to automatically learn and extract the most relevant features (or parameters) from raw data for a given classification problem through a set of non-linear transformations. \\
Though deep
learning architectures have existed since the early 80s
\citep{Fukushima80}, they involve complex technological problems
that only allowed their use in massive datasets in the last
decade. Several factors have contributed to the rise in their popularity: (i) the availability
of much larger training sets, with millions of labeled examples \footnote{
ConvNets are particularly sensitive to this since the risk of
over-fitting is large given the complexity of the models}; (ii)
powerful GPU implementations, making the training of very large models
practical; (iii) improved model regularization algorithms, which helped reducing the computing time.

ConvNets have been proven to perform extremely well in image
recognition tasks. For example, they have achieved an error rate of
0.23\% on the MNIST database, which is a collection of manuscript
numbers considered as a standard test for all new machine learning algorithms \citep{Ciresan2012}. When applied to facial recognition, they achieve a 97.6\% recognition rate on 5,600 images of more
than 10 subjects \citep{Matusugu13}. The ImageNet Large Scale Visual
Recognition Challenge is a benchmark in object classification and
detection, with millions of images and hundreds of object classes. In
\cite{Krizhevsky2012}, ConvNets were able to get an error rate of
15.3\% compared to 26.2\% achieved by the second best competitors (non
deep).  Also, the performance of convolutional neural networks on the ImageNet
tests is now close to a purely human based classification \citep{Russakovsky14}.

ConvNets were first applied to galaxy morphological classification
earlier this year in the framework of the Galaxy Zoo Challenge in the
Kaggle platform \footnote{\url{https://www.kaggle.com/c/galaxy-zoo-the-galaxy-challenge}.} . The aim of the challenge was to find an algorithm
able to predict the 37 votes of the Galaxy Zoo 2 release. The winner
of the competition used ConvNets to get a final RMS of $\sim 7\%$ on
the parameters \citep{D14}. This work clearly showed that ConvNets are a very
promising tool for automated morphological classifications.  

\begin{figure*}
\includegraphics[angle=0,scale=.50]{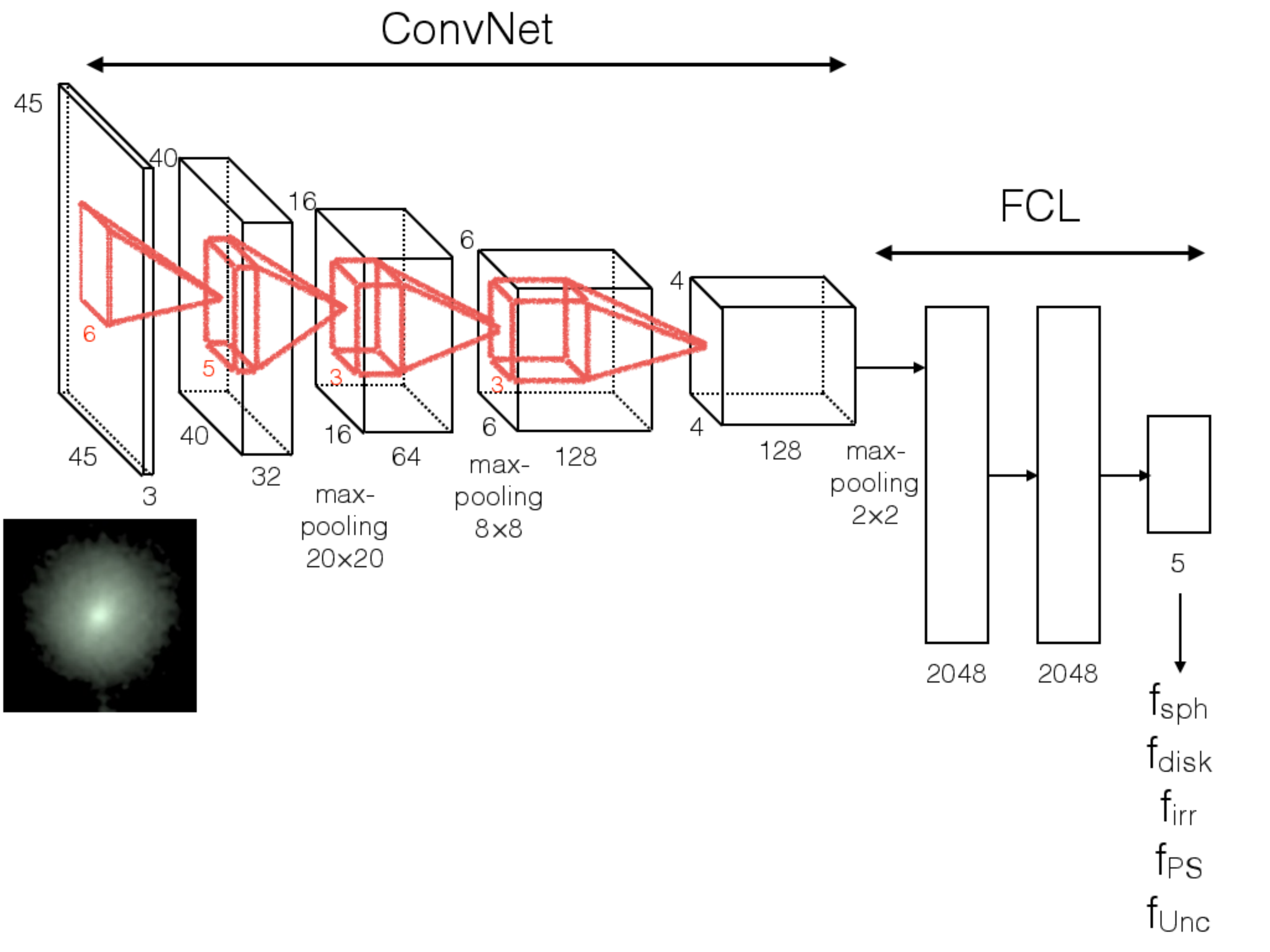}
\caption{Configuration of the Convolutional Neural Network used in this paper. The Network is based on the one used by~\protect\cite{D14} on SDSS galaxies. It is made of 5 convolutional layers followed by 2 fully connected perceptron layers. In the convolutional part there are also 3 max-pooling steps of different sizes. The input are \emph{SDDSized} CANDELS galaxies as explained in the text and the output (for this paper) is made of 5 real values corresponding to the fractions defined in the CANDELS classification scheme. }
\label{fig:convnet}
\end{figure*}

 There is no clear methodology to find the optimal convolutional neural network for a given problem except for trying different configurations and comparing the outputs. The one used for the Galaxy Zoo challenge provided excellent results for a similar problem to ours (fig.~\ref{fig:convnet}). We therefore decided to use the D15 configuration to classify the CANDELS sample. Given the different nature of SDSS and CANDELS images, our methodology, by design, requires specific pre-processing steps, as discussed in section~\ref{sec:pre}. This is certainly not the cleanest approach but it is sufficient for our classification purposes as discussed in subsequent sections.

\subsection{Training set}

The ConvNet is trained to reproduce the CANDELS visual morphological classification defined in \cite{2014arXiv1401.2455K}. This classification is based on the efforts of 65 individual classifiers who contributed to the visual inspection of all galaxies in the GOODS-S field (being 3-5 the average number of classifiers per galaxy). The classifiers were asked to provide a number of flags related to the galaxy' structure, morphological k-correction, interaction status and clumpiness. As a result, each galaxy in the catalog  has a number of flags, which measure the fraction of classifiers who selected a morphological feature. The classification was mainly performed in the H band (F160W), even though each classifier had access to the images of the same galaxy in other wavelengths.

\begin{figure*}
\includegraphics[angle=0,scale=.50]{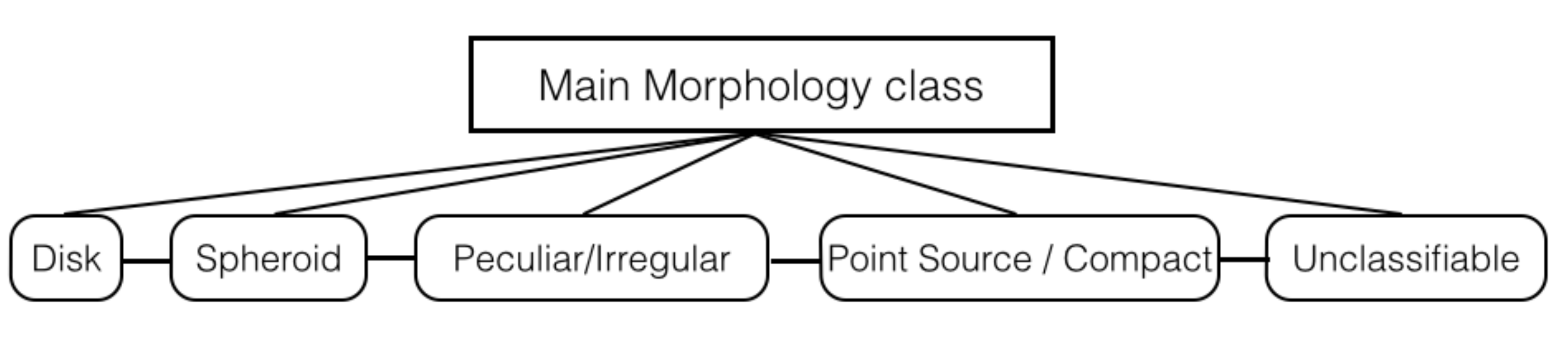}
\caption{CANDELS \emph{Main Morphology} visual classification scheme as described in \protect\cite{2014arXiv1401.2455K}. Each classifier (3-5 per galaxy on average) is asked to provide for each galaxy 5 flags corresponding to the main morphological properties of the galaxy as labeled in the figure. The flags are then combined to produce fractions of people that voted for a given feature. }
\label{fig:CANDELS_tree}
\end{figure*}

In this work, we will focus on the main \emph{classification tree} which defines the main morphological class (fig.~\ref{fig:CANDELS_tree}). For each galaxy there are therefore 5 parameters, $f_{spheroid}$, $f_{disk}$, $f_{irr}$, $f_{PS}$ and $f_{Unc}$ which refer respectively to the frequency at which human classifiers flagged a given galaxy as \emph{having a spheroid}, \emph{a disk}, \emph{some irregularities}, being a point-source (or unresolved) and \emph{unclassifiable}. It is important to notice that one flag does not exclude the other (except for the \emph{Unc} one) i.e. a galaxy can obviously have both a disk and a spheroid or have a disk and be irregular, so the sum of all frequencies for a given object is not one. \\

The main purpose of this work is to mimic the human behavior. In other words, we want the machine to be able to predict how many people will vote for a given feature given the galaxy image. Recall that the objective we consider here is to replace humans by computers, no to find the \emph{correct} morphology of a galaxy, which actually depends on the definition one wants to adopt. Hence if the visual classification is intrinsically biased, so will be the machine based one.  

The classification in GOODS-S contains $\sim 8000$ galaxies for which we know the visual classification done by (expert) humans so we can use part of this sample to train the machine learning algorithm and keep a fraction for an independent test. Recall also that during the preparation of the present work, the UDS field has also been finalized so it also represents an independent test for the classification as discussed in section~\ref{sec:all_CANDELS}. In the following, we describe the pre-processing done to the images before feeding the ConvNet.

\subsection{Pre-processing}
\label{sec:pre}

\begin{figure*}
\includegraphics[angle=0,scale=0.50]{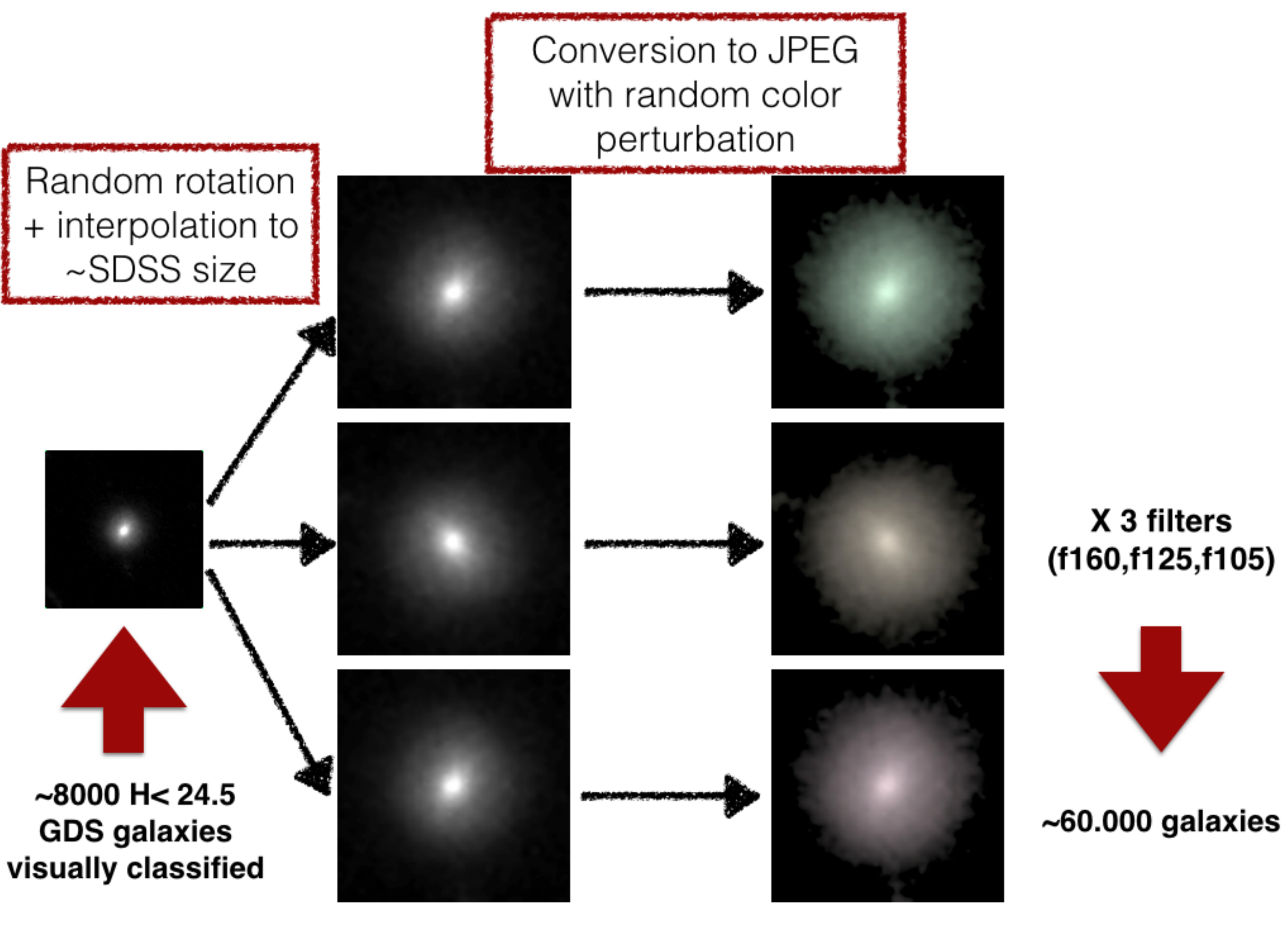}
\caption{Pre-processing of the CANDELS stamps before being fed to the convolutional neural network. Galaxies are first interpolated so that they all have similar sizes. In a second step, we add some redundancy to the data by performing random rotations in order to avoid over-fitting and finally converted the images to JPEG. This is repeated for 3 CANDELS filters. See text for details.}
\label{fig:preproc}
\end{figure*}

As previously discussed, we will use for this work, the ConvNet design shown in D15 optimized for the SDSS. There are some obvious problems related to this approach, since galaxies at high redshift are intrinsically smaller\footnote{typically 5-10 pixels - $\sim0.3^{"}$ - compared to 40 pixels -$\sim10^{"}$- for the SDSS galaxies} and fainter. Also the training set is made of only $\sim8000$ galaxies from GOODS-S with visual parameters, compared to the $60\times10^3$ galaxies used for the SDSS training. This last point is particularly critical since training the ConvNet with a significantly smaller sample can easily lead to over-fitting issues, i.e. too many parameters in the model we want to build compared with the number of data points \\
To overcome the latter potential issues, we pre-processed the training set before feeding it to the ConvNet applying the following steps (see fig.~\ref{fig:preproc}):

\begin{itemize}

\item All galaxies in the GOODS-S visual morphology catalog are interpolated to the typical SDSS size (i.e. $\sim$ 40 pixels). This is performed using a classical cubic interpolation. The procedure obviously introduces some redundancy in the data since we artificially reduce the pixel size, but ensures that the network \emph{sees} the same ratio of background vs. galaxy pixels than for the SDSS. It is important since the size of the convolution box is fixed.  An alternative approach would have been to adapt the network size to the typical size of CANDELS images. In any case, some interpolation is required given the wide redshift range probed by the CANDELS data ($z\sim0.1$ to $z\sim3$) which means that the length scale changes by more than a factor of 4. Therefore, even if the interpolation factor could be decreased, it is required at some level. In this work, since we are interested in broad morphologies, the impact of interpolation is not a major issue and therefore we decided to keep the original network.  

\item Each galaxy is randomly rotated 3 times before feeding it to the net. Since our dataset is significantly smaller than the one used in the GZOO competition, there is a clear risk of over-fitting in the classification process. We therefore introduce additional redundancy in the training set to increase the number of training points taking advantage of the fact that morphological classifications should be rotationally invariant \citep{D14}. As explained in D15, the algorithm itself will introduce additional redundancy by performing two more $90^{o}$ rotations. 

\item We then introduce some random Gaussian noise to each of the rotated images so that the pixel values of each realization are not exactly the same. The added noise is small enough not to affect the visual aspect of the galaxy but it slightly changes the pixel values. This ensures that the redundancy is actually efficient and that the network considers each rotated galaxy as a different object with very similar morphological parameters just as the human eye does. Finally, each of the rotated images is converted to JPEG with a power-law stretching optimized for astronomy \footnote{http://www.astromatic.net/software/stiff} \citep{2012ASPC..461..263B} and a 10\% compression. This is important to keep the number of possible pixel values reasonable and also to have a similar normalization for all galaxies.  We stress again that since we are here interested in broad morphologies (disk vs. bulge, irregular, compact) the impact of compression is not critical, as shown in subsequent sections. For more detailed morphologies (e.g. LSB features, bars etc), especially at high redshift, a careful investigation of the optimal compression will certainly be required. 

\item The previous steps were repeated in three CANDELS filters (f105, f125 and f160) to reach a final training set of $\sim 58.000$ galaxies ($8000\times3(rotations)\times3(filters)$), very close to the 60.000 SDSS object for which the net was designed. Note that the spatial coverage of all filters is not exactly the same which explains why we only reach $\sim60.000$ galaxies. The size of the dataset is enough to avoid over-fitting and reach satisfactory results as shown in the next sections. The use of the same galaxies in three different filters might introduce some biases since the morphology might look slightly different from one filter to the other. However, \cite{2014arXiv1401.2455K} show that the fraction of galaxies that actually change their morphology between these 3 filters is very small. In any case, we also tried the algorithm using only f160 images (reducing the training set by a factor of 3) leading to no significant changes in the final results ($\sim0.01$ change in the final RMSE value).

\item We finally introduce some \emph{noise} in the visual parameters of each galaxy ($f_{spheroid}$, $f_{disk}$, $f_{irr}$, $f_{PS}$ and $f_{Unc}$) by adding a random gaussian $10\%$ scatter. This is done, firstly to make sure that the ConvNet does not see exactly the same data points for different redundant images and force optimization. Second, because the CANDELS fractions are very discretized since the actual number of classifiers per galaxy is rather small and therefore the full range of values from 0 to 1 is not covered. The $10\%$ value is calibrated empirically and it is of the order of magnitude of the intrinsic noise of the labels (assuming that they follow a binomial distribution - see section~\ref{sec:all_CANDELS}). Below this value the effect is almost negligible and above, the original signal is diluted. As we will show in section~\ref{sec:all_CANDELS} this has also some important consequences on the final output.

\end{itemize}

The final dataset used for classification contains thus $\sim58.000$ redundant JPEG images of which 47.700 are used for training the machine (i.e. finding the best model), 5.300 are used for real-time evaluation during model training (validation dataset) and 5.000 galaxies are used to assess the final accuracy with the best final model (test dataset). These 5.000 galaxies constitute the test sample and are not used at all during the training process (but their visual morphology is known) so they can be independently used to study the behavior of the best trained model on an unknown dataset. The final model is taken at 2500 chunks. As described in \cite{D14}, to further improve the classification accuracy, averaging of 17 variants of the best model is applied as post-processing. These variants include modifications such as removal of dense layers, different filter size configurations, and different number of filters among others. We refer to \cite{D14} for more details. The best model followed by the averaging process is then used to classify the other 4 CANDELS fields in which the visual morphology is not yet available.  The classification is done at a rate of $\sim1000$ galaxies/hour on a TESLA M2090 GPU, which is compatible with the treatment of massive datasets expected in the near future (e.g EUCLID, WFIRST). \\

\begin{figure*}
\begin{center}
\includegraphics[angle=0,scale=1]{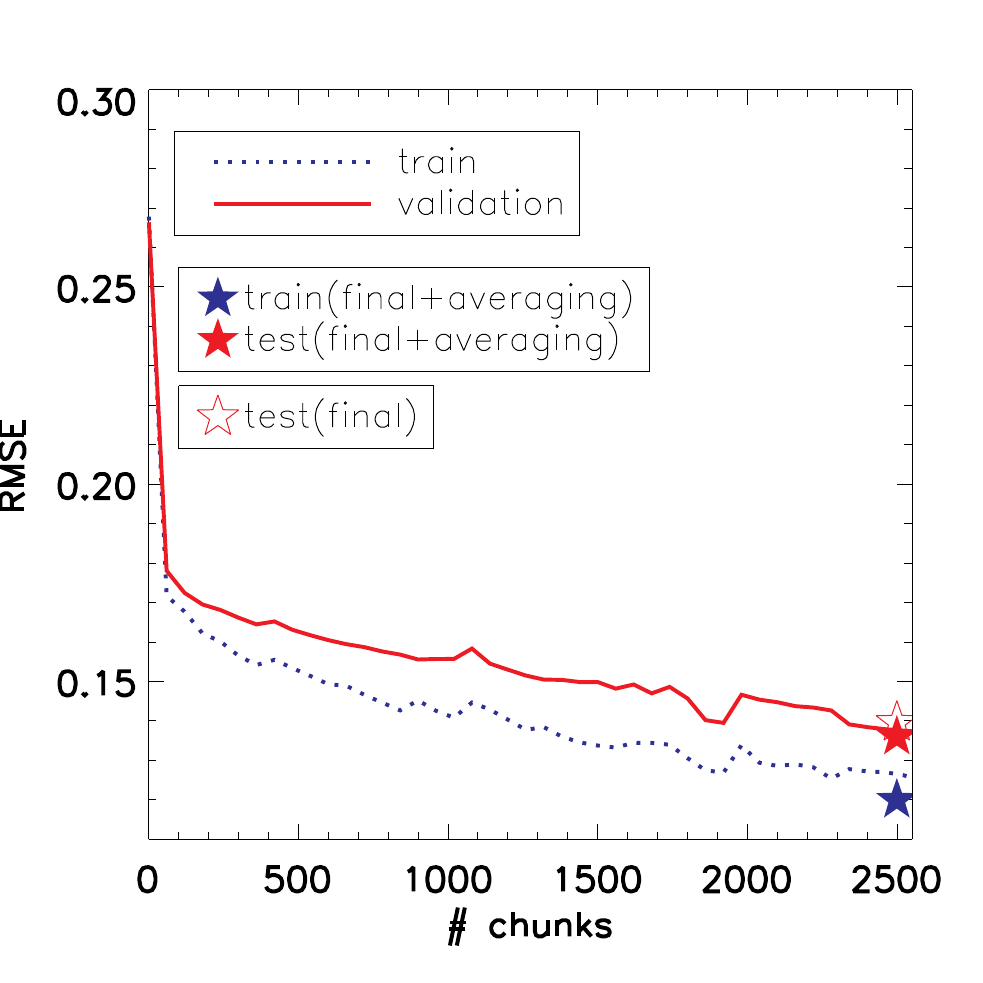}
\caption{Time trajectories for the training (dotted blue line) and validation (red solide line) sets (see text for details). The RMSE si computed every 60 chunks. The blue/red stars indicate the values computed with the final model (2500 chunks) on the training and test samples respectively after averaging and reported in Table~\protect\ref{tbl:tbl_bias_scatter_merged}. The empty star shows the RMSE on the test sample before averaging.}
\label{fig:time_traj}
\end{center}
\end{figure*}

The evolution of the root mean square error (RMSE) during the final learning process for the training and validation datasets is shown in figure~\ref{fig:time_traj}. The difference in RMSE on the validations dataset in the last 10 iterations is of the order of $10^{-4}$, confirming that the algorithm has converged. There is no significant over-fitting given the convergence of the validation set's RMSE. As expected, the RMSE for the training set is slightly smaller ($\sim0.01$), as this is the data directly used to fit our ConvNet model (recall that the validation data-set is used for real time evaluation of the model on unseen data). We also show in figure~\ref{fig:time_traj} the values of the RMSE for the test sample before and after averaging. As explained above, this third data-set is needed to assess the final RMSE of the model, as it may happen that the 2500 chunks we use for convergence are over-fitted to the validation data-set. The RMSE over the test set is very consistent with the one obtained on the validation dataset. Averaging, slightly reduces the RMSE by $\sim10^{-3}$, consistent with the values reported in \cite{D14}.

We made sure that the different pre-processing steps described above result always in a decrease of the average  root mean square error (RMSE) on the validation and test samples. More precisely, before any pre-processing, the average RMSE is $\sim0.25$. Adding noise to the labels decreases the error to $\sim0.22$. Interpolation makes it reach $\sim0.17$ and finally redundancy together with noise addition bring it to the final value of $\sim0.13$ (figure~\ref{fig:time_traj}). \\

\section{Accuracy}
\label{sec:acc}

\subsection{Recovering votes}

Figure~\ref{fig:visual_auto} shows the relation between the visual fractions for each galaxy provided in \cite{2014arXiv1401.2455K} once the random shifts have been applied and the predicted values, for the main classification tree ($f_{spheroid}$, $f_{disk}$, $f_{irr}$, $f_{PS}$ and $f_{Unc}$). We only plot in figure~\ref{fig:visual_auto} objects in the test sample (5000 objects) which were not used for training in order to assess the behavior of the machine with an unknown dataset. Results in terms of bias and scatter are also tabulated in table~\ref{tbl:tbl_bias_scatter}. There is a clear one-to-one correlation between the automatically derived quantities and the visual ones. Table~\ref{tbl:tbl_bias_scatter} shows that the typical bias and dispersion are lower than $10\%$. It is important to keep in mind that the distribution of frequencies is not homogenous between 0 and 1 (there are bins in which there are very few objects) and the machine is therefore optimized to minimize the global bias. In fact, the {median bias and scatter for all morphological frequencies are even smaller and range between $0-0.02$ and $0.03-0.1$ respectively as shown in table~\ref{tbl:tbl_bias_scatter_merged}. If we plot instead galaxies in the training set, the scatter is almost the same, as expected from the learning histories shown in figure~\ref{fig:time_traj}. This confirms that the model is well-optimized and that there is no over-fitting (fig.~\ref{fig:visual_auto_all}). \\

Despite of the scatter, it is important to notice that the tails in the distribution seen in fig.~\ref{fig:visual_auto} do not necessarily imply miss-classifications as we currently define them, i.e. galaxies which clearly fall in the wrong morphological class after visual inspection. As a matter of fact, a galaxy that might have a slightly larger bulge probability in the automated scheme than in the purely visual classification, will be however clearly classified as a disk since its probability is much higher. Figure~\ref{fig:max_f} shows the relation between the maximum visual frequency, defined as the maximum frequency irrespective of the morphology for each galaxy, and the maximum automatic frequency. Both quantities are correlated with the expected scatter with no tails even though there seems to be an increasing bias at low frequencies ($f_{max}<0.5$). This is not surprising since those are the most unclear objects of the visual catalog.

\begin{table*}
\begin{center}
\begin{tabular}{c c  c c c c}
\hline
        \hline
      
    \multicolumn{6}{c}{{\bf Test Sample}}\\
    \hline
& $0<f_{sph}<0.2$& $0.2<f_{sph}<0.4$& $0.4<f_{sph}<0.6$& $0.6<f_{sph}<0.8$& $0.8<f_{sph}<1.0$ \\
Bias & 0.03&-0.01& 0.00&-0.05&-0.10 \\
RMSE & 0.09& 0.15& 0.15& 0.17& 0.16 \\
Scatter & 0.07& 0.14& 0.14& 0.12& 0.09 \\
\hline
& $0<f_{disk}<0.2$& $0.2<f_{disk}<0.4$& $0.4<f_{disk}<0.6$& $0.6<f_{disk}<0.8$& $0.8<f_{disk}<1.0$ \\
Bias &-0.00& 0.11& 0.06& 0.06&-0.00 \\
RMSE & 0.09& 0.17& 0.16& 0.13& 0.09 \\
Scatter & 0.05& 0.17& 0.15& 0.10& 0.05 \\
\hline
& $0<f_{irr}<0.2$& $0.2<f_{irr}<0.4$& $0.4<f_{irr}<0.6$& $0.6<f_{irr}<0.8$& $0.8<f_{irr}<1.0$ \\
Bias & 0.01&-0.06&-0.10&-0.12&-0.14 \\
RMSE & 0.06& 0.13& 0.16& 0.20& 0.23 \\
Scatter & 0.05& 0.13& 0.15& 0.12& 0.12 \\
\hline
& $0<f_{PS}<0.2$& $0.2<f_{PS}<0.4$& $0.4<f_{PS}<0.6$& $0.6<f_{PS}<0.8$& $0.8<f_{PS}<1.0$ \\
Bias &-0.01&-0.11&-0.10&-0.04&-0.09 \\
RMSE & 0.04& 0.14& 0.21& 0.19& 0.16 \\
Scatter & 0.04& 0.15& 0.21& 0.15& 0.08 \\
\hline
& $0<f_{Unc}<0.2$& $0.2<f_{Unc}<0.4$& $0.4<f_{Unc}<0.6$& $0.6<f_{Unc}<0.8$& $0.8<f_{Unc}<1.0$ \\
Bias &-0.02&-0.17&-0.07& 0.19&-0.03 \\
RMSE & 0.03& 0.16& 0.12& 0.23& 0.09 \\
Scatter & 0.03& 0.21& 0.07& 0.22& 0.02 \\
  \hline
  \hline
  \multicolumn{6}{c}{{\bf Training sample}}\\
  \hline
& $0<f_{sph}<0.2$& $0.2<f_{sph}<0.4$& $0.4<f_{sph}<0.6$& $0.6<f_{sph}<0.8$& $0.8<f_{sph}<1.0$ \\
Bias & 0.03&-0.02&-0.02&-0.01&-0.07 \\
RMSE & 0.08& 0.13& 0.15& 0.13& 0.12 \\
Scatter & 0.06& 0.13& 0.13& 0.10& 0.07 \\
\hline
& $0<f_{disk}<0.2$& $0.2<f_{disk}<0.4$& $0.4<f_{disk}<0.6$& $0.6<f_{disk}<0.8$& $0.8<f_{disk}<1.0$ \\
Bias & 0.01& 0.07& 0.08& 0.05&-0.00 \\
RMSE & 0.09& 0.15& 0.14& 0.12& 0.08 \\
Scatter & 0.06& 0.13& 0.12& 0.09& 0.05 \\
\hline
& $0<f_{irr}<0.2$& $0.2<f_{irr}<0.4$& $0.4<f_{irr}<0.6$& $0.6<f_{irr}<0.8$& $0.8<f_{irr}<1.0$ \\
Bias & 0.00&-0.06&-0.08&-0.08&-0.11 \\
RMSE & 0.05& 0.12& 0.15& 0.16& 0.18 \\
Scatter & 0.05& 0.12& 0.13& 0.12& 0.10 \\
\hline
& $0<f_{PS}<0.2$& $0.2<f_{PS}<0.4$& $0.4<f_{PS}<0.6$& $0.6<f_{PS}<0.8$& $0.8<f_{PS}<1.0$ \\
Bias &-0.01&-0.11&-0.16&-0.07& 0.01 \\
RMSE & 0.04& 0.13& 0.18& 0.19& 0.13 \\
Scatter & 0.03& 0.15& 0.18& 0.14& 0.08 \\
\hline
& $0<f_{Unc}<0.2$& $0.2<f_{Unc}<0.4$& $0.4<f_{Unc}<0.6$& $0.6<f_{Unc}<0.8$& $0.8<f_{Unc}<1.0$ \\
Bias &-0.02&-0.10&-0.11&-0.01& 0.03 \\
RMSE & 0.03& 0.14& 0.19& 0.22& 0.22 \\
Scatter & 0.03& 0.14& 0.17& 0.15& 0.09 \\
\hline

\end{tabular}
\caption{Median bias ($\Delta_f=(f_{auto}-f_{visu})$), root mean square error (RMSE) and scatter as a function of the visual morphological frequencies for the test (top) and the training (bottom) sets.}
\label{tbl:tbl_bias_scatter}
\end{center}
\end{table*}

\begin{table*}
\begin{center}
\begin{tabular}{c c  c c}
\hline
        \hline
      
    \multicolumn{4}{c}{{\bf Test Sample}}\\
    \hline
Parameter & Bias & Scatter & RMSE\\
$f_{spheroid}$ & 0.03& 0.09 & 0.17\\
$f_{disk}$ & 0.03& 0.08 & 0.15\\
$f_{irr}$ &-0.01& 0.07 & 0.14\\
$f_{PS}$ &-0.01& 0.04 & 0.10\\
$f_{Unc}$ &-0.02& 0.03 & 0.07\\
{\bf ALL} & 0.00 & 0.05 & 0.13 \\

\hline
\hline
   \multicolumn{4}{c}{{\bf Training sample}}\\
    \hline
    Parameter & Bias & Scatter & RMSE\\
$f_{spheroid}$ & 0.02& 0.08 & 0.15\\
$f_{disk}$ & 0.02& 0.08 & 0.14\\
$f_{irr}$ &-0.01& 0.06 & 0.12\\
$f_{PS}$ &-0.01& 0.04 & 0.09\\
$f_{Unc}$ &-0.02& 0.03 & 0.05\\
{\bf ALL} & -0.01 & 0.05 & 0.12 \\
    \hline
\end{tabular}
\caption{Median bias ($\Delta_f=(f_{auto}-f_{visu})$) and scatter for each visual morphological frequency for the test and training samples.}
\label{tbl:tbl_bias_scatter_merged}
\end{center}
\end{table*}

We also explore in figure~\ref{fig:visual_auto_phys} how the performance of the classification depends on physical properties such as redshift, magnitude and size relative to the PSF FWHM. Interestingly, we do not observe any particular trend on the bias or the scatter with magnitude and redshift. The bias in the morphological fractions stays $<0.05$, and the scatter is rather constant at $0.1$ for all magnitudes and redshifts spanned by our sample. Only very small objects, close to the size of the PSF or very large ($>4$ times the PSF size), have a larger bias ($\sim0.05-0.1$). For large objects, this could be explained by the fact that part of the wings might be lost during the interpolation process at fixed size. Recall that this does not necessarily mean that the morphology can be assessed equally independently of brightness, redshift or size, but that the algorithm is able to reproduce the visual classification (with its eventual biases) with the same accuracy.

\begin{figure*}
\begin{center}
$\begin{array}{c c}
\includegraphics[width=0.40\textwidth]{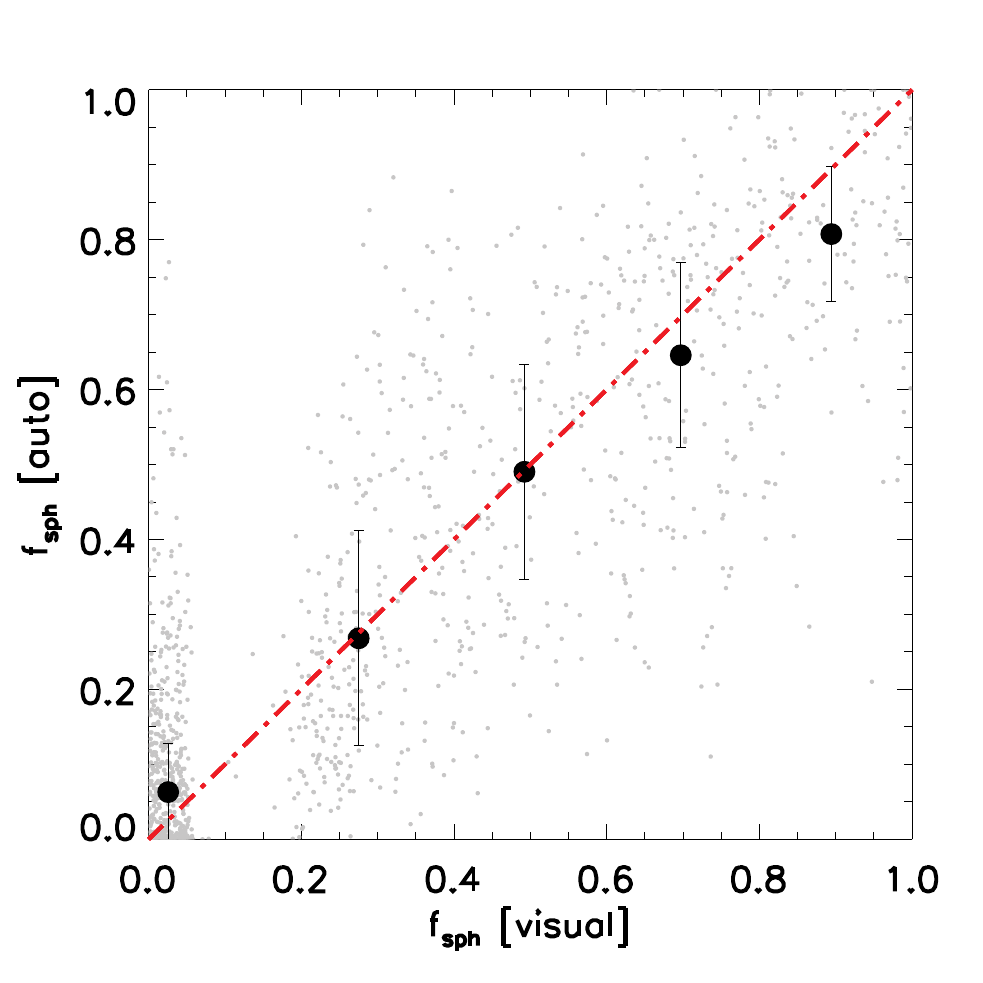} & \includegraphics[width=0.40\textwidth]{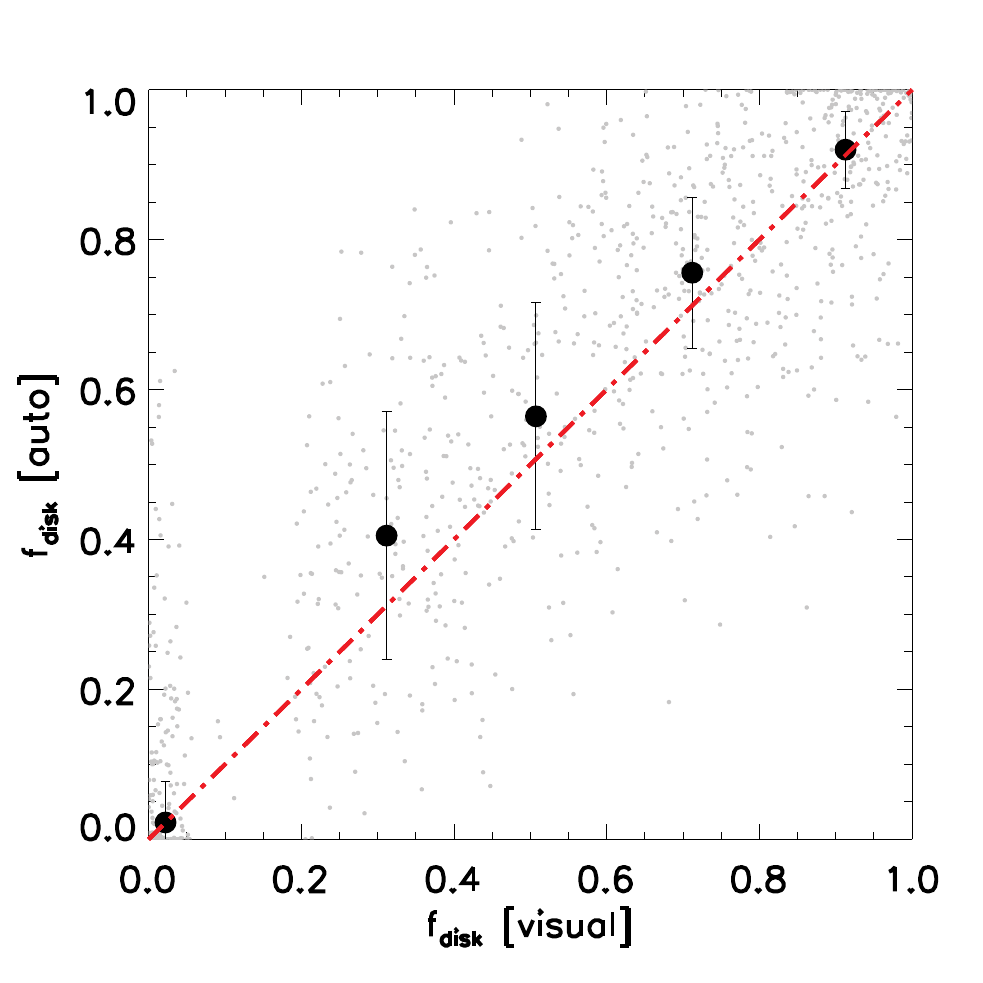}\\
\includegraphics[width=0.40\textwidth]{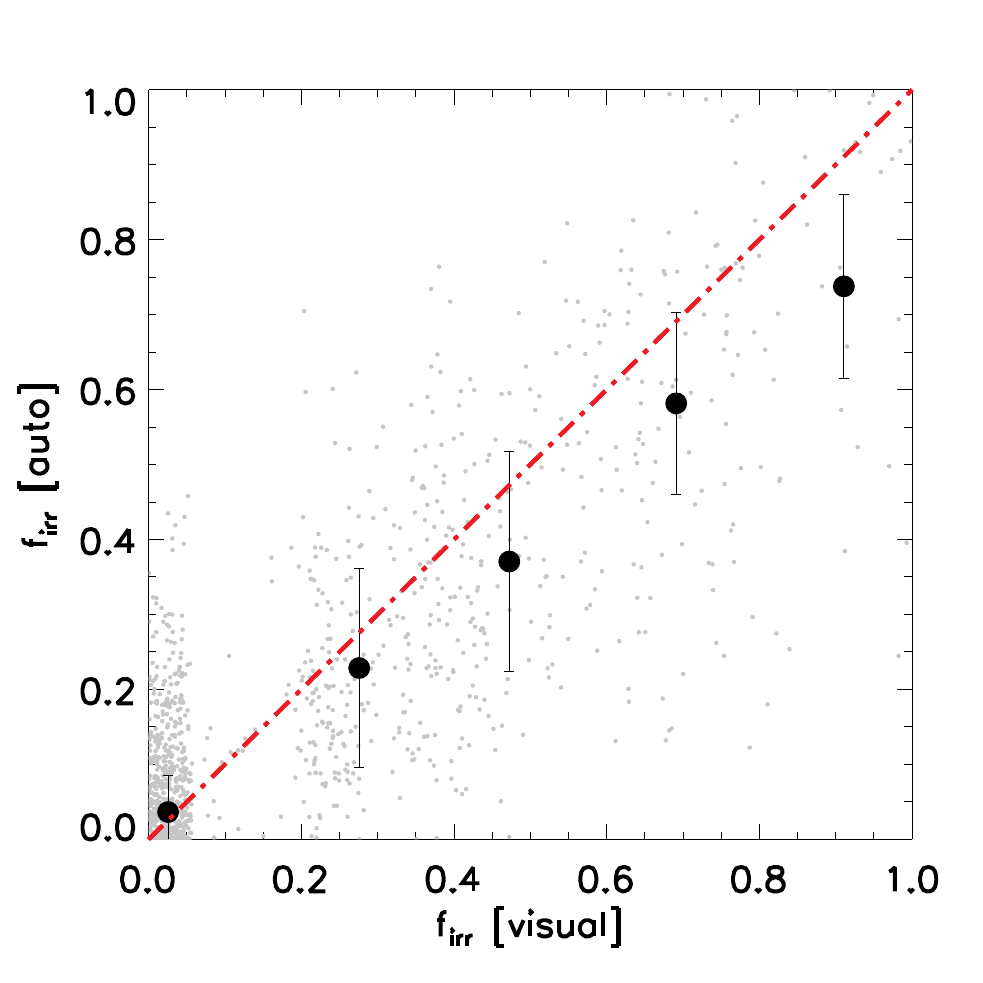} & \includegraphics[width=0.40\textwidth]{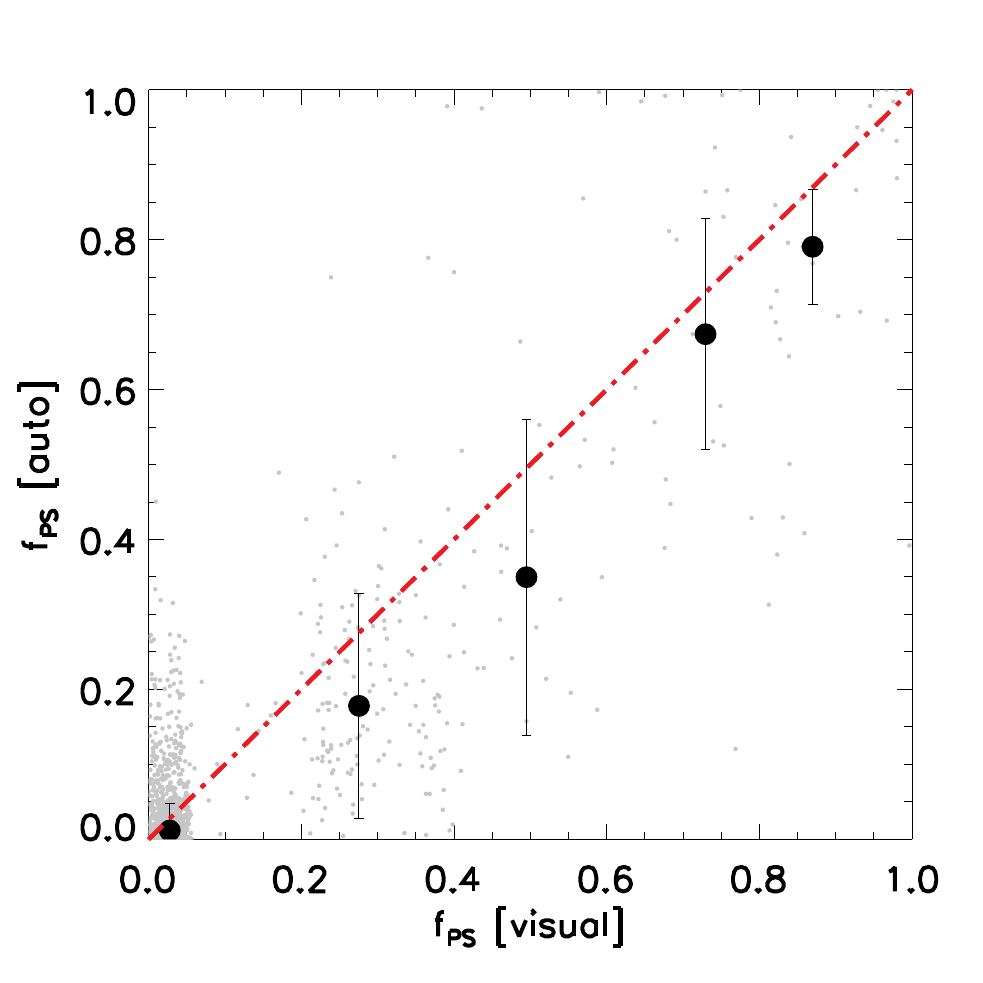}\\
\includegraphics[width=0.40\textwidth]{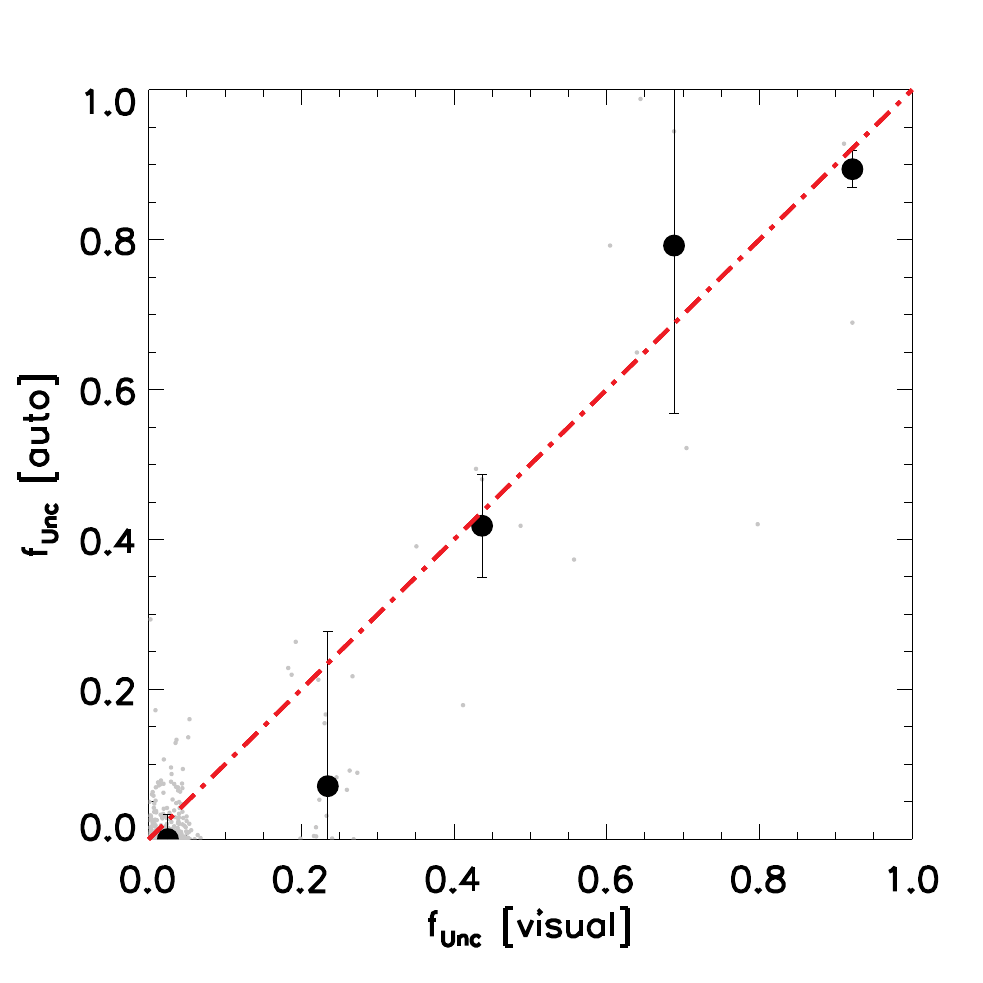} & \\
\end{array}$
\caption{Correlation between the fractions of classifiers voting for a given feature (spheroid (top left), disk (top right), irregular (middle left), point source (middle right) and unclassifiable (bottom left)) and the predictions of the ConvNet based classification on a test dataset. Detailed quantifications of the bias and the dispersion are shown in table~\protect\ref{tbl:tbl_bias_scatter}. } 
\label{fig:visual_auto}
\end{center}
\end{figure*}

\begin{figure*}
\begin{center}
$\begin{array}{c c}
\includegraphics[width=0.40\textwidth]{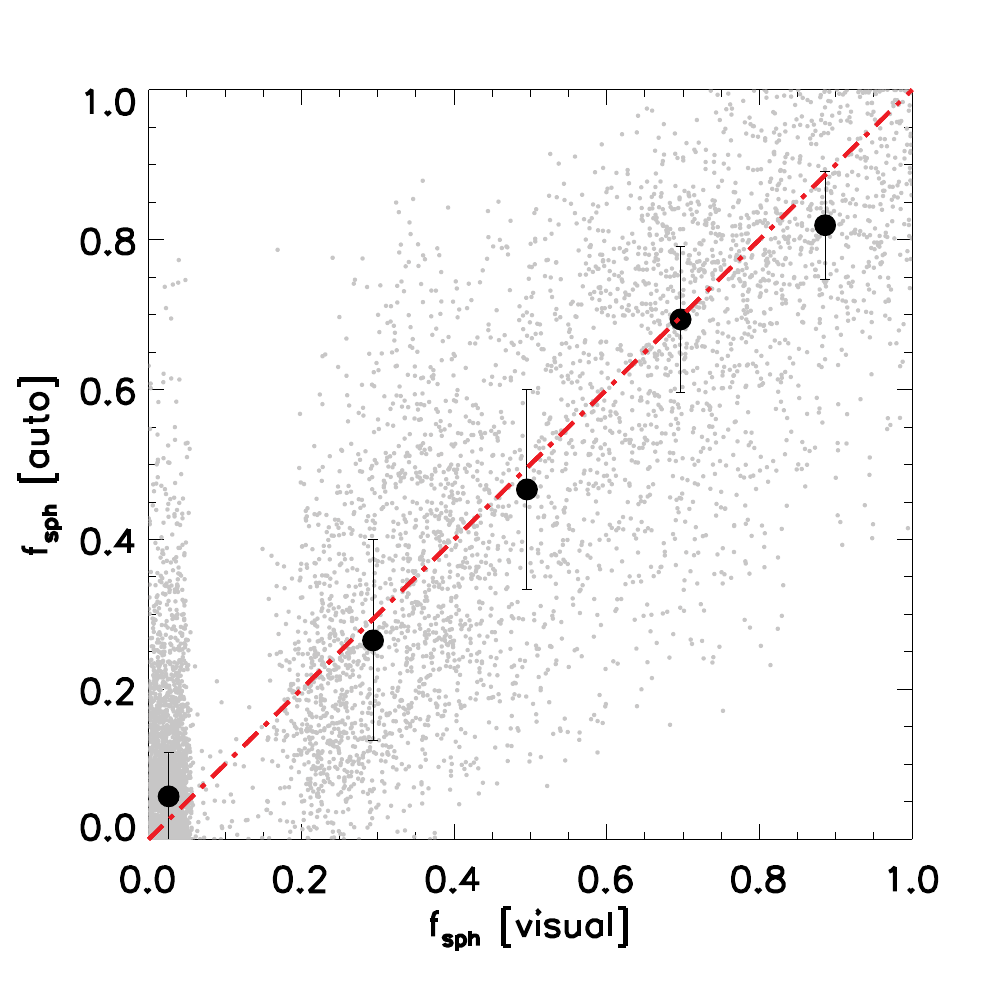} & \includegraphics[width=0.40\textwidth]{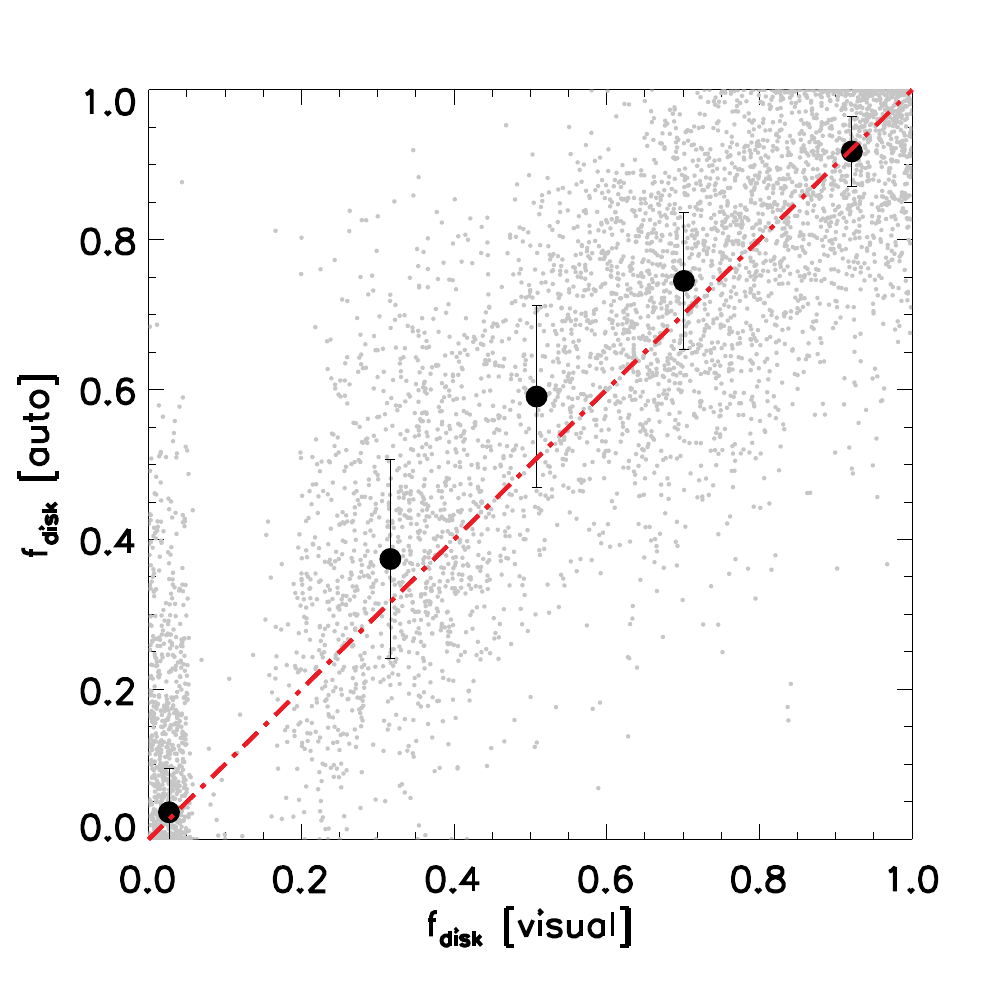}\\
\includegraphics[width=0.40\textwidth]{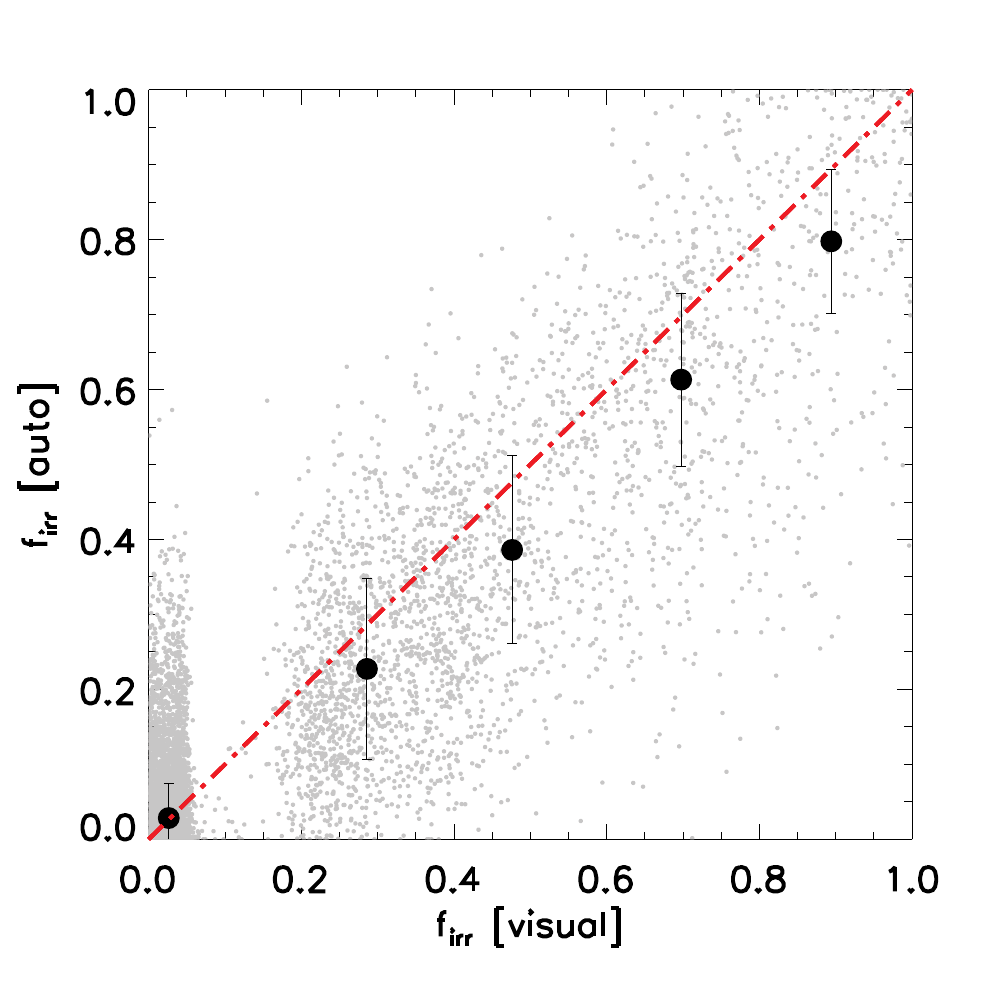} & \includegraphics[width=0.40\textwidth]{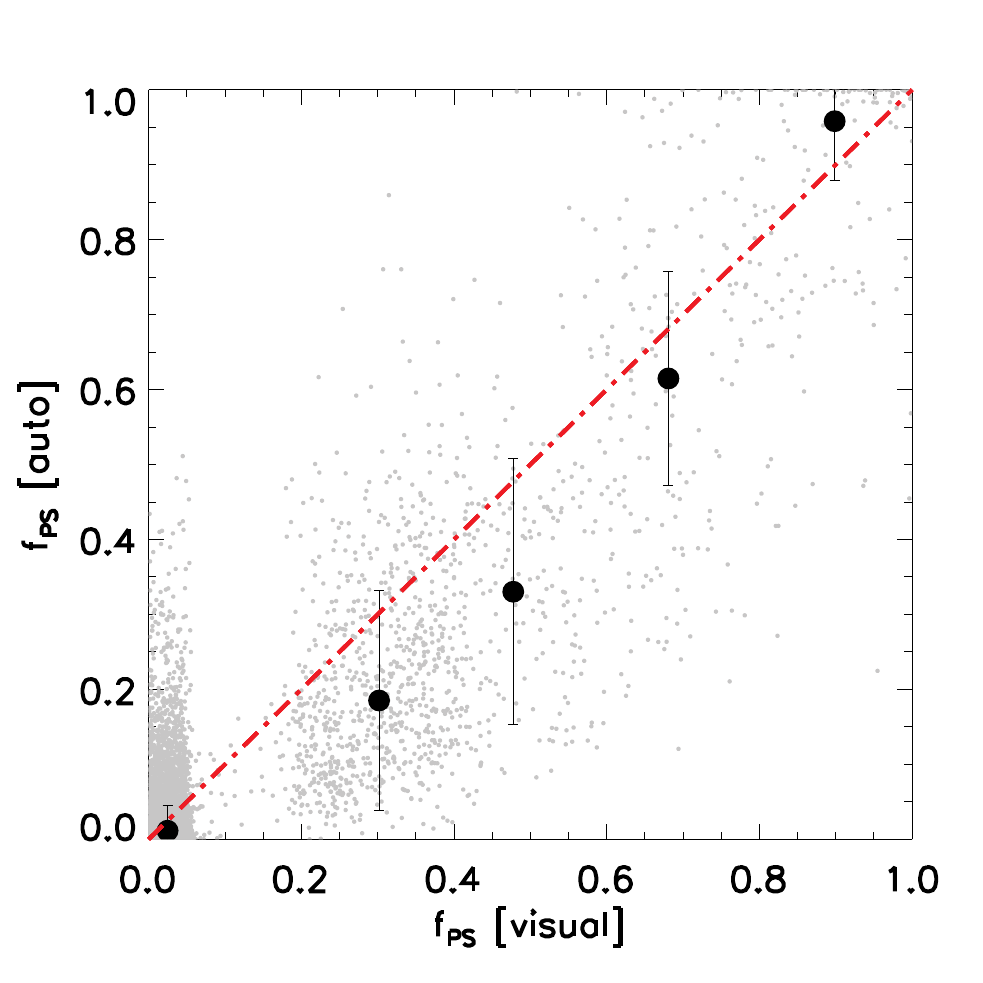}\\
\includegraphics[width=0.40\textwidth]{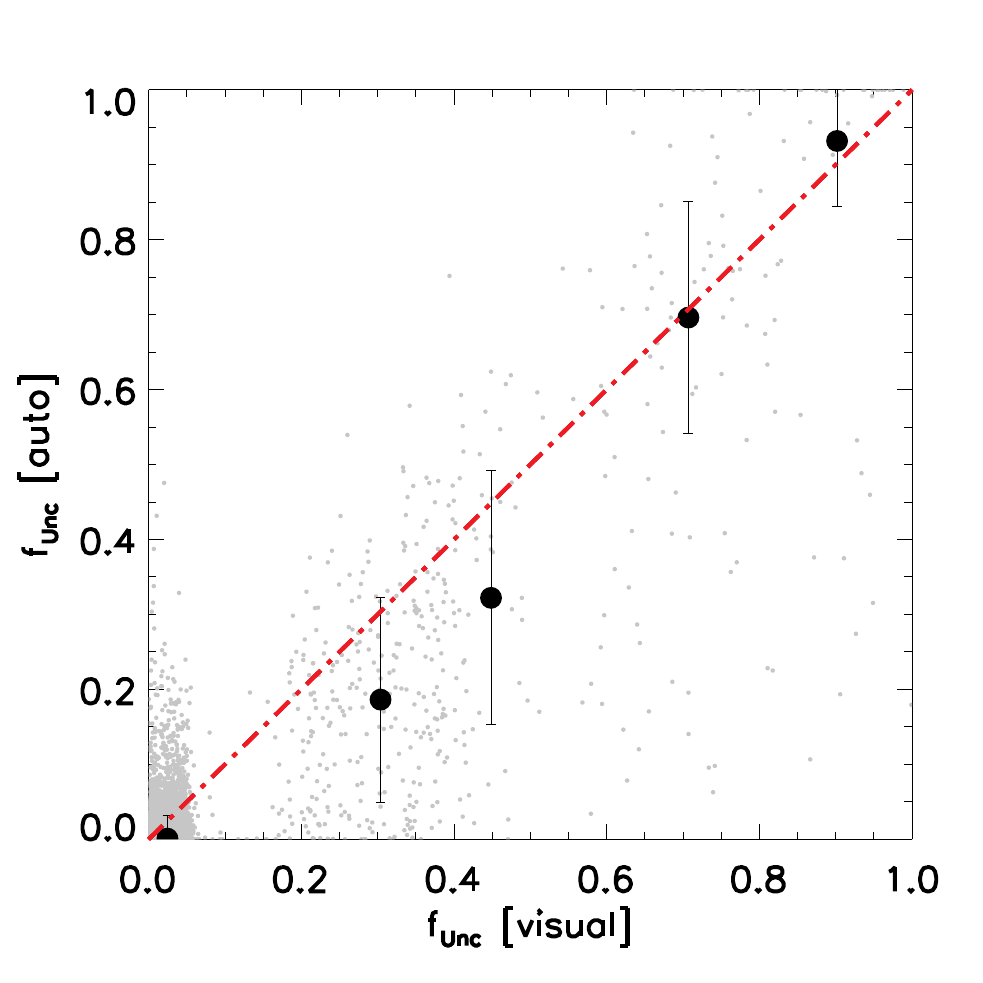} & \\
\end{array}$
\caption{Same as figure~\ref{fig:visual_auto} but for objects used for the training. } 
\label{fig:visual_auto_all}
\end{center}
\end{figure*}

\begin{figure*}
\begin{center}
\includegraphics[width=0.40\textwidth]{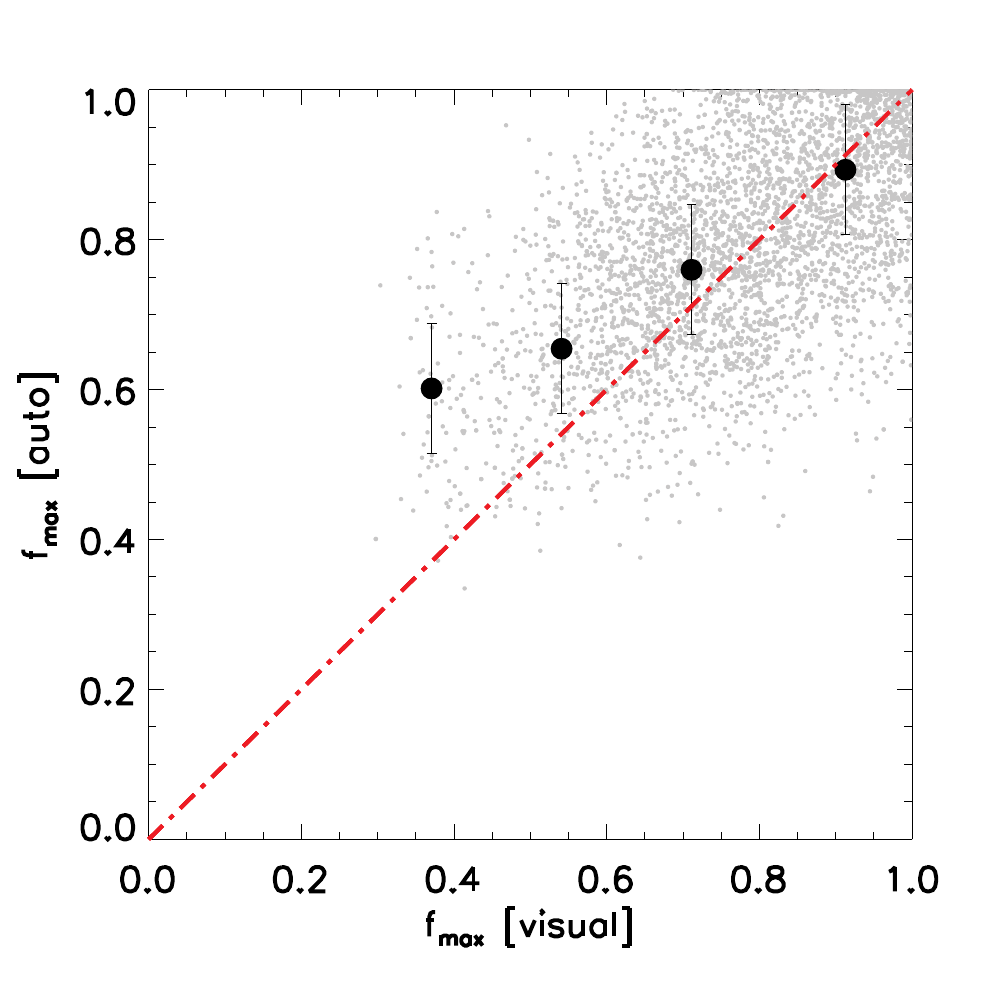}
\caption{Relation between the maximum fraction in the visual and the automatic classifications.}
\label{fig:max_f}
\end{center}
\end{figure*}

\begin{figure*}
\begin{center}
$\begin{array}{c c c}
\includegraphics[width=0.30\textwidth]{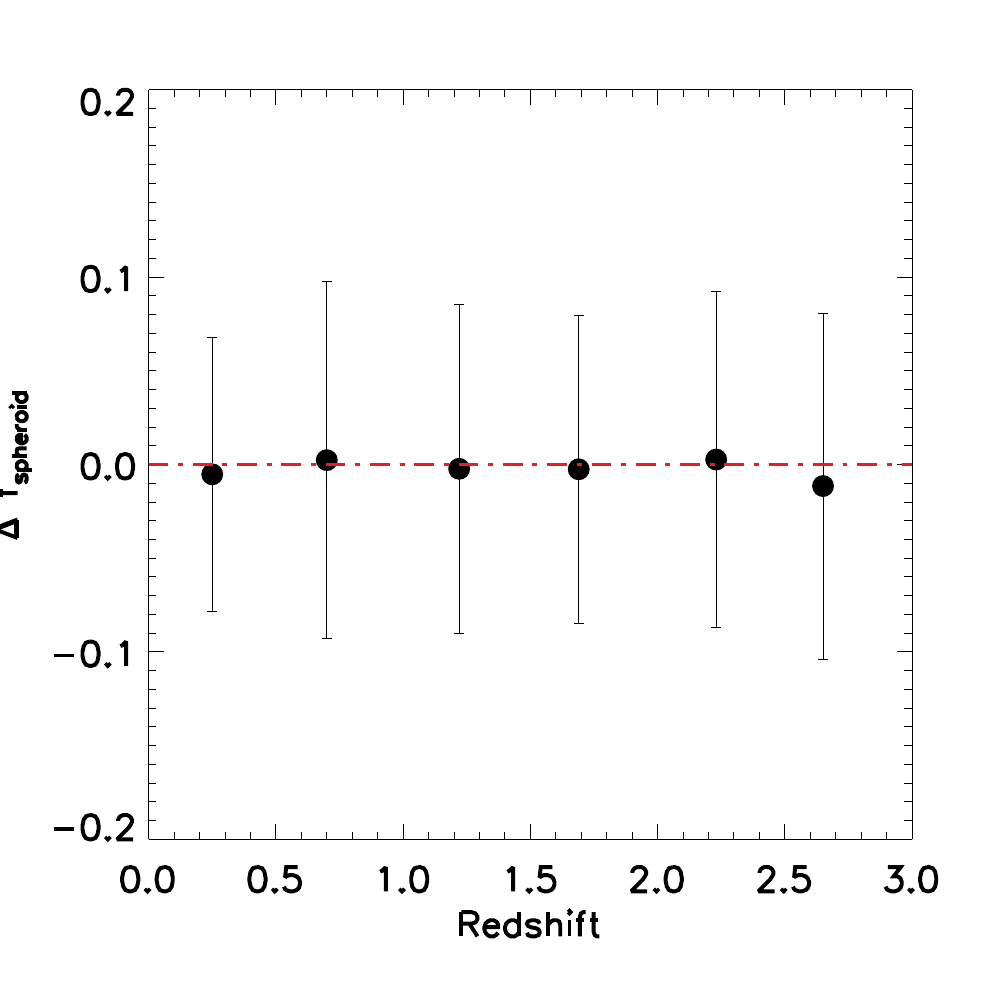} & \includegraphics[width=0.30\textwidth]{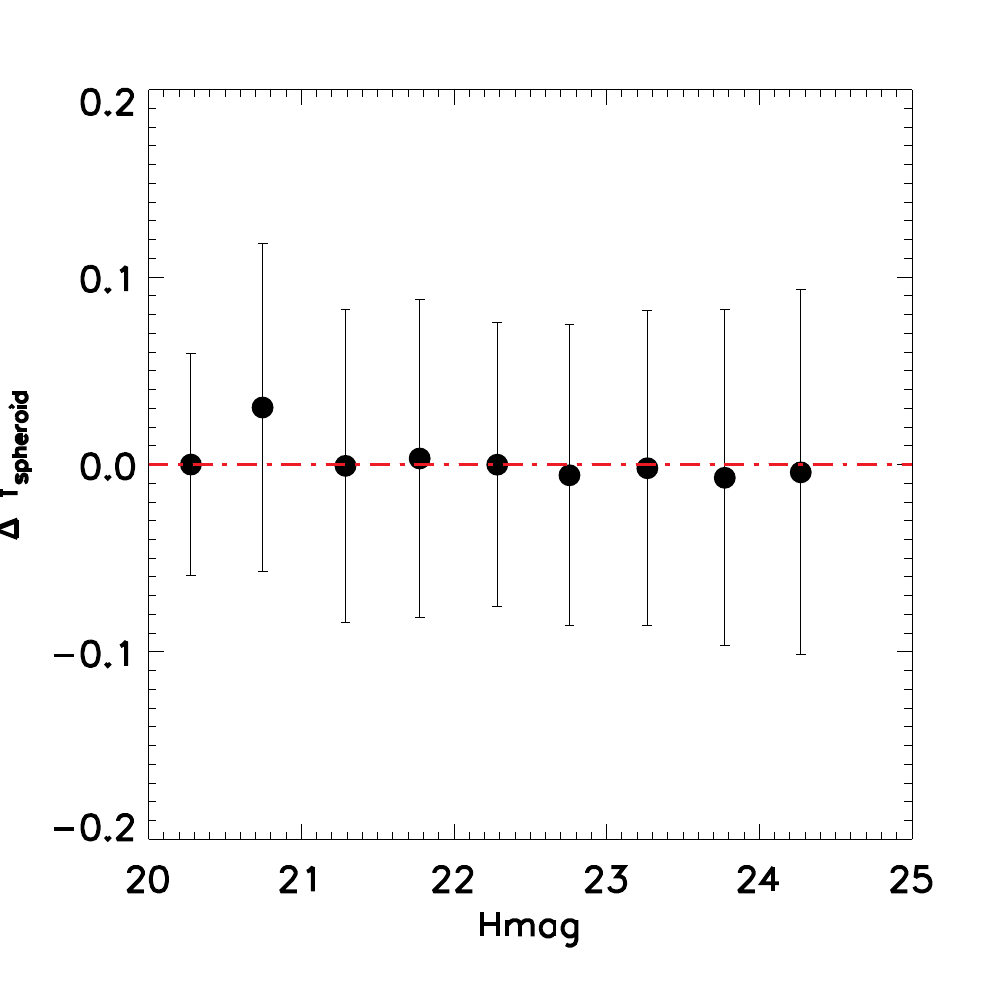} &  \includegraphics[width=0.30\textwidth]{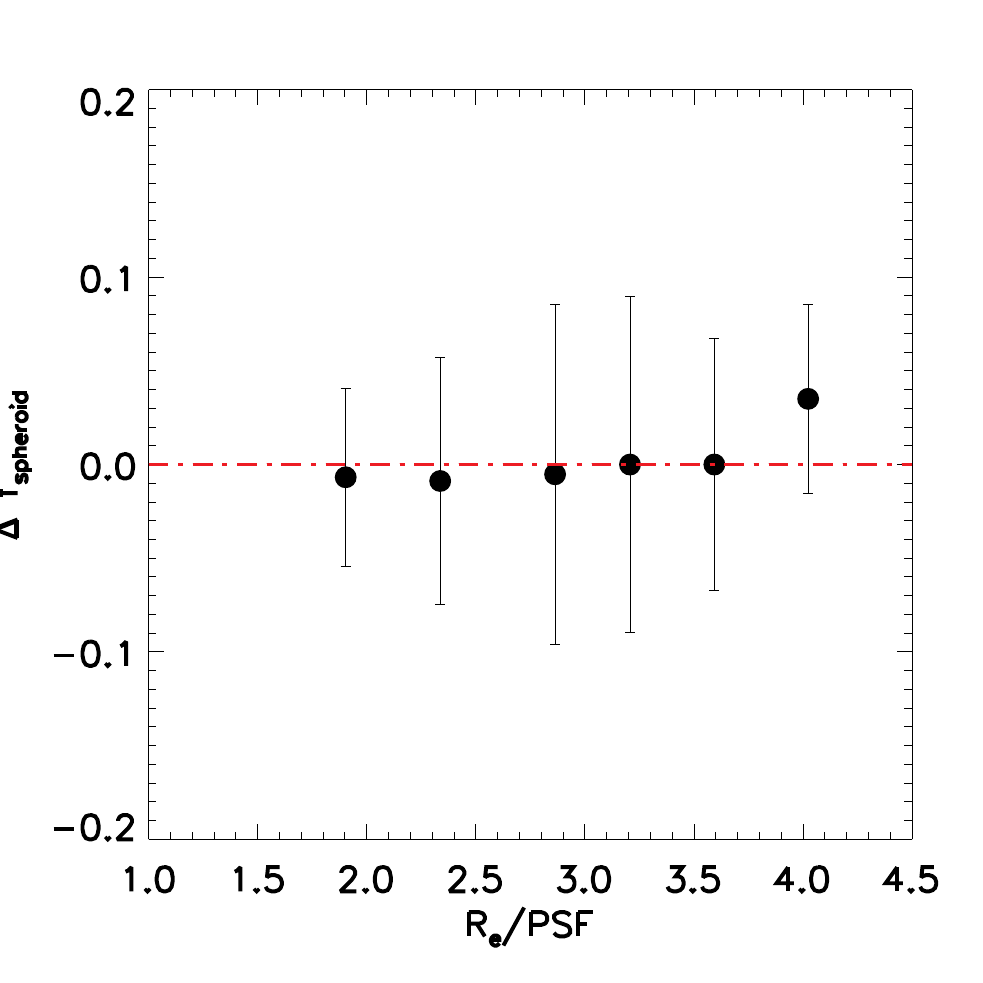} \\

\includegraphics[width=0.30\textwidth]{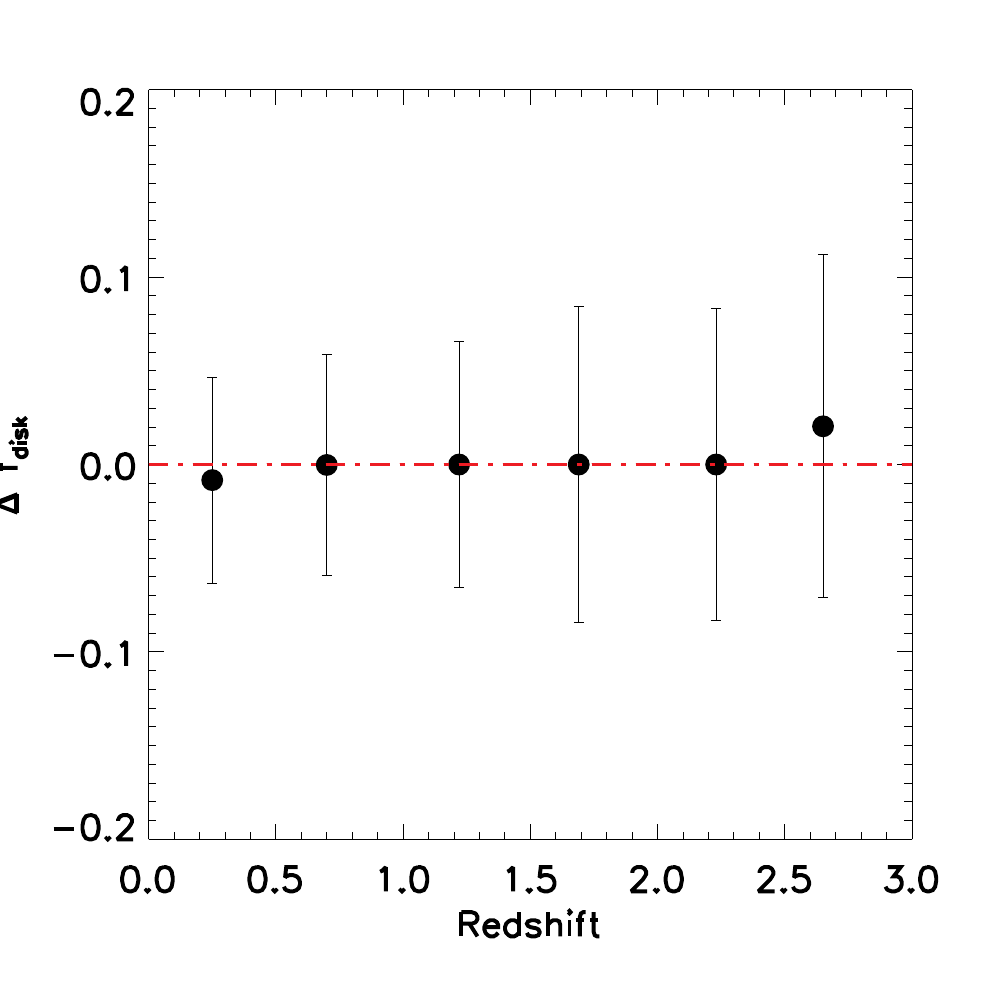} & \includegraphics[width=0.30\textwidth]{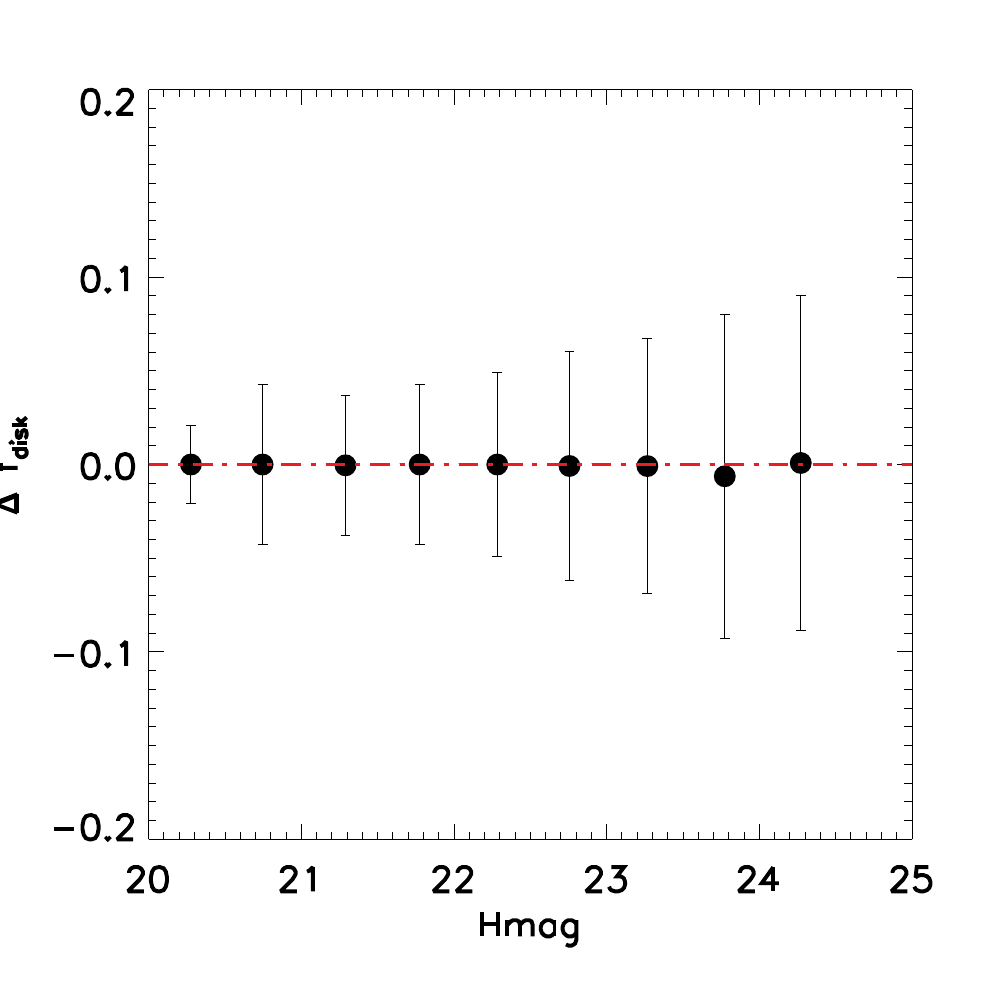} &  \includegraphics[width=0.30\textwidth]{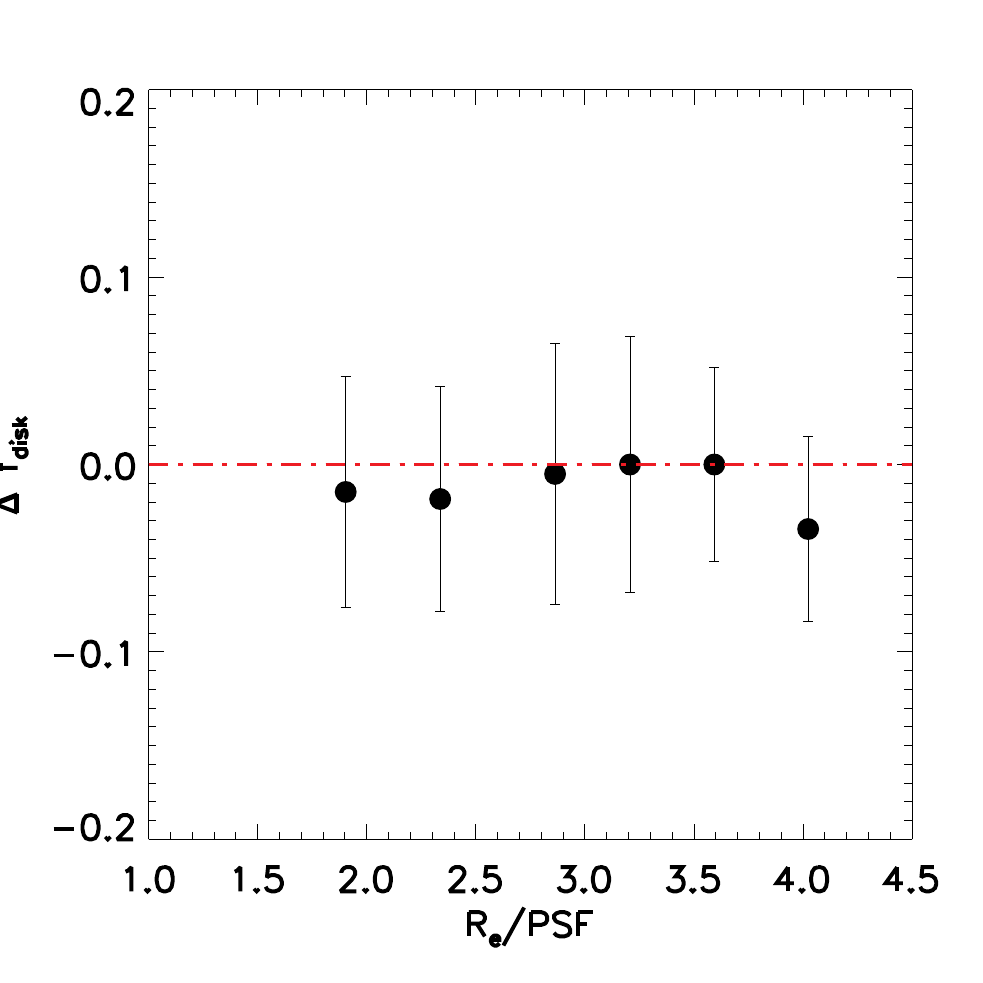} \\

\includegraphics[width=0.30\textwidth]{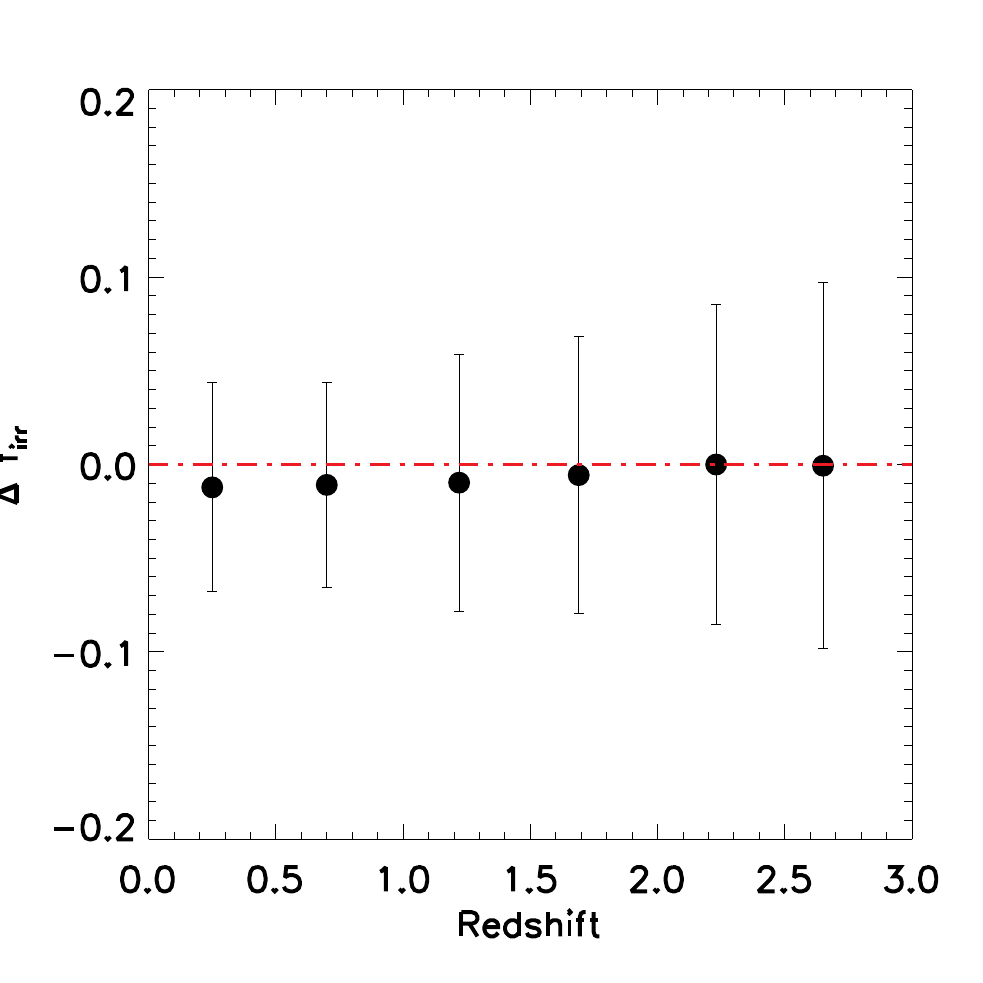} & \includegraphics[width=0.30\textwidth]{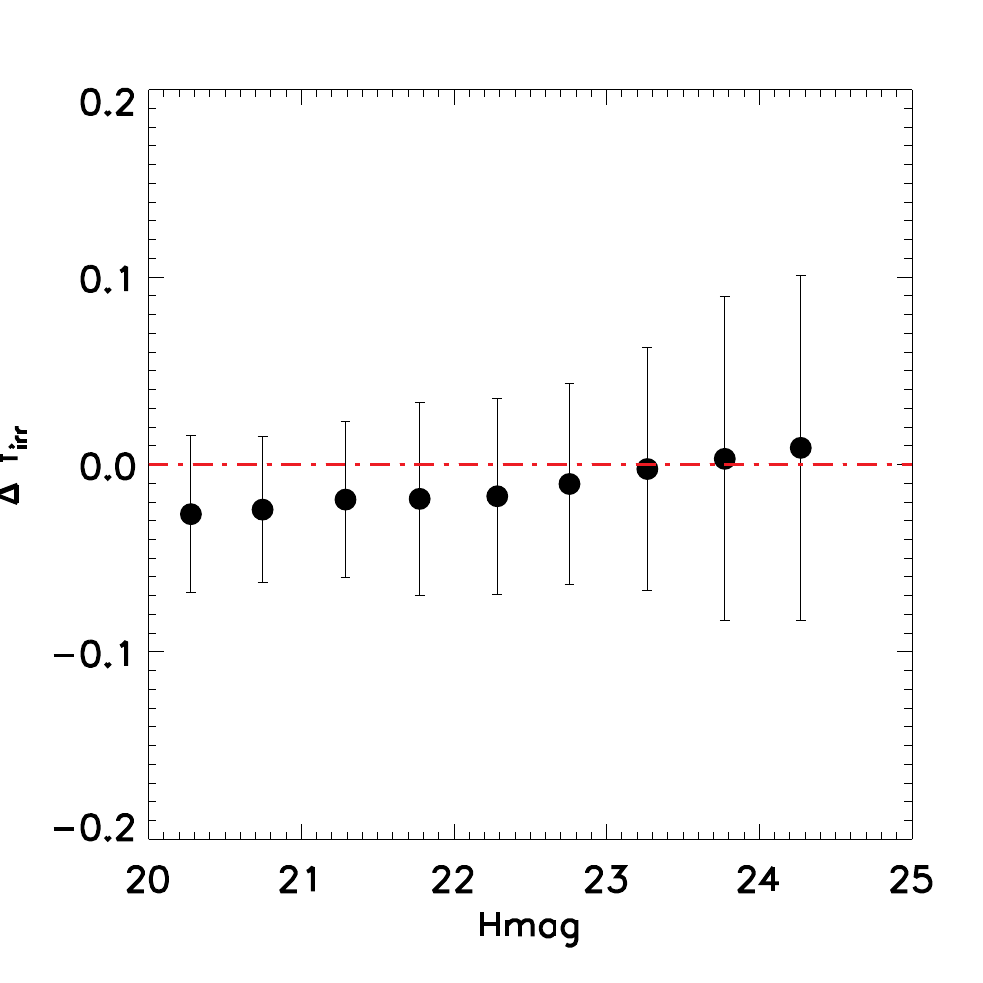} &  \includegraphics[width=0.30\textwidth]{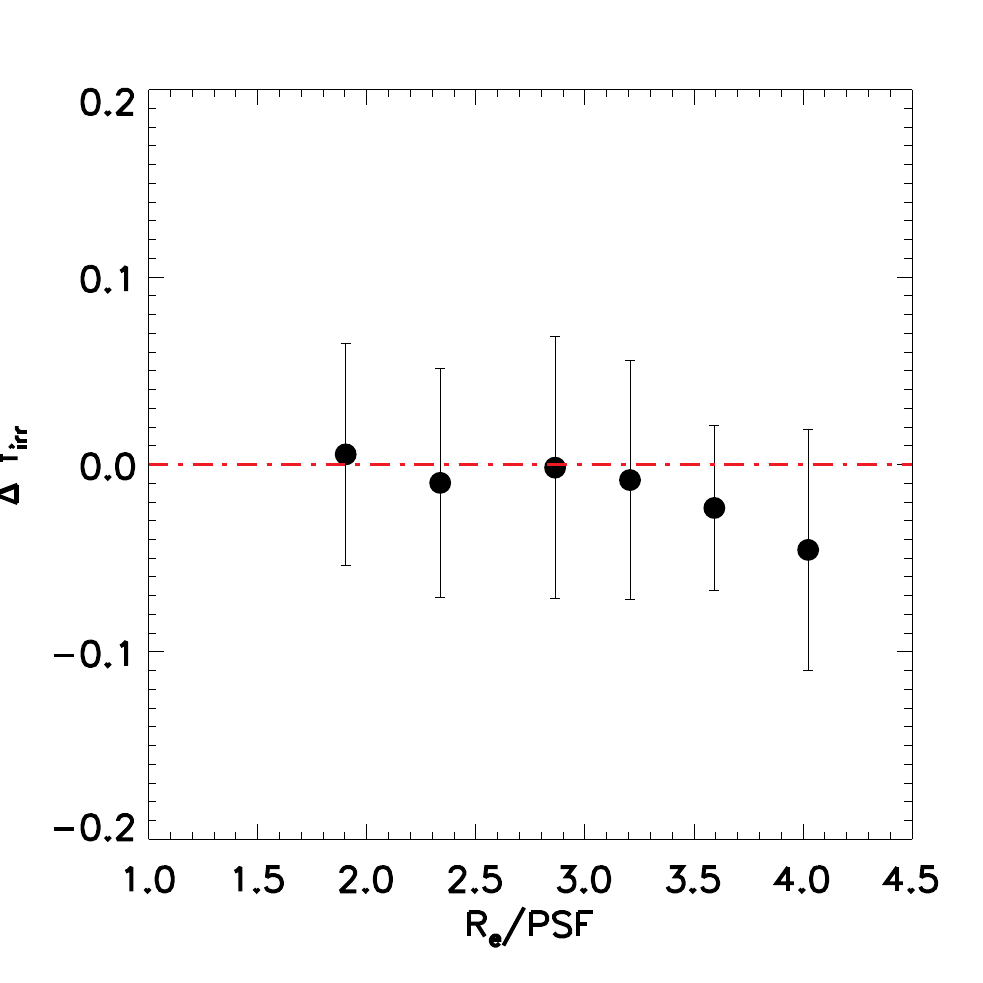} \\

\end{array}$
\caption{Mean Bias ($\Delta f=f_{auto}-f_{visu}$) and scatter ($\sqrt{VAR(\Delta _f)}$) of the three main morphological fractions (spheroid, disk and irregular from top to bottom) as a function of redshift, magnitude and resolution (from left to right). } 
\label{fig:visual_auto_phys}
\end{center}
\end{figure*}

\subsection{Recovering dominant classes and miss-classifications}
\label{sec:classes}

An important measurement in any automated classification scheme is the fraction of objects which are miss-classified, i.e. objects that will fall in a different morphological class in the automated classification compared to the visual one. Since both classifications are continuous in the sense that each galaxy has 5 real numbers associated to it, the answer to this question will strongly depend on the \emph{boxes} one considers and on how these boxes are defined. \\

In order to provide an estimate of this miss-classification rate that can be compared to previous classification methods, we select objects which do have a clear dominant class (DC) in the automatic and visual classifications. We define a galaxy with a dominant class if at least one frequency is considerably larger than the other 4. We then compare how both dominant classes match. \\
We adopt here a conservative offset value of 0.5 between the highest frequency and the second highest i.e. if $f_{max}>0.75$ then the second largest probability has to be smaller than 0.25, as a criterion to identify galaxies with a clear dominant morphology. There are therefore 5 dominant classes, i.e. \emph{dominant spheroid}, \emph{dominant disk}, \emph{dominant irregular}, \emph{dominant point source} and \emph{dominant unclear}. The results of such a comparison are shown in figure~\ref{fig:dom_comp_secure}. The degree of agreement in the identification of the main morphology of a galaxy is $\sim97-100\%$.  \\

\begin{figure*}
\begin{center}
\includegraphics[angle=0,scale=.8]{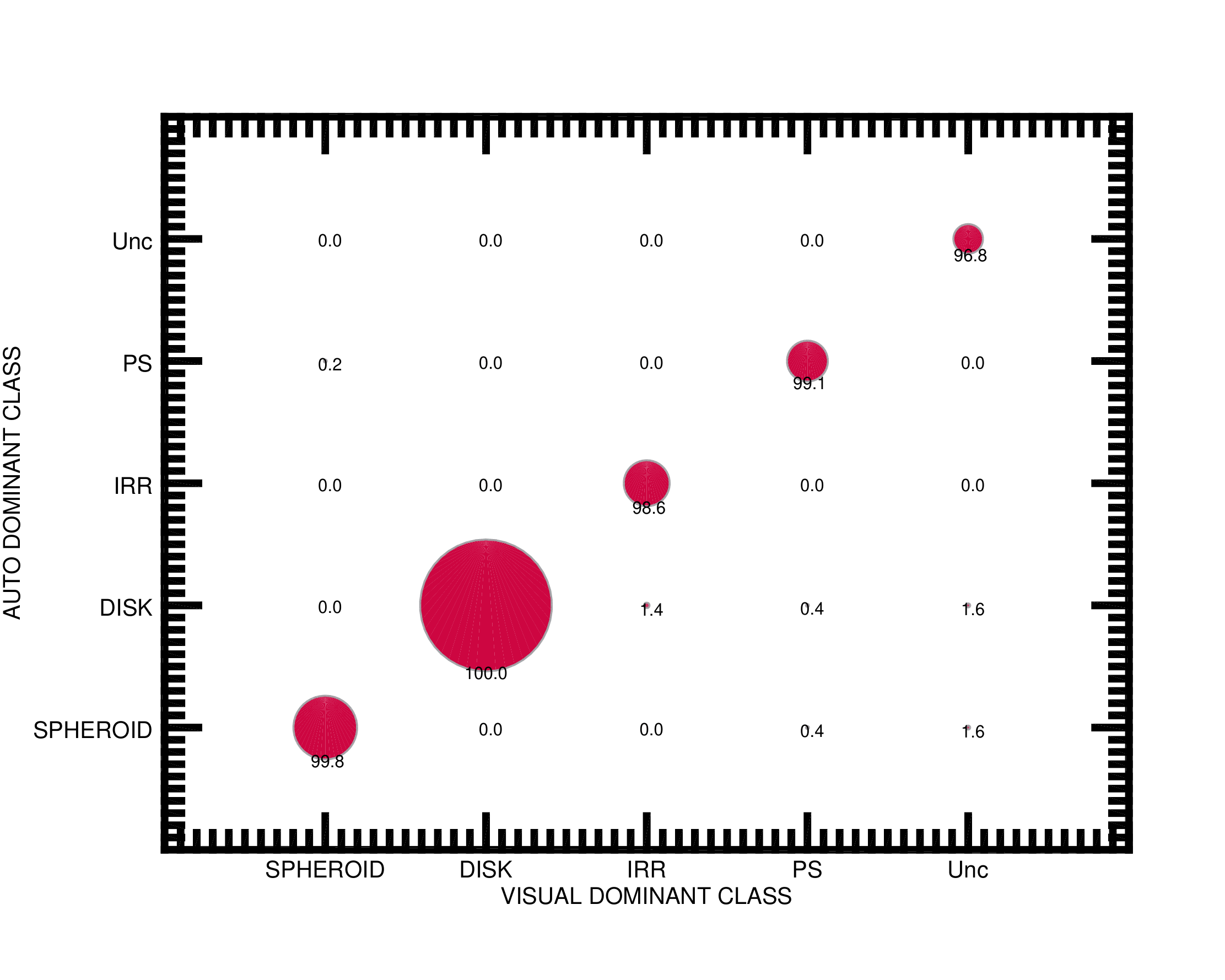}
\caption{Relation between the visual and automatic dominant morphological classes for well defined objects. The sizes of the symbols are proportional to the number of objects. The level of agreement is $>95\%$.  }
\label{fig:dom_comp_secure}
\end{center}
\end{figure*}

In a more general way, we can also investigate how the global classification accuracy depends on the level of agreement between the classifiers. As shown in \cite{D14} for the SDSS classification, objects for which a high number of people provided the same classification are better recovered than the ones that present a uniform distribution in their frequencies. This is simply reflecting the fact that galaxies that are not easily classified by humans are also hardly recovered by the classification model. Following the same approach as D15, we define the level of agreement $a$ between classifiers for a 5 class problem:
$$ a=1-H(f)/log(5)$$
where $H(f)$ is the entropy defined as:
\begin{equation}
\begin{split}
H(f)=&-f_{spheroid}log(f_{spheroid})\\
&-f_{disk}log(f_{disk})\\
&-f_{irr}log(f_{irr})\\
&-f_{PS}log(f_{PS})\\
&-f_{Unc}log(f_{Unc})\\
\end{split}
\end{equation}

The agreement parameter $a$ ranges between 0 and 1, with large values indicating high level of agreement (most of the classifiers selected the same class) and low values associated to objects with low levels of agreement (the votes are distributed uniformly between the different classes). \\
Figure~\ref{fig:entropy} reports the mean classification accuracy defined as the match between the automatic dominant class and the visual dominant class, as a function of $a$. The agreement parameter $a$ is computed using the automatic and visual classifications. As expected, the accuracy increases when the level of agreement increases. Well defined objects reach an accuracy $>90\%$ but it drops to $\sim50\%$ for galaxies with $a<0.2$. This behavior is very similar to the one reported in figure~9 of D15, which confirms the similar behavior of the classifier at high redshift. 

\begin{figure*}
\begin{center}
\includegraphics[angle=0,scale=.8]{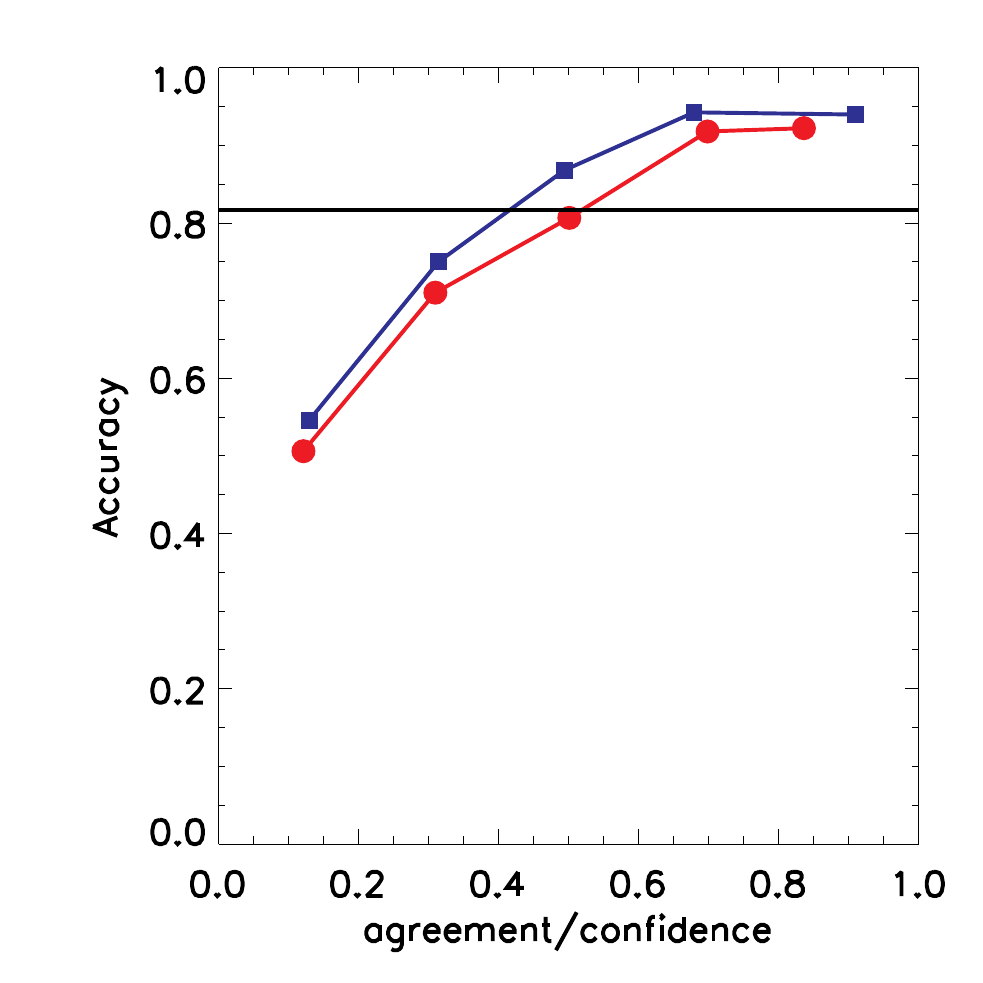}
\caption{  Classification accuracy as a function of the level of agreement between classifiers ($a$). The red line shows the relation when $a$ is computed using the visual classification. The blue line indicates the same relation but $a$ computed form the automated classification.  The horizontal line indicates the average accuracy. }
\label{fig:entropy}
\end{center}
\end{figure*}

The results above clearly represent a major step forward compared to other CAS-based methods. Firstly, CAS methods are not able to clearly distinguish between unclassifiable objects and galaxies since the morphological parameters for unclassifiable objects can have any unpredictable value. ConvNets identify them without ambiguity. \\
A similar issue affects point/compact sources which will usually fall in the early-type galaxy (ETG) class in CAS methods, unless a previous cleaning is performed. The most important thing is however that, even for the distinction of dominant spheroids from dominant disks, advanced CAS-based methods such as galSVM do show a tail of dominant disks with high ETG probability and vice-versa (fig.~\ref{fig:DC_galSVM}) yielding to a $\sim20\%$ miss-classification rate \citep{2014arXiv1406.1175H}. The situation is more dramatic for the distinction of dominant irregulars from dominant disks. It is almost impossible with CAS-based approaches, given that at high redshift many of the disks present high asymmetric values \citep{2014arXiv1406.1175H}. This is clearly shown in the right panel of figure~\ref{fig:DC_galSVM} where dominant disks have a very wide irregular probability distribution. ConvNets here do provide a huge improvement by perfectly separating both classes. \\

Figure~\ref{fig:dom_stamps} shows some example stamps of these 5 DCs selected in the COSMOS field where no visual morphologies are available. Objects are fully randomly selected. Clearly the visual aspect of all objects matches the dominant class in which they fall in the ConvNet classification, confirming the low miss-classification rate estimated in figure~\ref{fig:dom_comp_secure} for GOODS-S.

\begin{figure*}
\begin{center}
$\begin{array}{c c}
\includegraphics[width=0.40\textwidth]{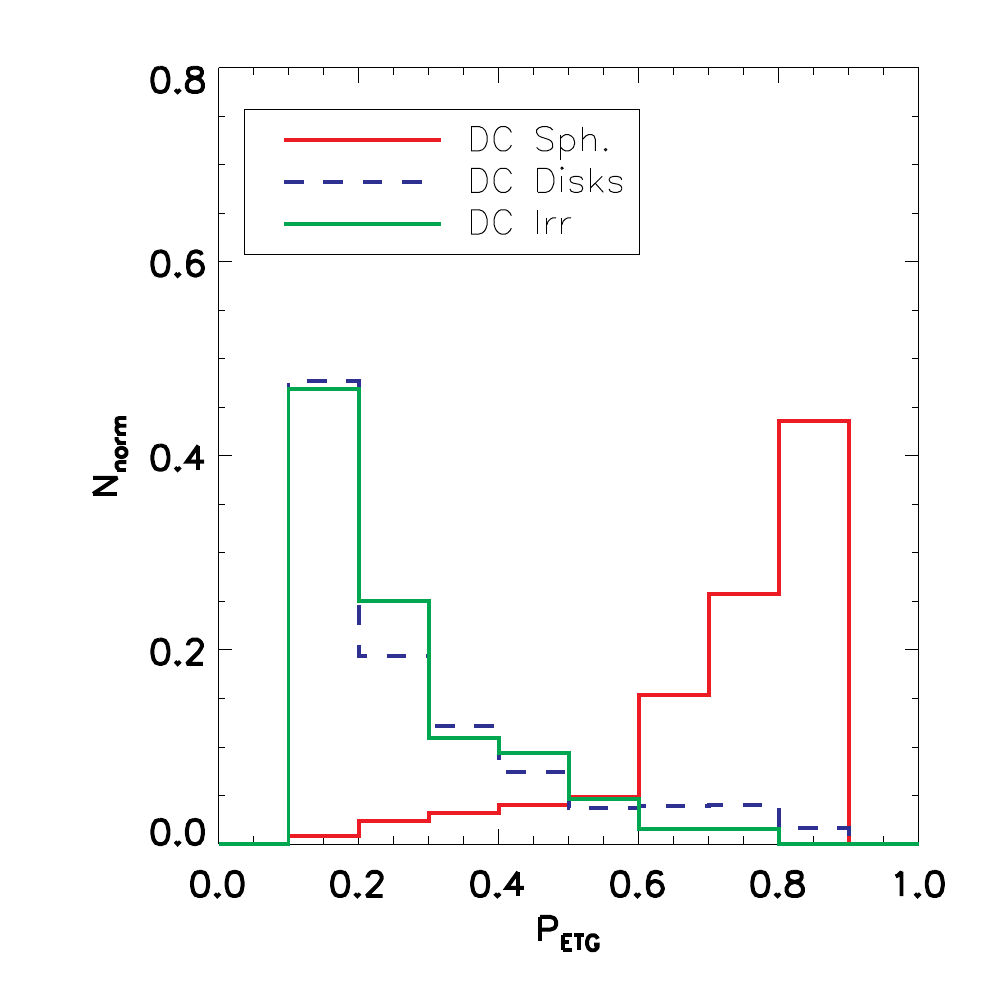} & \includegraphics[width=0.40\textwidth]{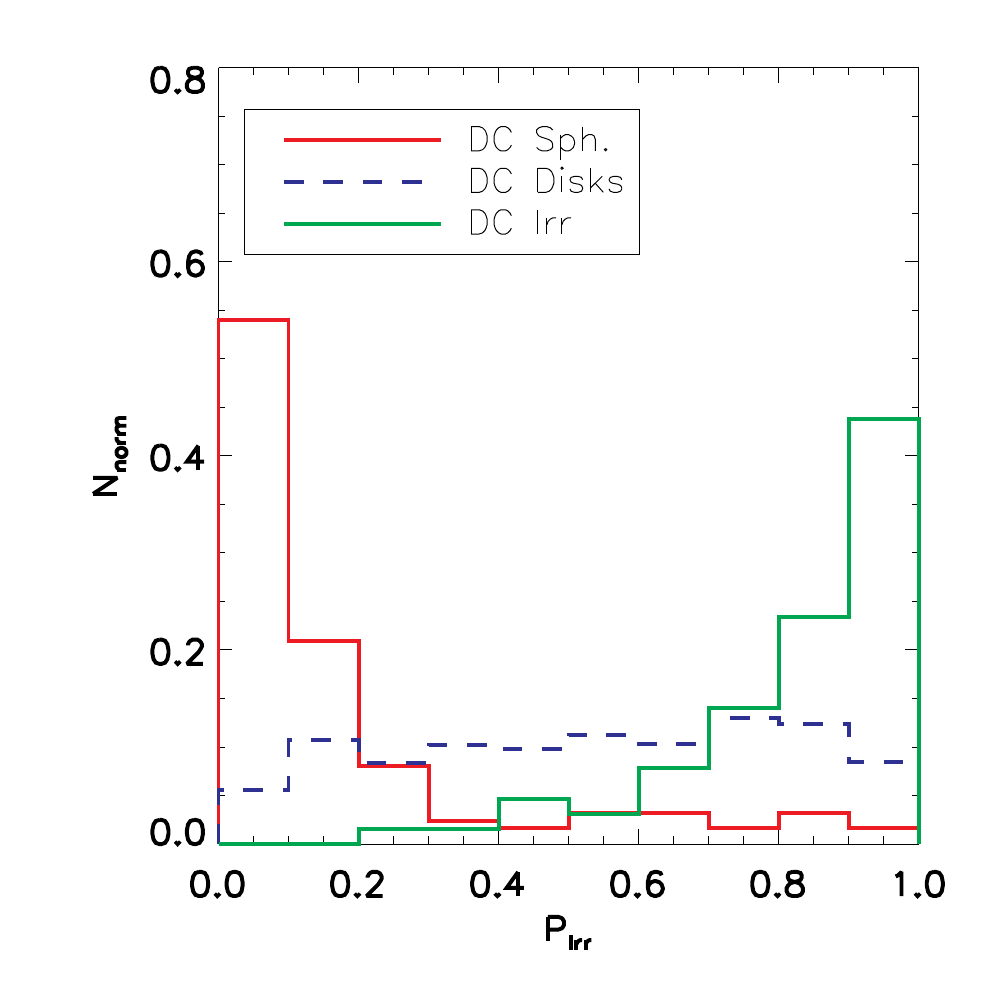}\\

\end{array}$
\caption{Probability distributions of being early-type (left panel) and irregular (right panel) estimated by galSVM (see~\protect\citealp{2014arXiv1406.1175H}) for three dominant classes in the CANDELS visual classification as labelled. Dominant disks cannot reliably separated from dominant irregulars using this approach.} 
\label{fig:DC_galSVM}
\end{center}
\end{figure*}

\begin{figure*}
\begin{center}
\includegraphics[width=0.99\textwidth]{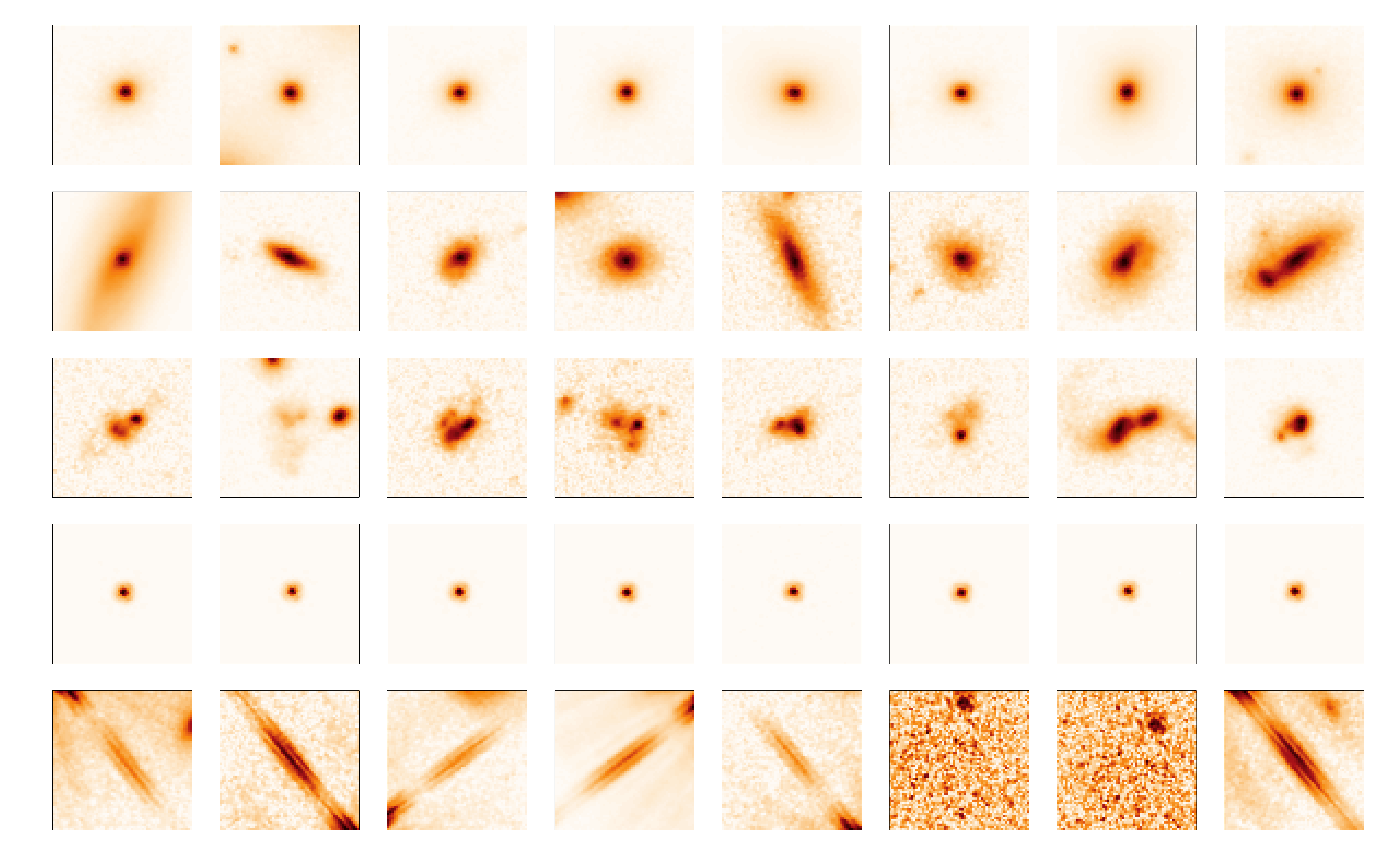}
\caption{Examples stamps of the 5 \emph{dominant} morphological classes in the COSMOS/CANDELS field. From top to bottom we show dominant spheroids, dominant disks, dominant irregulars, dominant point/sources-compact and dominant unclassifiable. The selection of these stamps is done fully randomly. Recall that COSMOS galaxies have not been used for training the algorithm, therefore they are completely new for the best model. The size of the stamps is $3.8^{"}\times3.8^{"}$. }
\label{fig:dom_stamps}
\end{center}
\end{figure*}

\subsection{Secondary classes - multi-component objects}

Also important are the galaxies that have a composition of different structures. We use 2 parameters to identify these objects, which are simply the value of the maximum frequency  ($f_{max}$) and the difference between the largest and the second largest frequency ($\Delta_{f1-f2}$). A galaxy with a fairly high $f_{max}$ value and a low $\Delta_{f1-f2}$ should be a galaxy with two clear components. For the purpose of this test, we define these galaxies as the ones that have $f_{max}> 0.5$ and a $\Delta_{f1-f2}< 0.5$. 

We then look for the three different possible combinations of primary and secondary classes (Disk+Spheroid (DS), Disk+Irregular (DI), Spheroid+Irregular (SI). Figure~\ref{fig:comp_sec} shows the relation between the 3 defined 2-component classes from the visual and the automatic classifications. The agreement is again close to $95\%$ for DSs and DIs which means that the algorithm is not only able to identify the primary class but also the secondary one, whenever the galaxy has two clear morphological components. The agreement for the SI class is poor. However this is a very marginal class since very few objects have both a dominant bulge with an irregular structure. They are usually associated to bulges with some kind of structure in the surroundings in the automatic classification (fig.~\ref{fig:stamps_sec}).

\begin{figure*}
\begin{center}
\includegraphics[angle=0,scale=.8]{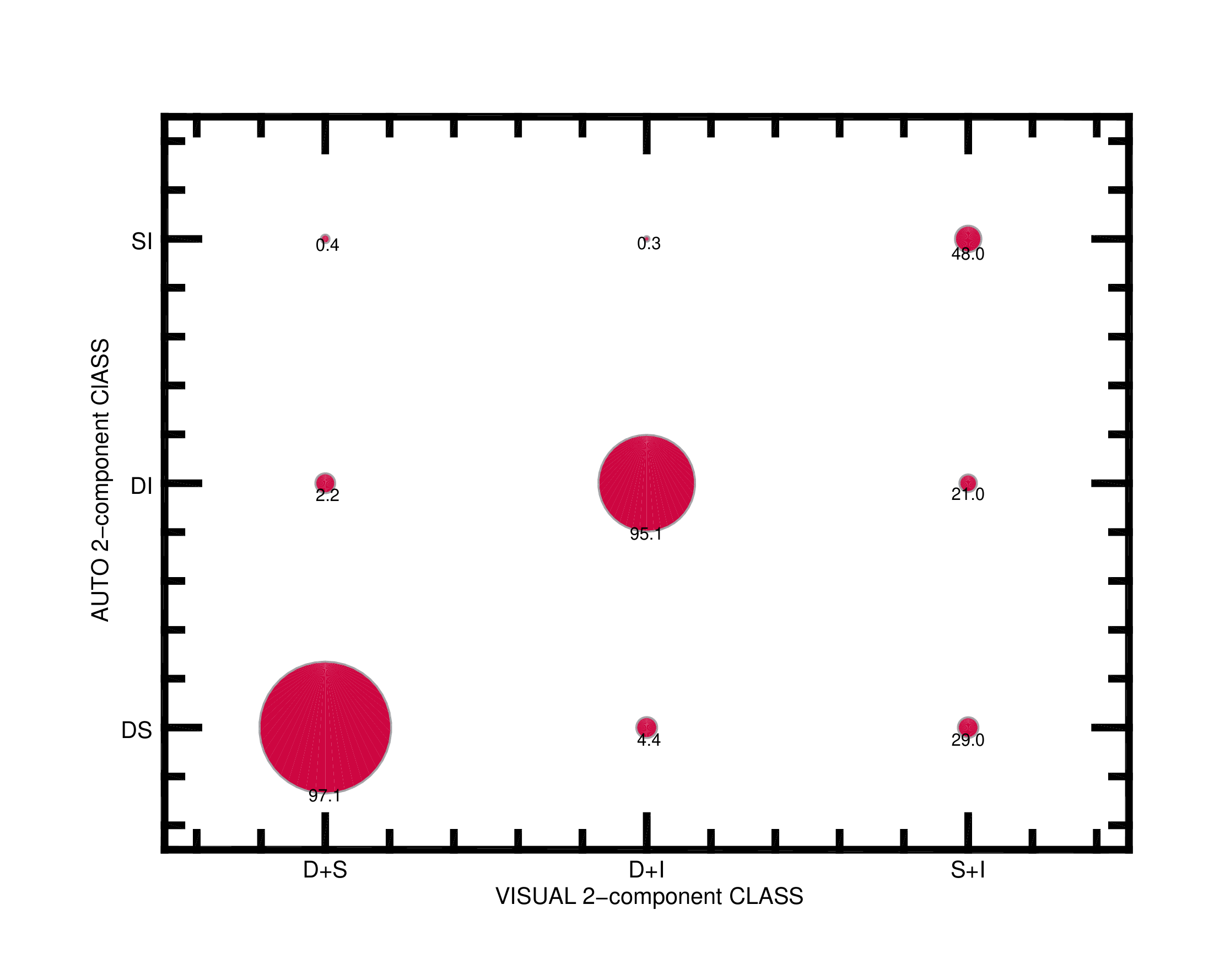}
\caption{Relation between the visual and automatic 2 component classes. The level of agreement is $>95\%$.  }
\label{fig:comp_sec}
\end{center}
\end{figure*}

\begin{figure*}
\begin{center}
\includegraphics[angle=0,width=.99\textwidth]{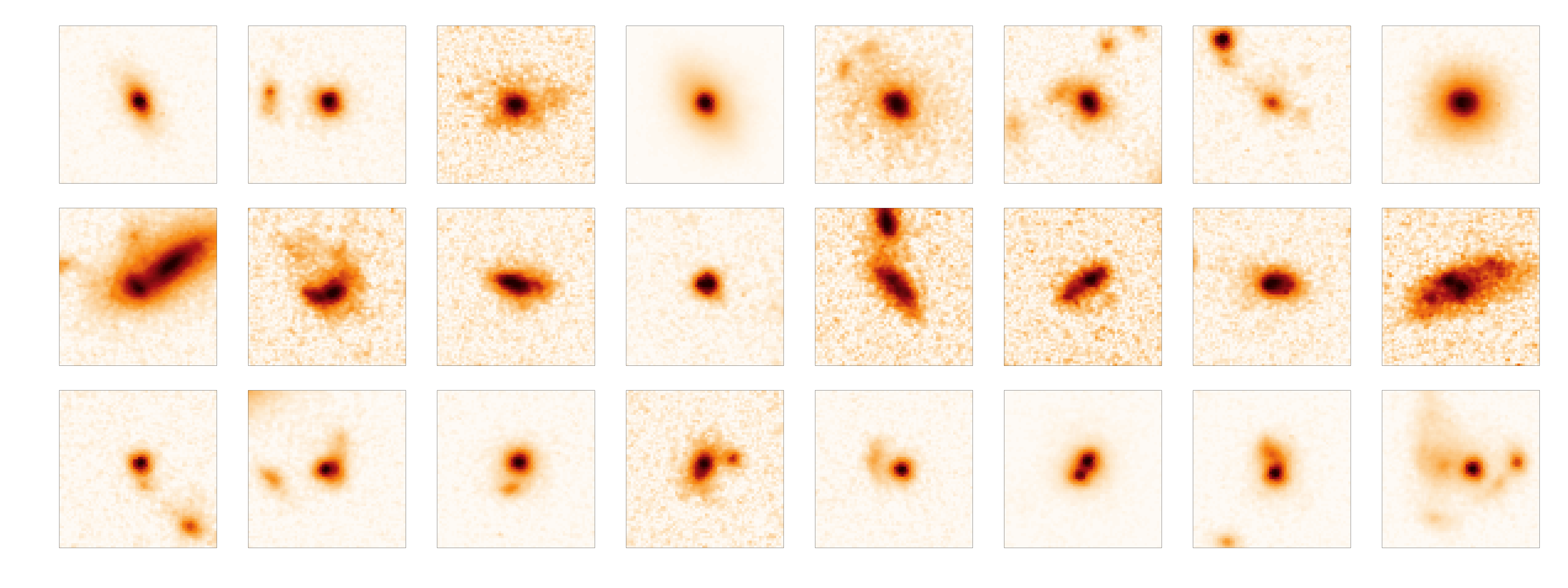}
\caption{Examples stamps of objects with two main morphological classes in the COSMOS/CANDELS field. From top to bottom we show spheroids+disks, disks+irregular, spheroids+irregular. The selection of these stamps is done fully randomly. Recall that COSMOS galaxies have not been used for training the algorithm, therefore they are completely new for the best model. The size of the stamps is $3.8^{"}\times3.8^{"}$.}
\label{fig:stamps_sec}
\end{center}
\end{figure*}

\subsection{Uncertain objects - Limitations}

 A galaxy with none of the 5 associated frequencies large enough (none of the available flags was clearly selected by the majority of the classifiers) should correspond to an object which has an uncertain morphology. The identification of these objects can help in understanding the limits of the morphological classification.  \\
 Figure~\ref{fig:uncertain} shows how the fraction of uncertain objects changes with magnitude, redshift and stellar mass for different $f_{max}$ thresholds, starting at  $f_{max}<0.4$ and finishing at  $f_{max}<0.7$, i.e. objects for which their maximum frequency is less than 0.4 and 0.7 respectively. 
 
 The number of objects with $f_{max}$ lower than 0.4-0.5 is very small ($<5\%$) for both the visual and automatic classifications which reflects the fact that the magnitude limit imposed ($H<24.5$) allows to identify a main morphology in most of the cases. \\
 When the threshold is increased, the expected trends are observed, i.e. the number of defined \emph{uncertain} objects increases with magnitude, redshift and is also higher for lower stellar masses. Interestingly, the trends are very similar for the visual and automatic morphologies. The automated classification is therefore reproducing the same uncertainties than the human eye encounters when classifying a galaxy. 

 In the bottom row of figure~\ref{fig:uncertain}, we also show the median value of $a$, the level of agreement between classifiers, in bins of magnitude, redshift and stellar mass. The level of agreement of the classification decreases for faint, distant and low mass objects as expected. The strongest correlation is however with magnitude indicating that that the main limitation to properly classify a galaxy is the signal-to-noise-ratio. Notice also that the median level of agreement is always $>0.4$ which according to figure~\ref{fig:entropy}, corresponds to an accuracy $>80\%$ for all objects.

\begin{figure*}
\begin{center}
$\begin{array}{c c c}
\includegraphics[width=0.30\textwidth]{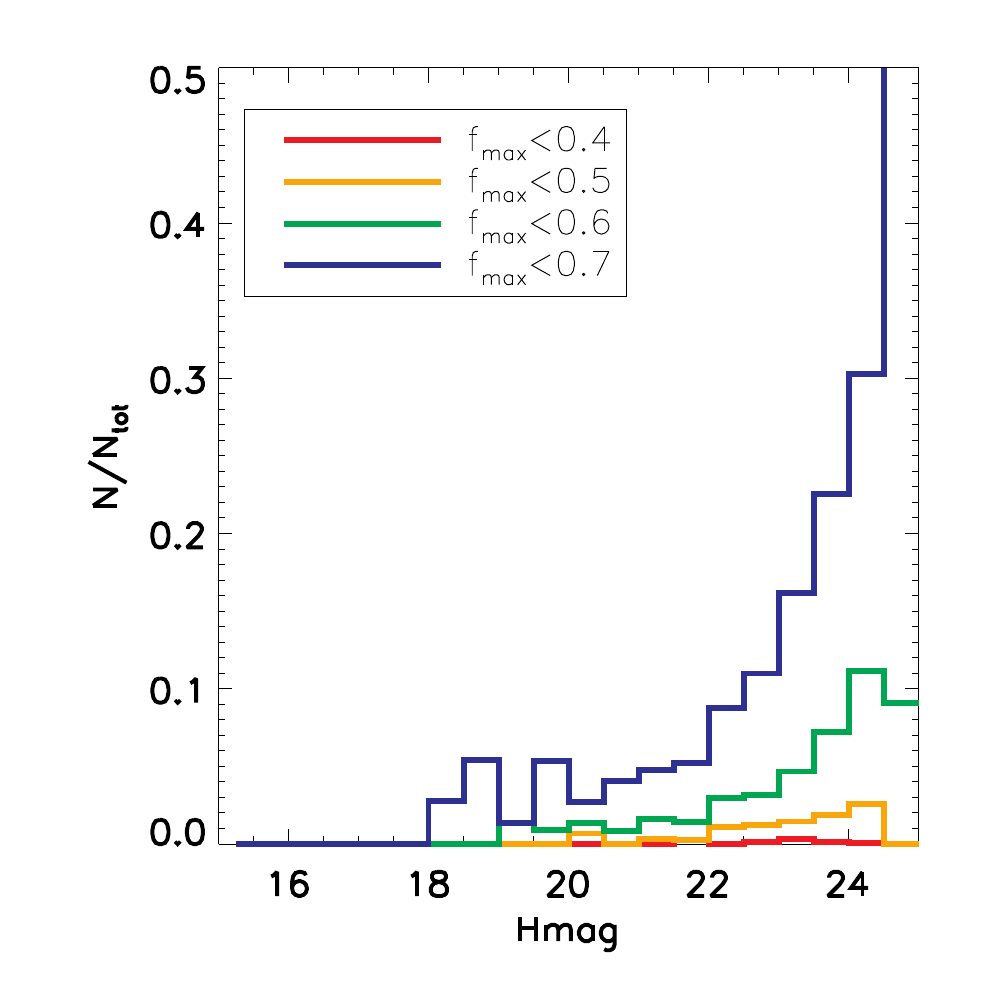} & \includegraphics[width=0.30\textwidth]{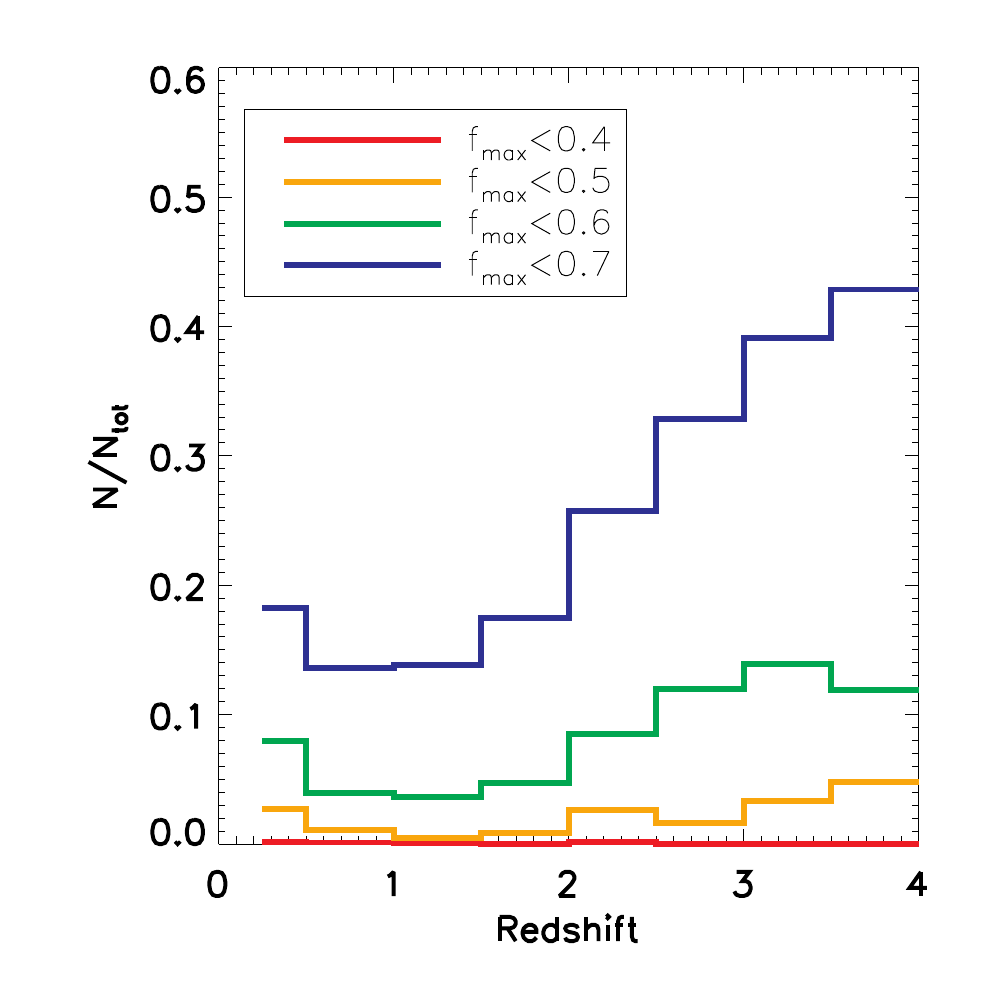} &  \includegraphics[width=0.30\textwidth]{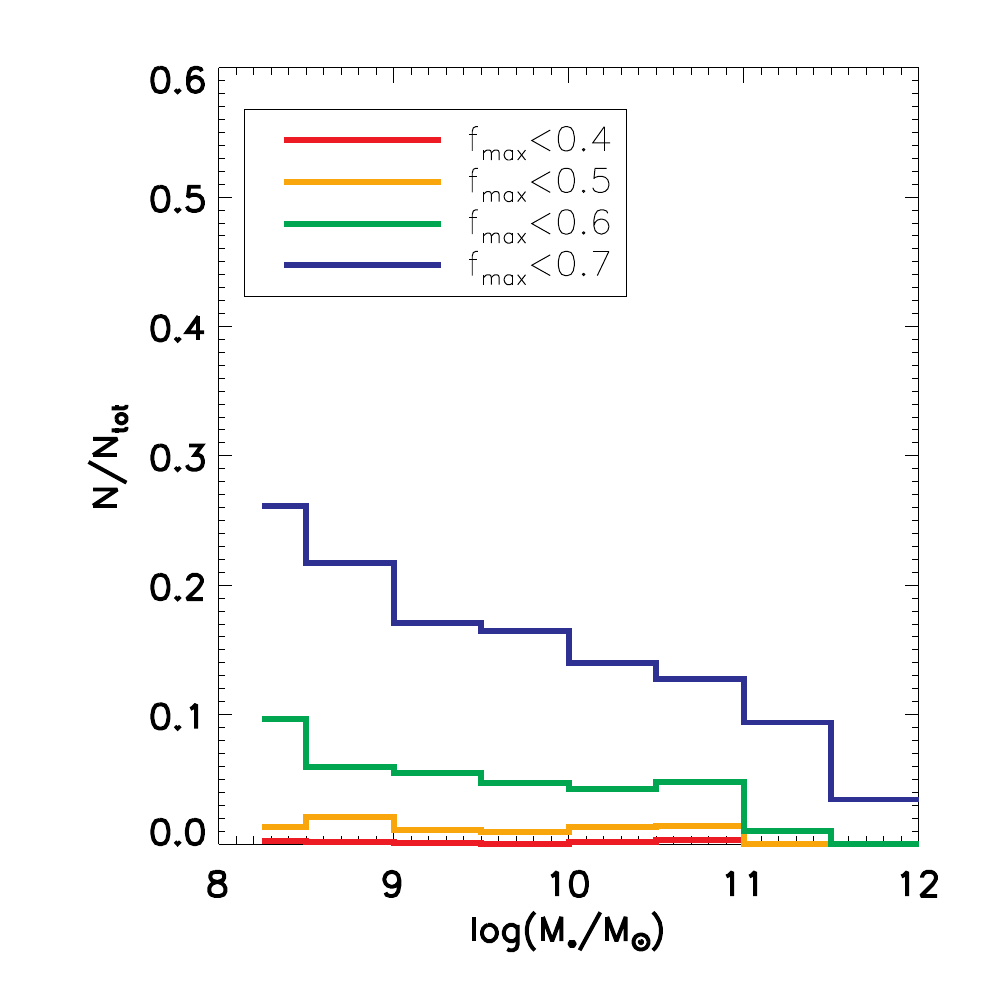} \\
\includegraphics[width=0.30\textwidth]{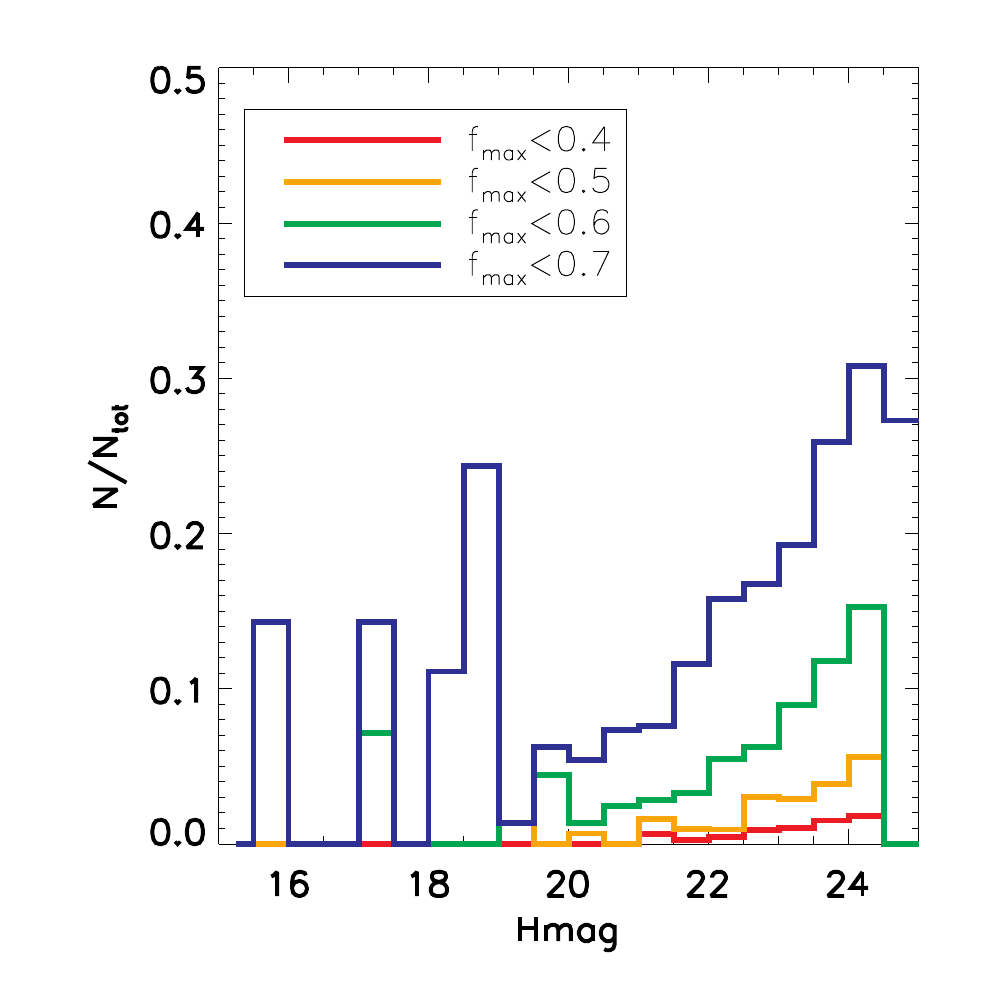} & \includegraphics[width=0.30\textwidth]{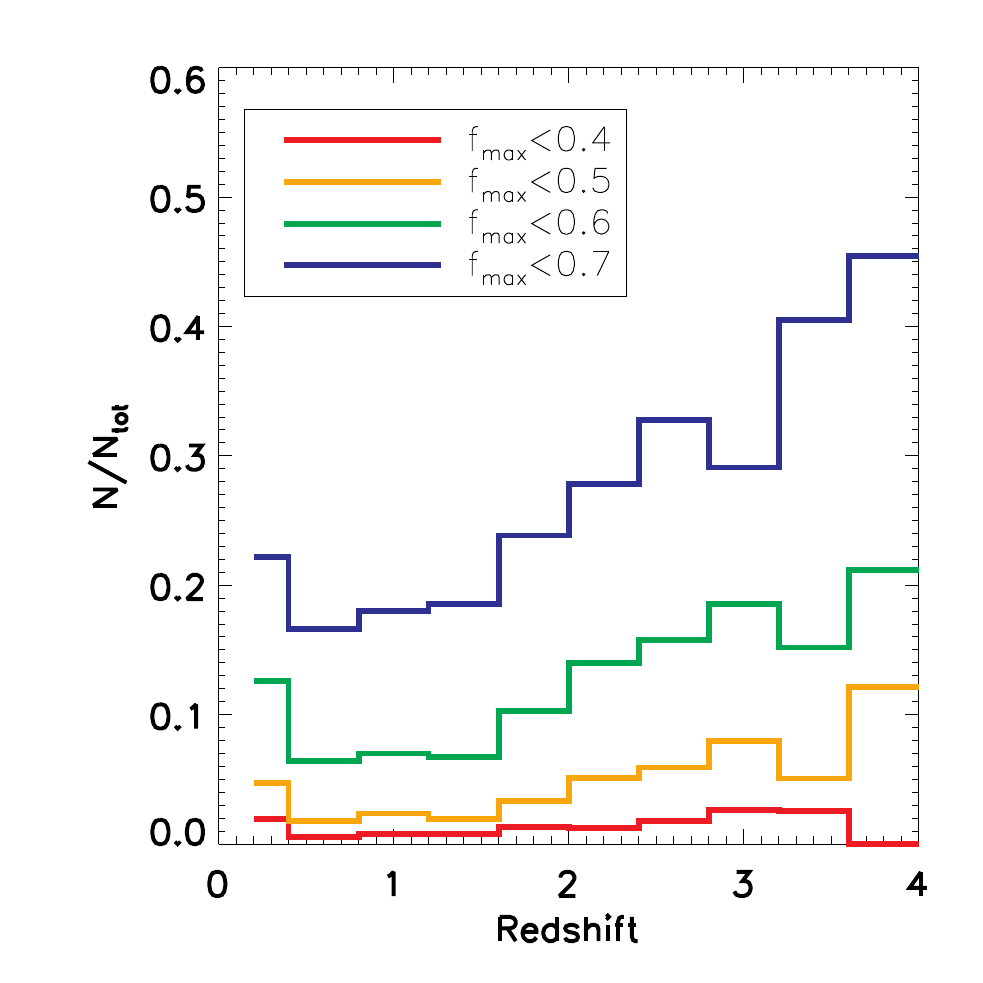} &  \includegraphics[width=0.30\textwidth]{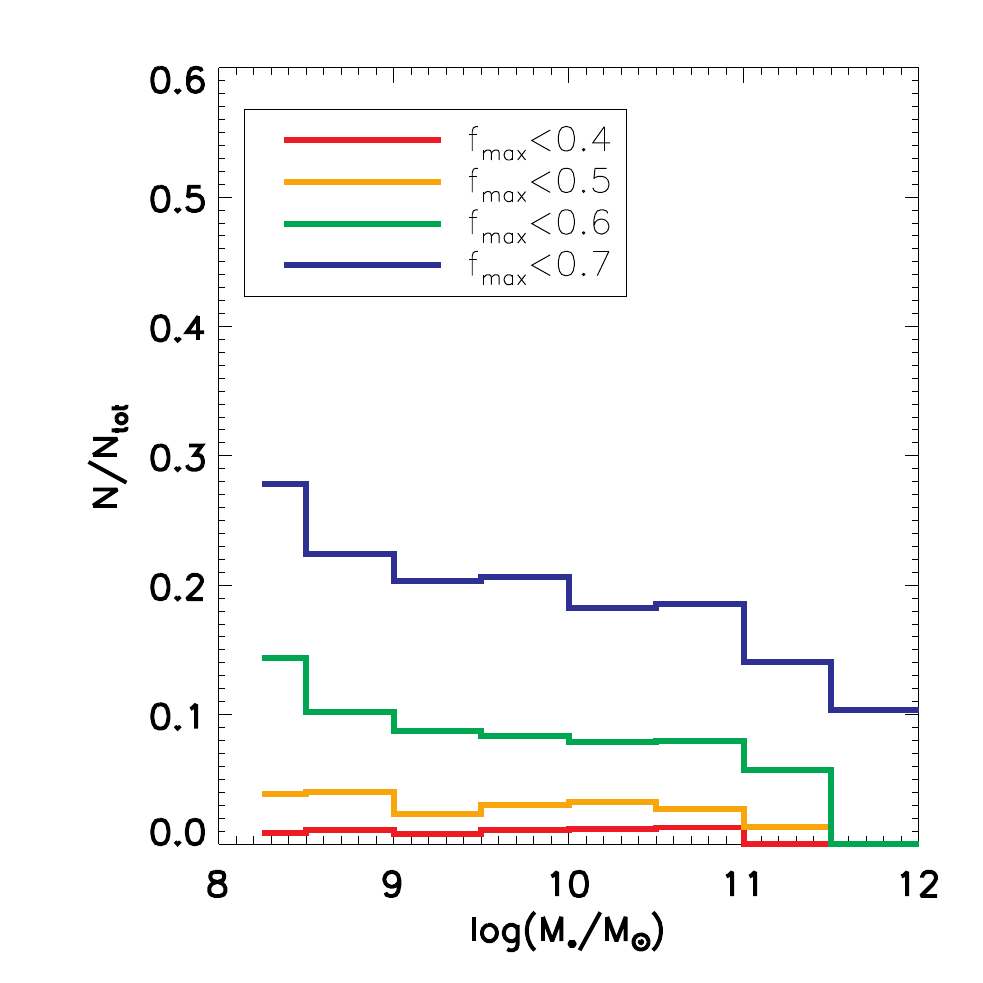} \\
\includegraphics[width=0.30\textwidth]{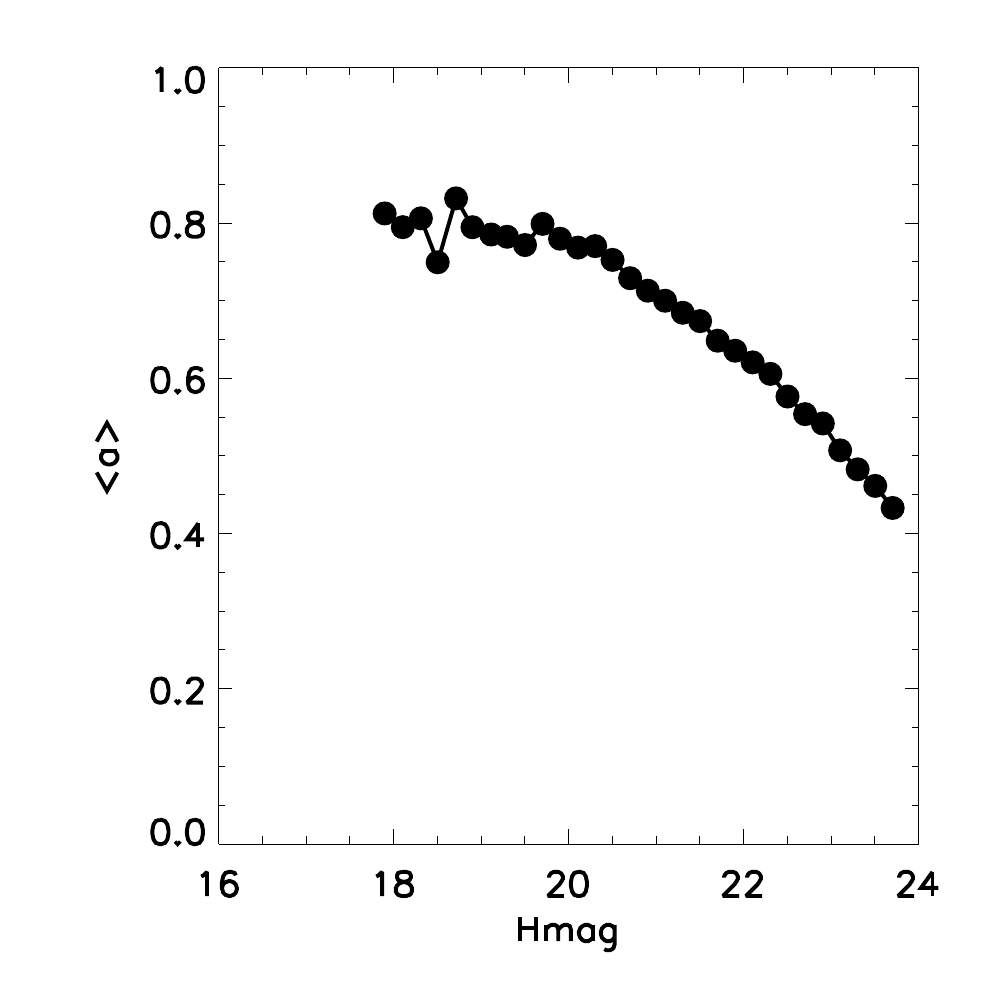} & \includegraphics[width=0.30\textwidth]{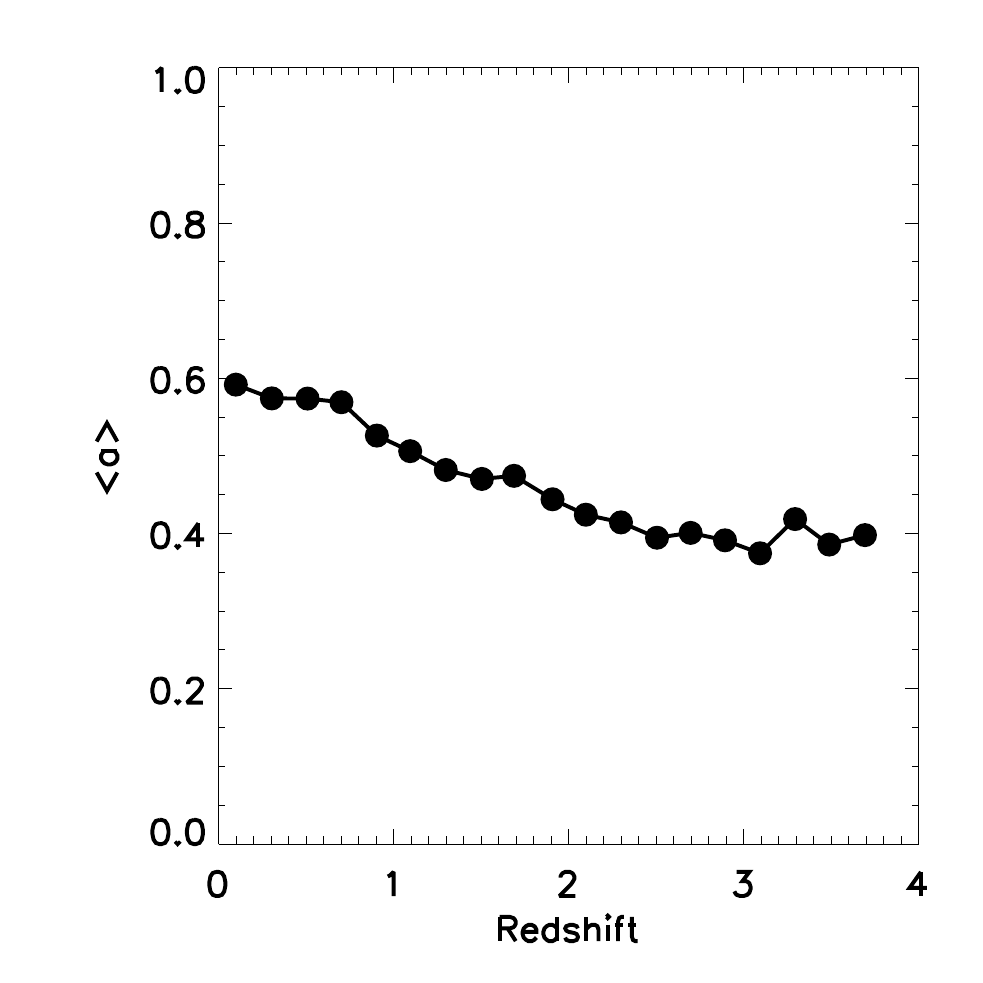} &  \includegraphics[width=0.30\textwidth]{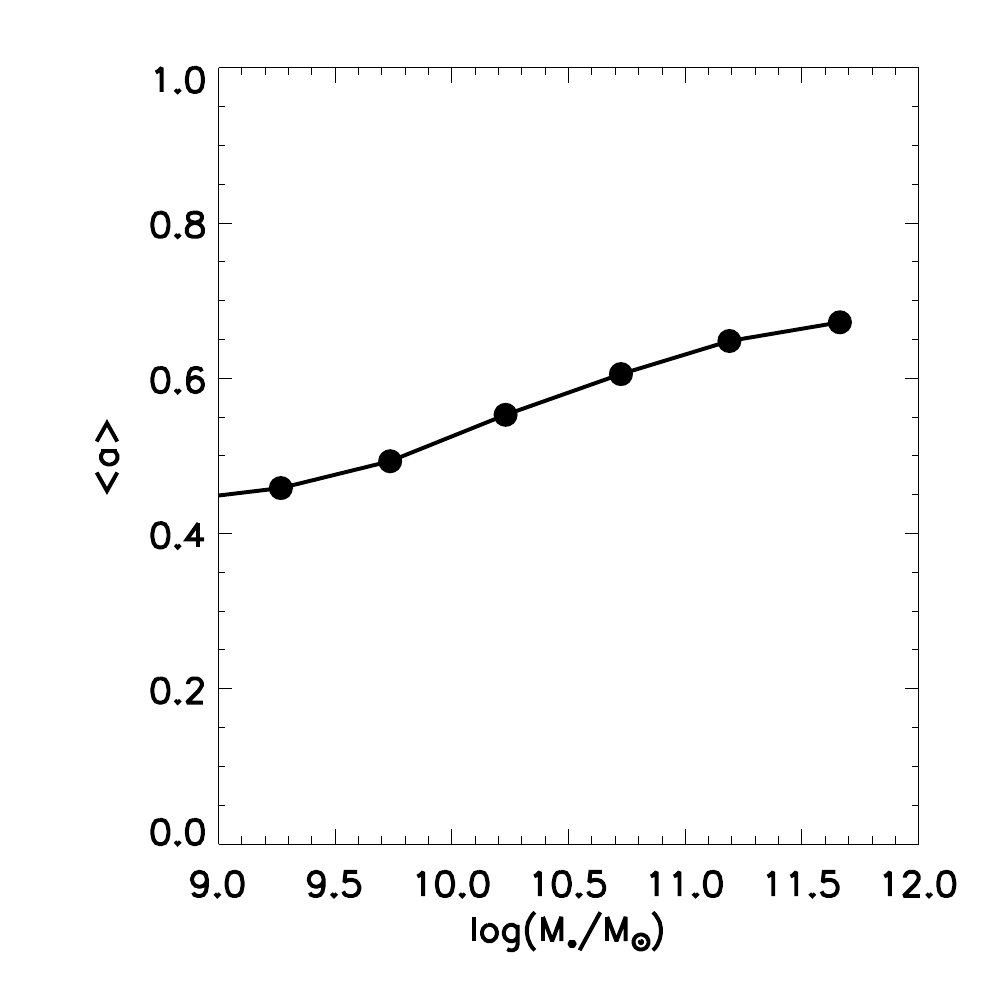} \\
\end{array}$
\caption{Fraction of uncertain objects defined for different $f_{max}$ thresholds as labelled in the automatic (top) and visual (middle) classifications. The fraction of uncertain objects increase for fainter objects, high redshifts and low masses. Similar trends are recovered in both classifications. The bottom line shows the relation between the level of agreement $a$ (see text) and magnitude (left), redshift (middle) and stellar mass (right). } 
\label{fig:uncertain}
\end{center}
\end{figure*}

\section{Accuracy in all CANDELS fields}
\label{sec:all_CANDELS}

All previous results are based on GOODS-S where visual classifications are available for training and testing. The main purpose of the present work is to extend the classification to all CANDELS fields where visual inspection is not yet available. It is therefore important to give an estimate of how the algorithm is behaving in these \emph{blank} fields. 

\subsection{Field-to-field homogeneity}

One quick sanity check consists in making sure that there are no significant statistical differences among the morphological distributions in the different fields. We do expect indeed that all fields should have similar fractions of all morphologies within cosmic variance since they have similar depths and are selected randomly. It is true that the CANDELS surveys has some \emph{deep} and \emph{wide} areas which are observed at different depths. However, we are imposing in this work a magnitude cut much brighter than the magnitude limit of the survey so our classification should not affected by these different depths. Therefore, eventual significant differences could be a sign of biases in the derived morphological classifications in a given field and an eventual signature of over-fitting problems. \\
Figure~\ref{fig:CDFs} shows the cumulative distribution functions (CDFs) of the different frequencies ($f_{sph}$, $f_{disk}$, $f_{irr}$)  in the 5 fields. We do not observe significant differences from field to field in the distribution of frequencies, suggesting that the algorithm is behaving in a similar way independently of the field. Recall however that the machine tends to \emph{smooth} the distribution compared to the visual one. In other words, it removes any gap or abrupt changes. Gaps are instead present in the visual classifications given the reduced number of classifiers per object (even after noise addition).

\begin{figure*}
\begin{center}
$\begin{array}{c c c}

\includegraphics[width=0.30\textwidth]{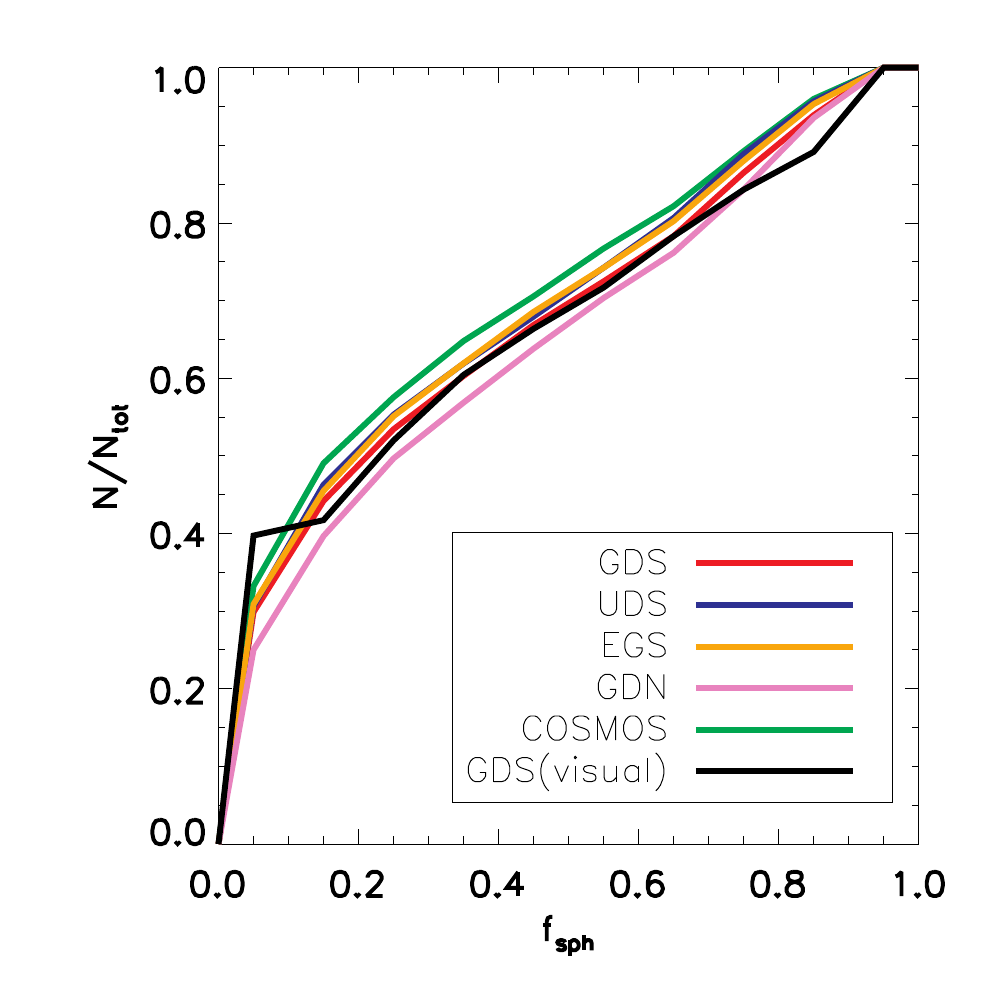} & \includegraphics[width=0.30\textwidth]{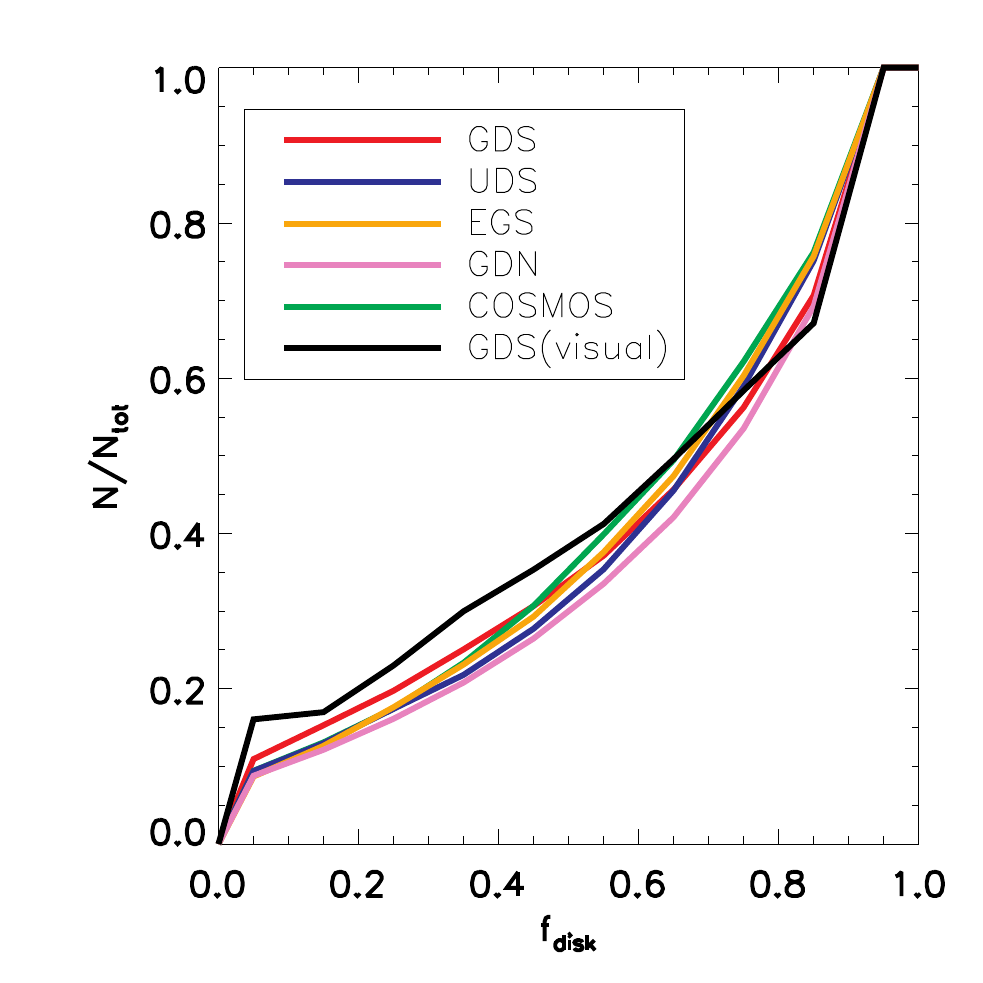} &  \includegraphics[width=0.30\textwidth]{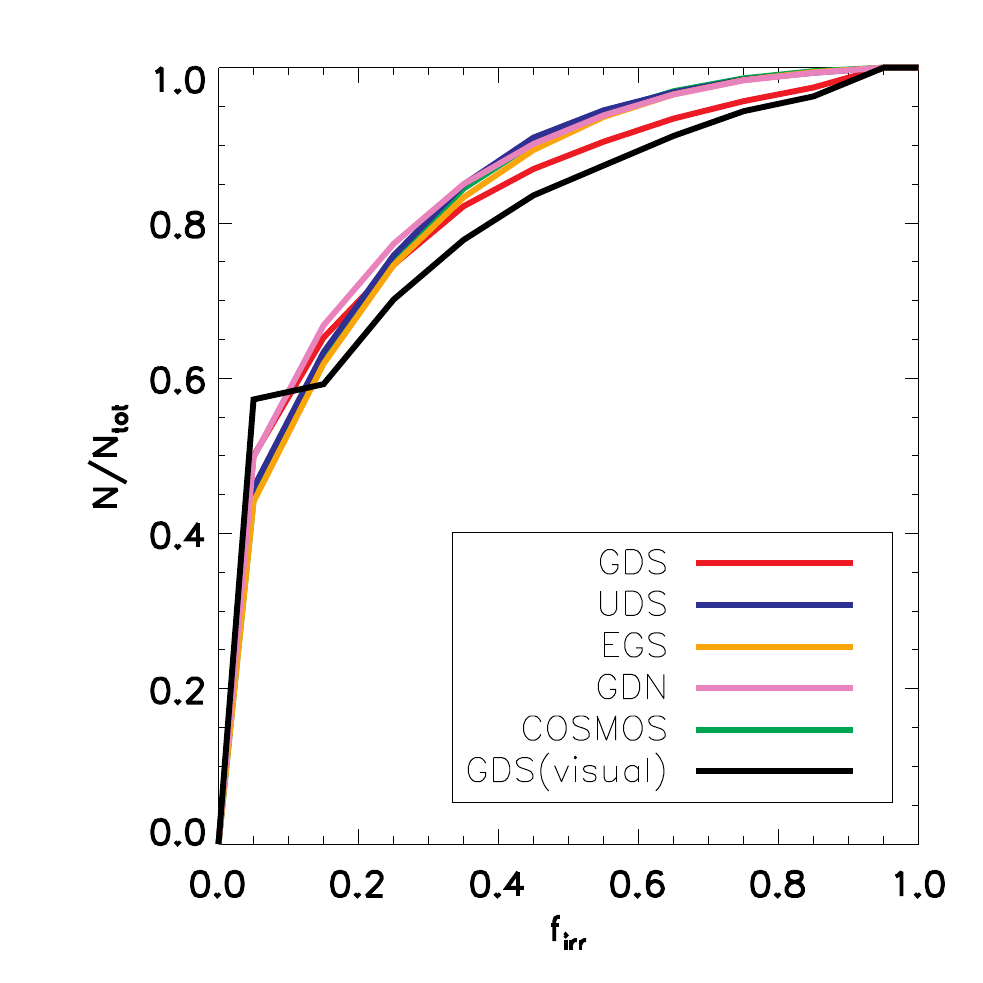} \\

\end{array}$
\caption{Cumulative distribution functions (CDFs) of $f_{sph}$ (left), $f_{disk}$ (middle) and $f_{irr}$ (right) derived in all 5 CANDELS fields as labelled. We also show in black the CDF of visual classifications in GDS (after addition of random noise). There are no major differences between the fields and the distributions follow the distributions of the visual classification. } 
\label{fig:CDFs}
\end{center}
\end{figure*}

\subsection{UDS visual classification }

During the production of the automated classification presented in this work, the visual classification for the UDS field has been finalized using the same classification scheme. Comparing the resulting parameters with the automated results on this field is therefore a fully independent test of the morphologies released in this work and a definitive test to rule out any over-fitting issues.\\
There are unfortunately important differences between the visual classifications in GOOD-S and UDS that need to be taken into account before performing a fair comparison. \\
As a matter of fact, as shown in figure~\ref{fig:CDFs}, the distribution of the morphological parameters for the ConvNets classification is similar in all fields and mimics the distribution of the visual GOODS-S classification as expected. The problem is that while in GOODS-S the number of classifiers per galaxy is roughly homogeneously distributed between 3-5 with some galaxies classified by $\sim 50$ people, in UDS $\sim90\%$ of the galaxies are only classified by 3 people and the remaining $5\%$ by 4 (see fig~\ref{fig:nclass}). This difference results in a different distribution of the visual morphological frequencies between UDS and GOODS-S (i.e. frequencies in UDS only have 4 possible values for most of the galaxies) which persists even after addition of random noise for smoothing (fig.~\ref{fig:nclass}). Since the automated classification necessarily follows the distribution for which it was trained, the comparison with UDS visual classifications will  have a larger scatter which is not due to a failure in the algorithm but to a difference in the inputs. 

\begin{figure*}
\begin{center}
$\begin{array}{c c c}

\includegraphics[width=0.30\textwidth]{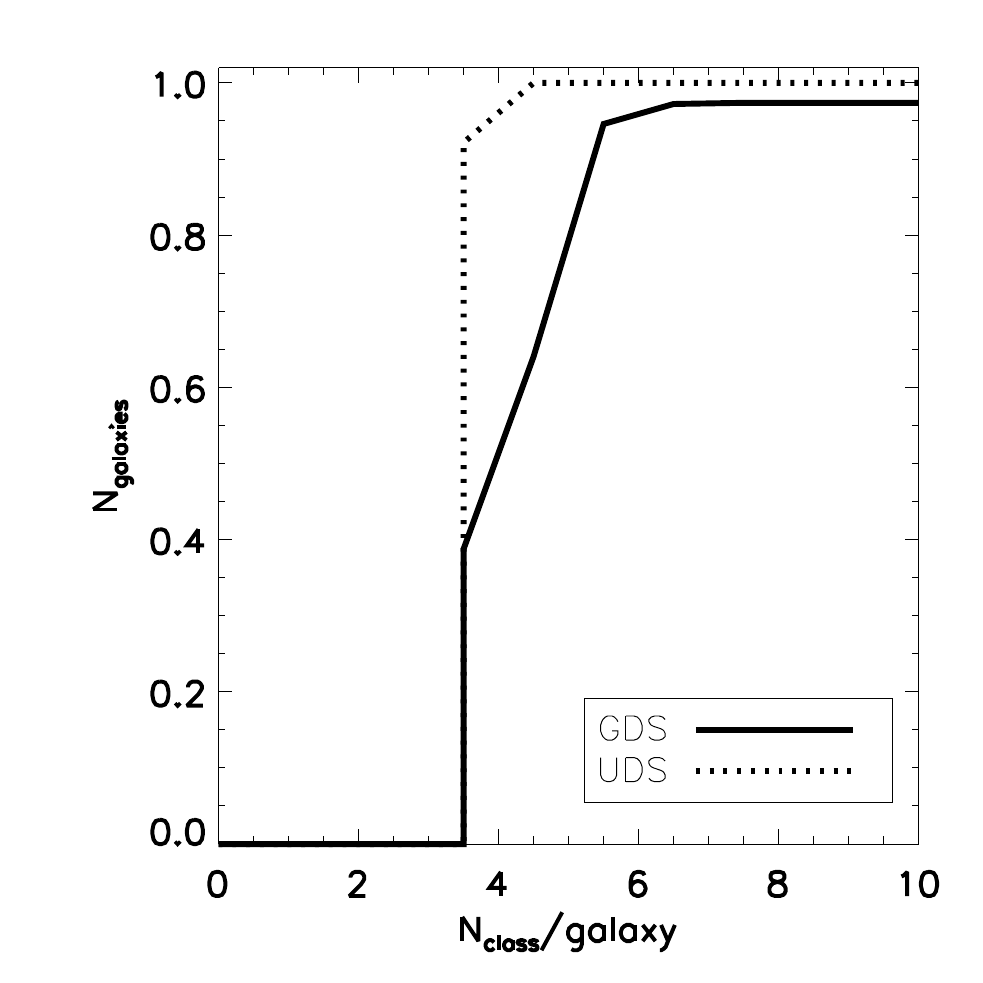} & \includegraphics[width=0.30\textwidth]{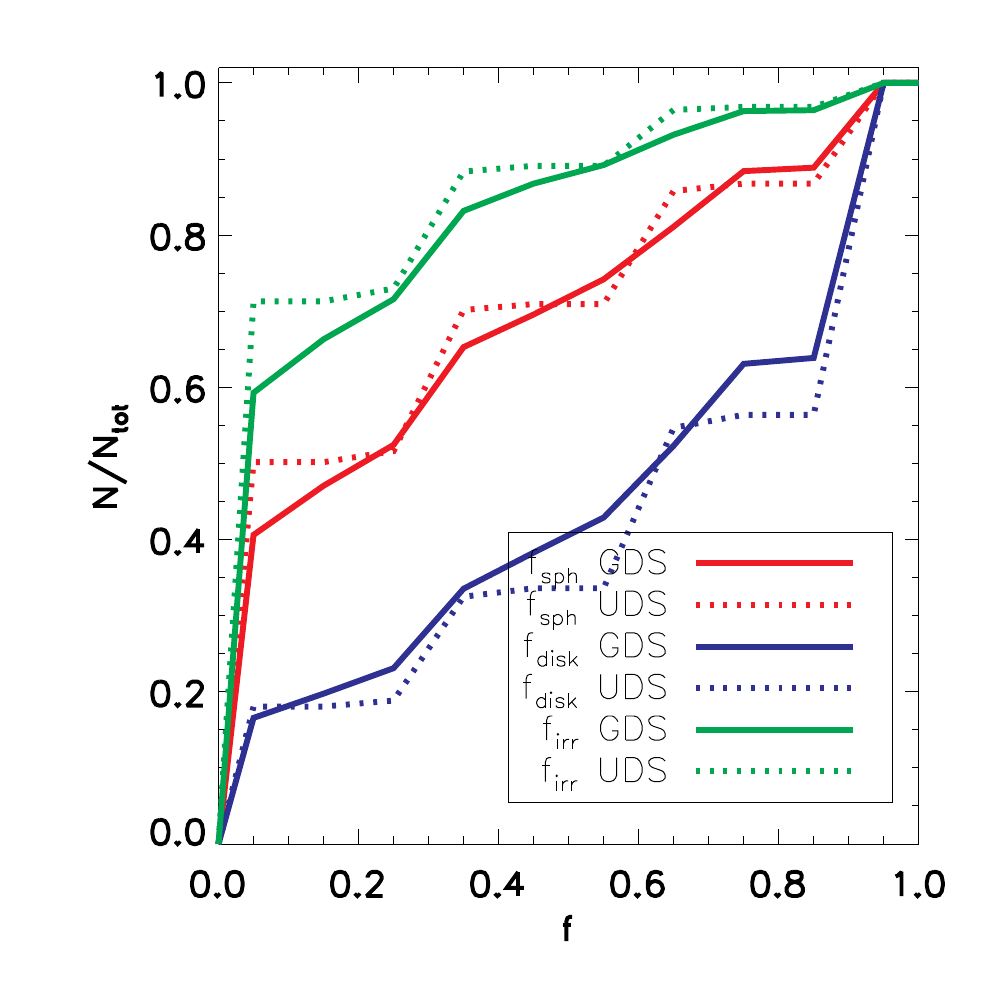} &  \includegraphics[width=0.30\textwidth]{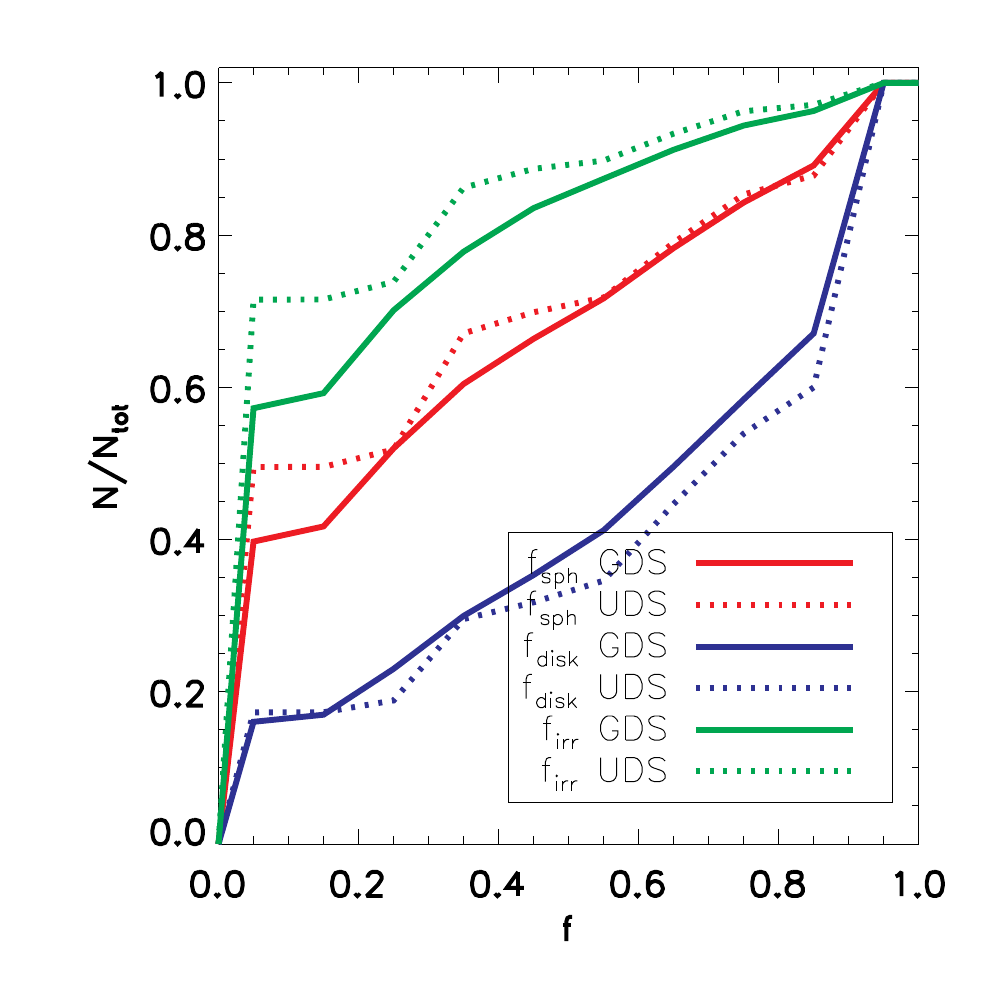} \\

\end{array}$
\caption{ Left: Number of visual classifiers per galaxy in the UDS and the GOODS-S fields. $90\%$ of the galaxies are classified by only 3 people in UDS. Middle: CDFs of the main morphological parameters in UDS and GOOD-S. Right: Same CDFs after addition of gaussian noise.} 
\label{fig:nclass}
\end{center}
\end{figure*}

In order to estimate how much this will affect the comparison in the UDS, we recomputed the GOOD-S frequencies by randomly taking only 3 classifiers per galaxy (i.e. ignoring the classifications whenever there are more than 3 classifiers) and compared with the automated classification as done in figure~\ref{fig:visual_auto_all}. \\
The results of such an exercise are shown in figure~\ref{fig:UDS_GDS}. In the left column we plot the comparison when all classifiers are taken into account (as in fig.~\ref{fig:visual_auto_all}) and in the middle column the same comparison but only with 3 classifiers. There is a clear increase of the scatter and the bias which is only caused by the change of the distribution of the input values (the output is exactly the same). Interestingly, the trends are very similar to what is observed in the comparison with the UDS (right column) which suggests that the worsening of the results in the UDS is not due to a bad behavior of the algorithm on this field, but simply to a different distribution of the inputs. 

The latter effect can also be understood if we consider that, at first level, the process of having $n$ classifiers visually selecting between two labels (binary classification) follows a binomial distribution. Let us assume for example that an image has an intrinsic probability $p$ to be classified as a spheroid. It follows that the variance of the distribution of the number of people labeling it as "yes" from a total of $n$ is $np(1-p)$. Therefore the deviation of the visually classified fractions is $\sqrt{p(1-p)/n}$. The deviation of the fractions will depend on the intrinsic probability $p$ and the number of annotations.  The less amount of annotators we have, the higher the variance on the fractions, i.e. less reliable the probabilities of each class will become (compared to the intrinsic one). So training a machine with a noisier training set will also result in a noisier classification.

This issue emphasizes one main advantage of the automated classifications with respect to the visual when a small number of classifiers is involved. Namely the results are by definition homogeneous for all datasets. The fact that the UDS and the GOODS-S with only 3 classifiers look very similar also suggests that the algorithm has a similar accuracy in both fields, confirming that the classification is not severely affected by over-fitting.

\begin{figure*}
\begin{center}
$\begin{array}{c c c}

\includegraphics[width=0.30\textwidth]{f_sph_visu_auto_all.pdf} & \includegraphics[width=0.30\textwidth]{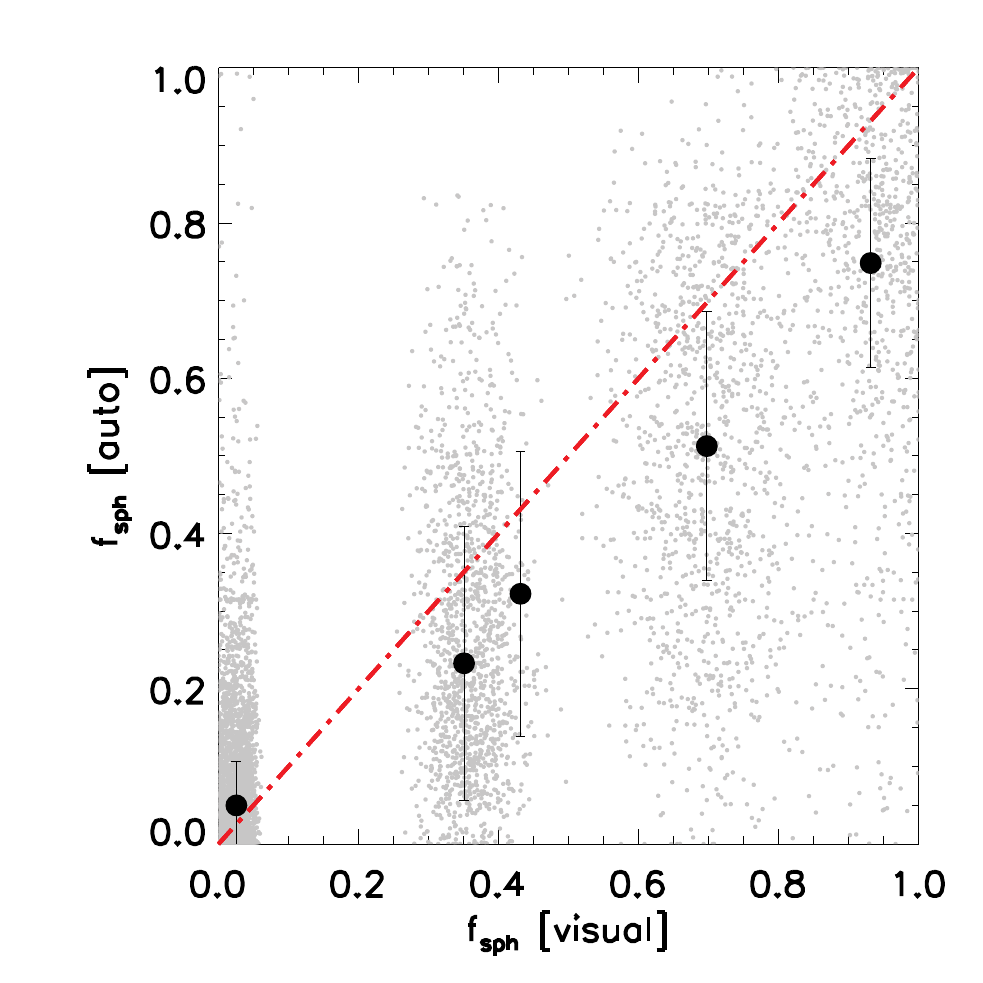} &  \includegraphics[width=0.30\textwidth]{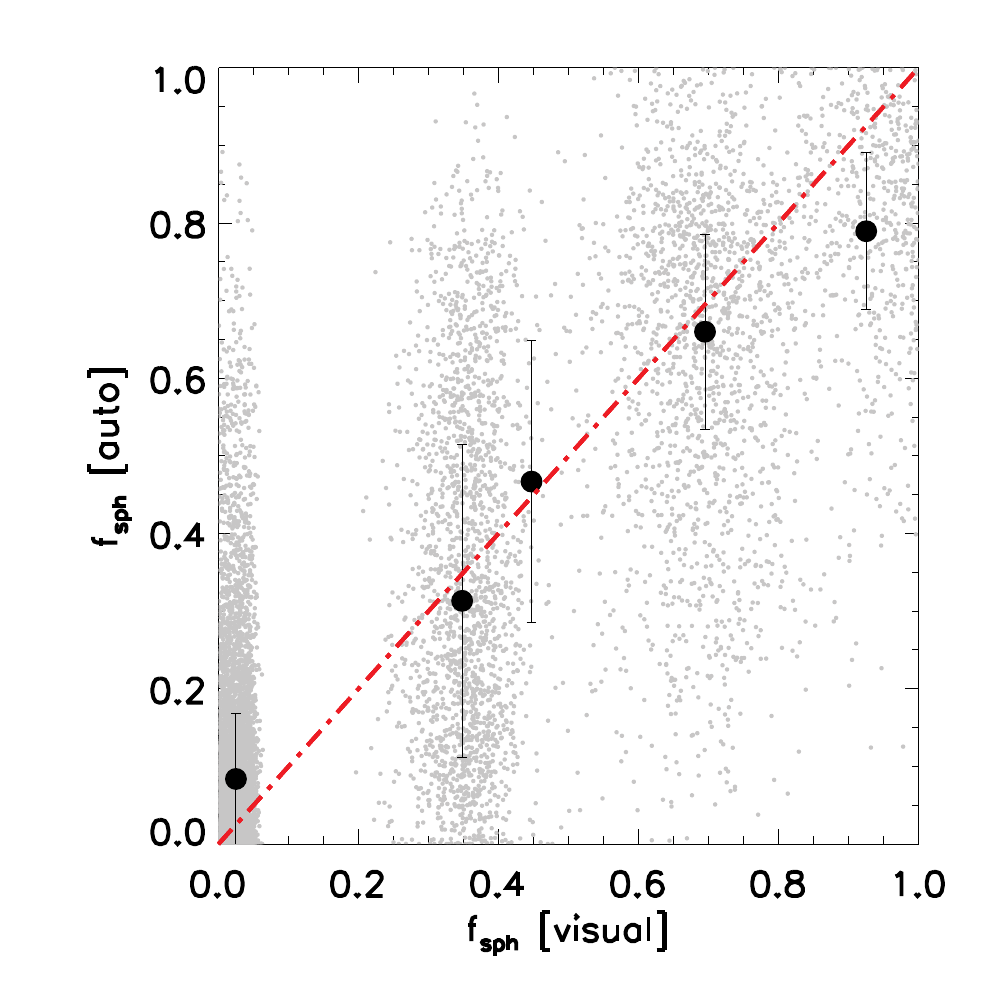} \\
\includegraphics[width=0.30\textwidth]{f_disk_visu_auto_all.pdf} & \includegraphics[width=0.30\textwidth]{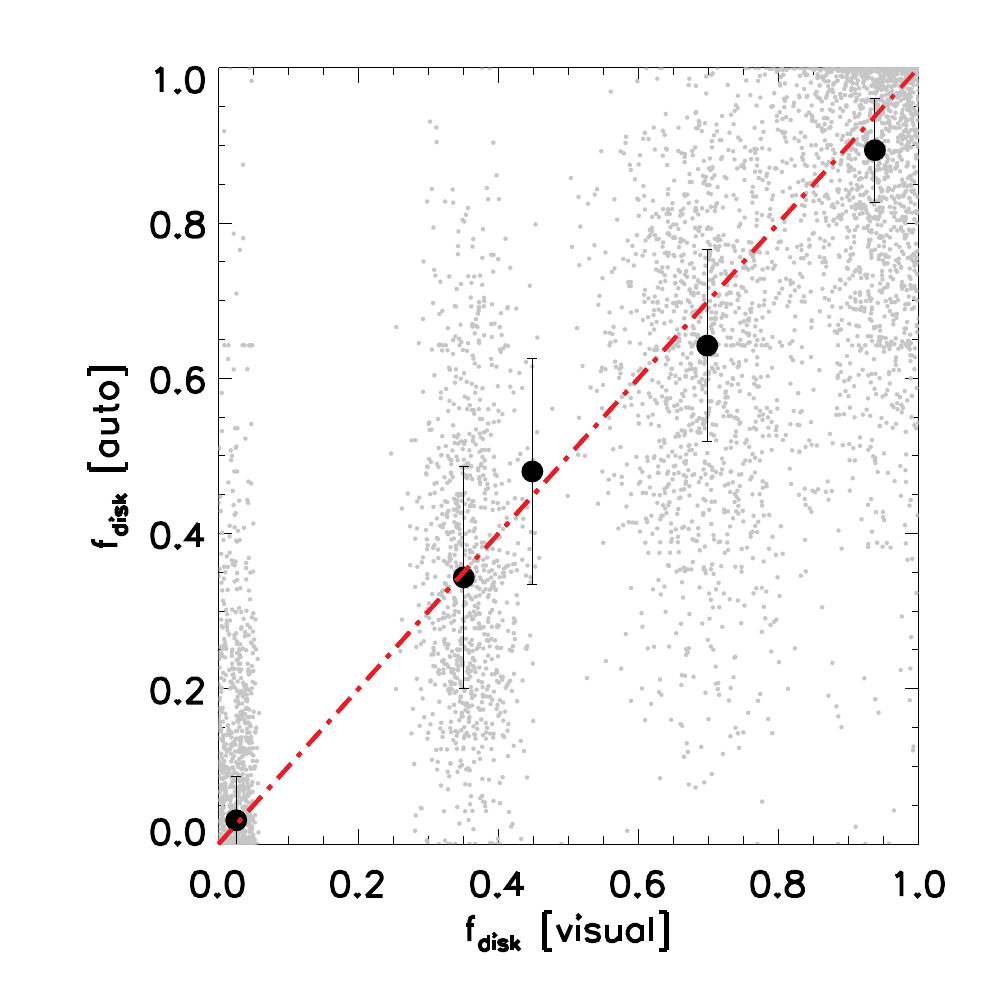} &  \includegraphics[width=0.30\textwidth]{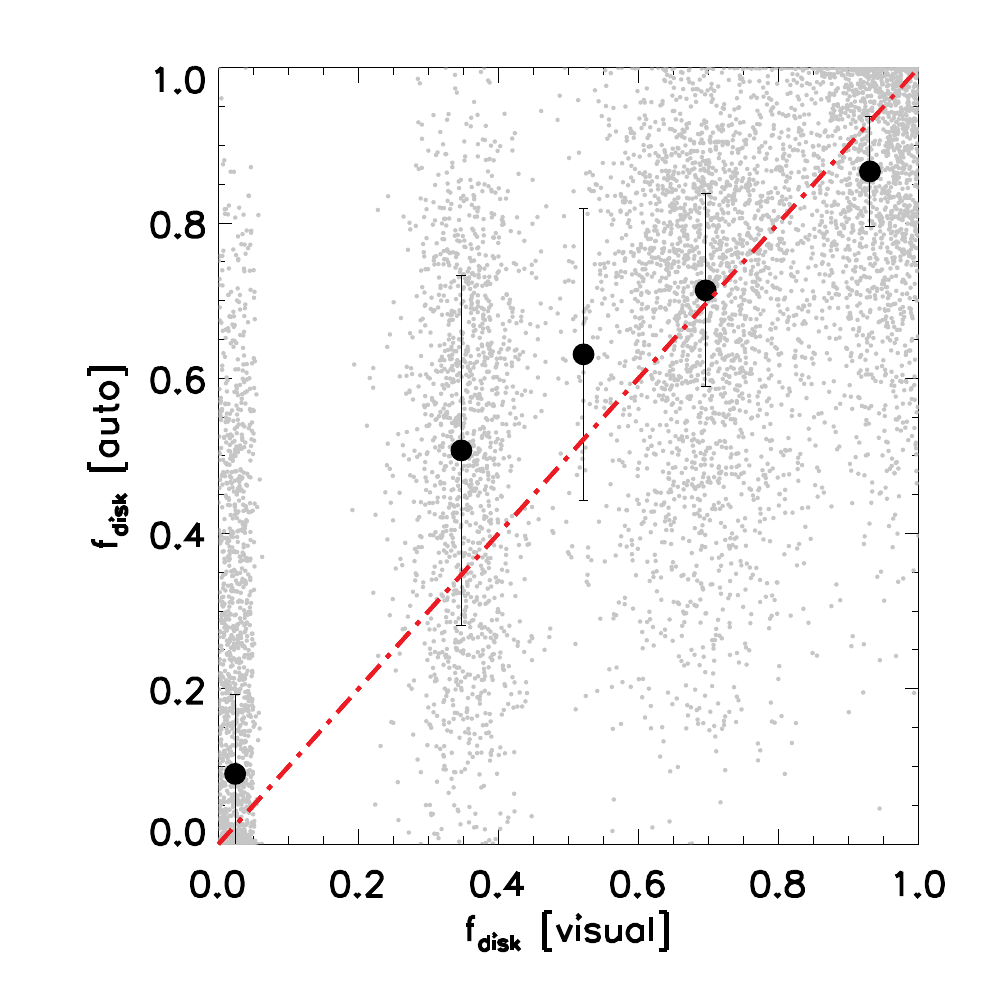} \\
\includegraphics[width=0.30\textwidth]{f_irr_visu_auto_all.pdf} & \includegraphics[width=0.30\textwidth]{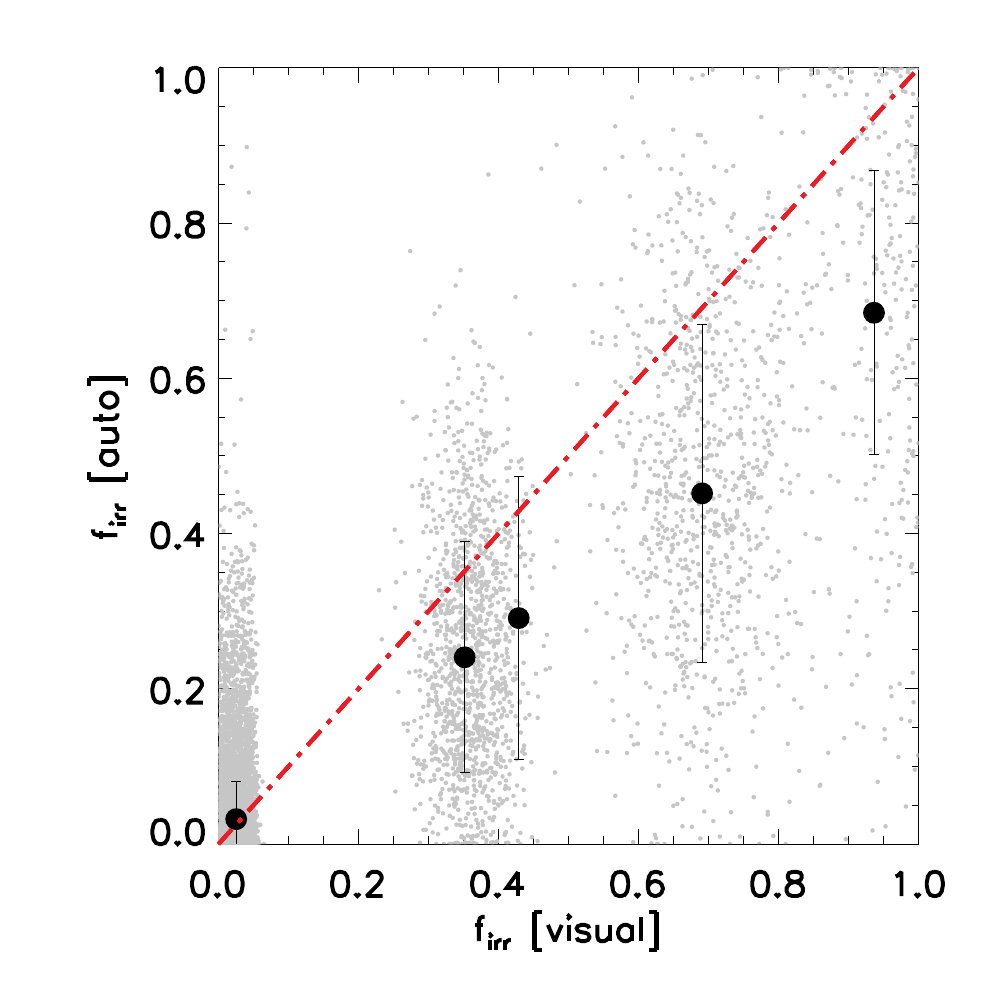} &  \includegraphics[width=0.30\textwidth]{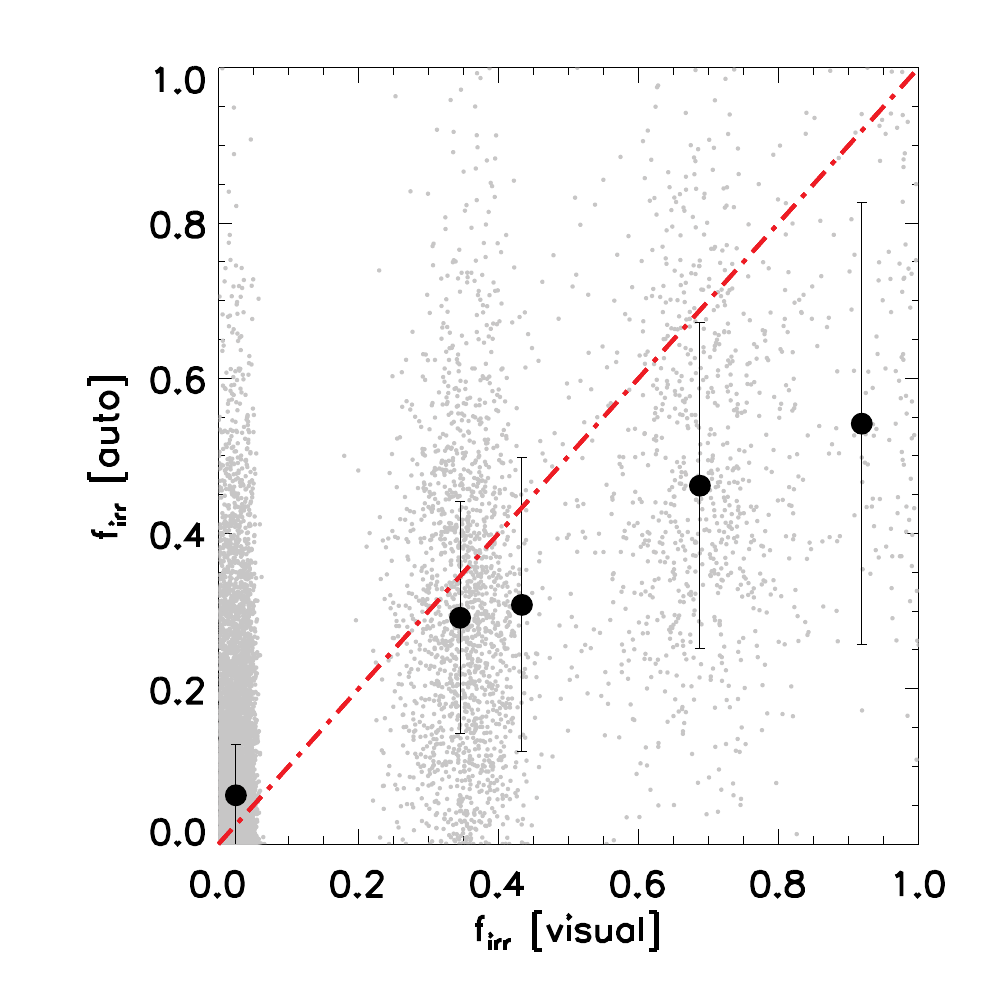} \\
\end{array}$
\caption{ Correlation between the fractions of classifiers voting for a given feature. Left: GOOD-S when all classifiers are considered. Middle: GOOD-S with only 3 classifiers. Right: UDS where $90\%$ of galaxies are classified by 3 people. The trends observed in the middle and right columns for all parameters are similar suggesting that the worsening of the results observed in the UDS are due to a difference in the input catalog.} 
\label{fig:UDS_GDS}
\end{center}
\end{figure*}

\section{Catalog  }
\label{sec:catalog}

The paper is accompanied by the public release of the morphology of all galaxies in the CANDELS fields brighter than $H_{F160W}=24.5$.  In addition to the 5 morphological parameters, we also provide in the catalog a 2 measurements of the quality of the classification discussed in the text ($a$ and $\Delta_{f_1-f_2}$) as well as the dominant class and the maximum frequency $f_{max}$. Table~\ref{tbl:cat} shows the first few lines of the catalog. The catalog is released through the Rainbow database: \url{http://rainbowx.fis.ucm.es/Rainbow\_navigator\_public/} \\

\begin{table*}
\begin{center}
\resizebox{\textwidth}{!}{
\begin{tabular}{|r|l|l|r|r|l|r|r|r|r|r|r|r|r|r|}
\hline
  \multicolumn{1}{|c|}{ID} &
  \multicolumn{1}{c|}{IAU\_NAME} &
  \multicolumn{1}{c|}{RA} &
  \multicolumn{1}{c|}{DEC} &
  \multicolumn{1}{c|}{Filter} &
  \multicolumn{1}{c|}{$f_{spheroid}$} &
  \multicolumn{1}{c|}{$f_{disk}$} &
  \multicolumn{1}{c|}{$f_{irr}$} &
  \multicolumn{1}{c|}{$f_{PS}$} &
  \multicolumn{1}{c|}{$f_{Unc}$} &
  \multicolumn{1}{c|}{$f_{max}$} &
  \multicolumn{1}{c|}{$\Delta_f$} &
  \multicolumn{1}{c|}{DOM\_CLASS} &
  \multicolumn{1}{c|}{$a$} \\
\hline
  1 & HCPG  J142112.26+5303004.5 & 215.3011017 & 53.051239 & f160 & 0.1 & 0.1 & 0.17 & 0.0 & 0.72 & 0.72 & 0.54 & 4 & 0.38\\
  1000 & HCPG  J142051.15+5300016.8 & 215.2131348 & 53.0046539 & f160 & 0.73 & 0.12 & 0.08 & 0.37 & 0.0 & 0.73 & 0.36 & 0 & 0.34\\
  10001 & HCPG  J141955.98+5253037.2 & 214.9832611 & 52.8936768 & f160 & 0.11 & 1.0 & 0.01 & 0.0 & 0.0 & 1.0 & 0.89 & 1 & 0.82\\
  10002 & HCPG  J142044.89+5301059.4 & 215.187027 & 53.0331574 & f160 & 0.57 & 1.0 & 0.0 & 0.01 & 0.0 & 1.0 & 0.43 & 1 & 0.78\\
  10003 & HCPG  J142013.52+5256044.1 & 215.0563202 & 52.9455872 & f160 & 0.25 & 0.88 & 0.22 & 0.01 & 0.03 & 0.88 & 0.63 & 1 & 0.4\\
  10004 & HCPG  J141924.91+5248004.0 & 214.8538055 & 52.8011017 & f160 & 0.84 & 0.16 & 0.06 & 0.24 & 0.01 & 0.84 & 0.59 & 0 & 0.39\\
  10005 & HCPG  J142025.18+5258045.7 & 215.1049042 & 52.9793701 & f160 & 0.34 & 0.92 & 0.16 & 0.0 & 0.0 & 0.92 & 0.58 & 1 & 0.53\\
  10010 & HCPG  J141906.89+5244043.3 & 214.778717 & 52.7453613 & f160 & 0.34 & 1.0 & 0.09 & 0.0 & 0.0 & 1.0 & 0.66 & 1 & 0.64\\
  10015 & HCPG  J141859.26+5243018.4 & 214.746933 & 52.7217865 & f160 & 0.19 & 0.97 & 0.18 & 0.0 & 0.0 & 0.97 & 0.78 & 1 & 0.59\\
  10017 & HCPG  J142009.87+5256005.7 & 215.0411224 & 52.934906 & f160 & 0.33 & 0.95 & 0.09 & 0.02 & 0.0 & 0.95 & 0.62 & 1 & 0.55\\
  10018 & HCPG  J141927.56+5248031.8 & 214.8648376 & 52.8088379 & f160 & 0.0 & 0.95 & 0.14 & 0.0 & 0.0 & 0.95 & 0.81 & 1 & 0.78\\
  10019 & HCPG  J141952.59+5253001.8 & 214.9691162 & 52.8838196 & f160 & 0.05 & 0.16 & 0.98 & 0.0 & 0.0 & 0.98 & 0.82 & 2 & 0.71\\
  10020 & HCPG  J142037.78+5301000.3 & 215.1574097 & 53.0167541 & f160 & 0.34 & 0.62 & 0.42 & 0.1 & 0.0 & 0.62 & 0.2 & 1 & 0.22\\
  10024 & HCPG  J141917.09+5246040.1 & 214.8211975 & 52.7778015 & f160 & 0.84 & 1.0 & 0.0 & 0.01 & 0.0 & 1.0 & 0.16 & 1 & 0.89\\
  10026 & HCPG  J141922.45+5247042.5 & 214.8435364 & 52.7951355 & f160 & 0.47 & 0.94 & 0.13 & 0.0 & 0.01 & 0.94 & 0.47 & 1 & 0.55\\
  10027 & HCPG  J141938.69+5250035.2 & 214.9111938 & 52.8431091 & f160 & 0.11 & 0.9 & 0.16 & 0.04 & 0.05 & 0.9 & 0.74 & 1 & 0.44\\
  10029 & HCPG  J142055.91+5304013.0 & 215.2329407 & 53.070282 & f160 & 0.78 & 0.13 & 0.02 & 0.34 & 0.01 & 0.78 & 0.44 & 0 & 0.42\\
  1003 & HCPG  J142011.48+5253015.9 & 215.0478363 & 52.8877411 & f160 & 0.58 & 0.85 & 0.15 & 0.04 & 0.01 & 0.85 & 0.27 & 1 & 0.45\\
  10032 & HCPG  J142027.07+5259005.7 & 215.112793 & 52.9849091 & f160 & 0.18 & 0.66 & 0.56 & 0.0 & 0.0 & 0.66 & 0.1 & 1 & 0.44\\
  10035 & HCPG  J141938.21+5250030.9 & 214.9091949 & 52.841919 & f160 & 0.22 & 0.91 & 0.17 & 0.04 & 0.0 & 0.91 & 0.69 & 1 & 0.48\\
  10036 & HCPG  J141939.83+5250048.2 & 214.9159393 & 52.8467102 & f160 & 0.44 & 0.21 & 0.02 & 0.44 & 0.25 & 0.44 & 0.0 & 0 & 0.08\\

\hline\end{tabular}
}
\caption{Sample of the morphological catalog released with the paper. In addition to the 5 main morphological indicators, we provide for each galaxy two measurements of the level agreement between classifiers (a, linked to the entropy - see text for details) and $\Delta_f$, the difference between the two largest frequencies. DOM\_CLASS gives the dominant class (class which has the maximum frequency), being 0, spheroid, 1, disk, 2, irregular, 3, point-source and 4 unclassifiable. The catalog can be downloaded form the Rainbow database: \protect\url{http://rainbowx.fis.ucm.es/Rainbow\_navigator\_public/}  }
\label{tbl:cat}
\end{center}
\end{table*}

The classification provided is by definition continuous, since each galaxy has 5 parameters spanning from 0 to 1. The use of these parameters to actually define morphological classes strongly depends on the science purposes and the galaxy properties one would like to highlight. Establishing thresholds in the different fractions necessarily implies a trade-off between pure and complete samples.

For illustration purposes on how to use the catalog, we propose here one possible classification in 5 different morphological classes based on establishing thresholds in the different frequencies  (see Huertas-Company et al. 2015a):

\begin{itemize}
\item \emph{pure bulges [SPH]:}~~$f_{sph}>2/3$ AND $f_{disk}<2/3$ AND $f_{irr}<1/10$
\item \emph{pure disks [DISK]:}~~$f_{sph}<2/3$ AND $f_{disk}>2/3$ AND $f_{irr}<1/10$
\item \emph{disk+sph [DISKSPH]:}~~$f_{sph}>2/3$ AND $f_{disk}>2/3$ AND $f_{irr}<1/10$
\item \emph{irregular disks [DISKIRR]:}~~$f_{disk}>2/3$ AND $f_{sph}<2/3$ AND $f_{irr}>1/10$
\item \emph{irregulars/mergers[IRR]:}~~$f_{disk}<2/3$ AND $f_{sph}<2/3$ AND $f_{irr}>1/10$

\end{itemize}

The thresholds are  obviously arbitrary but have been calibrated through visual inspection to make sure that they result in different morphological classes. The SPH class contains galaxies fully dominated by the bulge component with little or no disk at all. The DISK class is made of galaxies in which the disk component dominates over the bulge. Between both classes, lies the DISKSPH class in which we put galaxies with no clear dominant component. Then we distinguish 2 types of irregulars: DISKIRR, i.e. disk dominated galaxies with some asymmetric features and IRR, which are irregular galaxies with no clear dominant disk component (including mergers).

Some random example stamps in the COSMOS field are shown in figure~\ref{fig:stamps_pure}. Also for illustration purposes, we show in figures~\ref{fig:S\'ersic} to~\ref{fig:uvplaneirr} the S\'ersic index distributions and UVJ planes for galaxies with $M_*/M_\odot>10^{10}$ split in different morphological types and for several redshift bins. The expected trends are observed in both figures and are also very similar to the distributions shown by \citealp{2014arXiv1401.2455K} on which our classification is based. \\
We observe indeed that the different morphological types have very different S\'ersic index distributions. Objects with a clear bulge component according to their visual inspection (spheroids and bulge+disk systems), tend to have larger S\'ersic indices and also tend to be located in the passive zone of the UVJ plane. Disk-dominated objects peak at $n\sim1$ and are star-forming based on their locus on the UVJ plane. \\
One interesting class is the bulge+spheroid class (i.e. objects with no clear dominant disk or spheroidal component) since they do not have a clear locus in the UVJ diagram. Roughly half of them are passive and the other half are star-forming. Any selection based on star-formation activity will therefore split this population in two groups. Having a pure morphological classification enables to isolate objects that are difficult to identify with colors and/or single profile fitting. It is also interesting to notice that the large morphological catalog put together in this paper, allows to study objects which deviate from the general trends (i.e. passive disks, star-forming bulges) with reasonable statistics (see fig.~\ref{fig:strange}).

\begin{figure*}
\begin{center}
\includegraphics[width=0.99\textwidth]{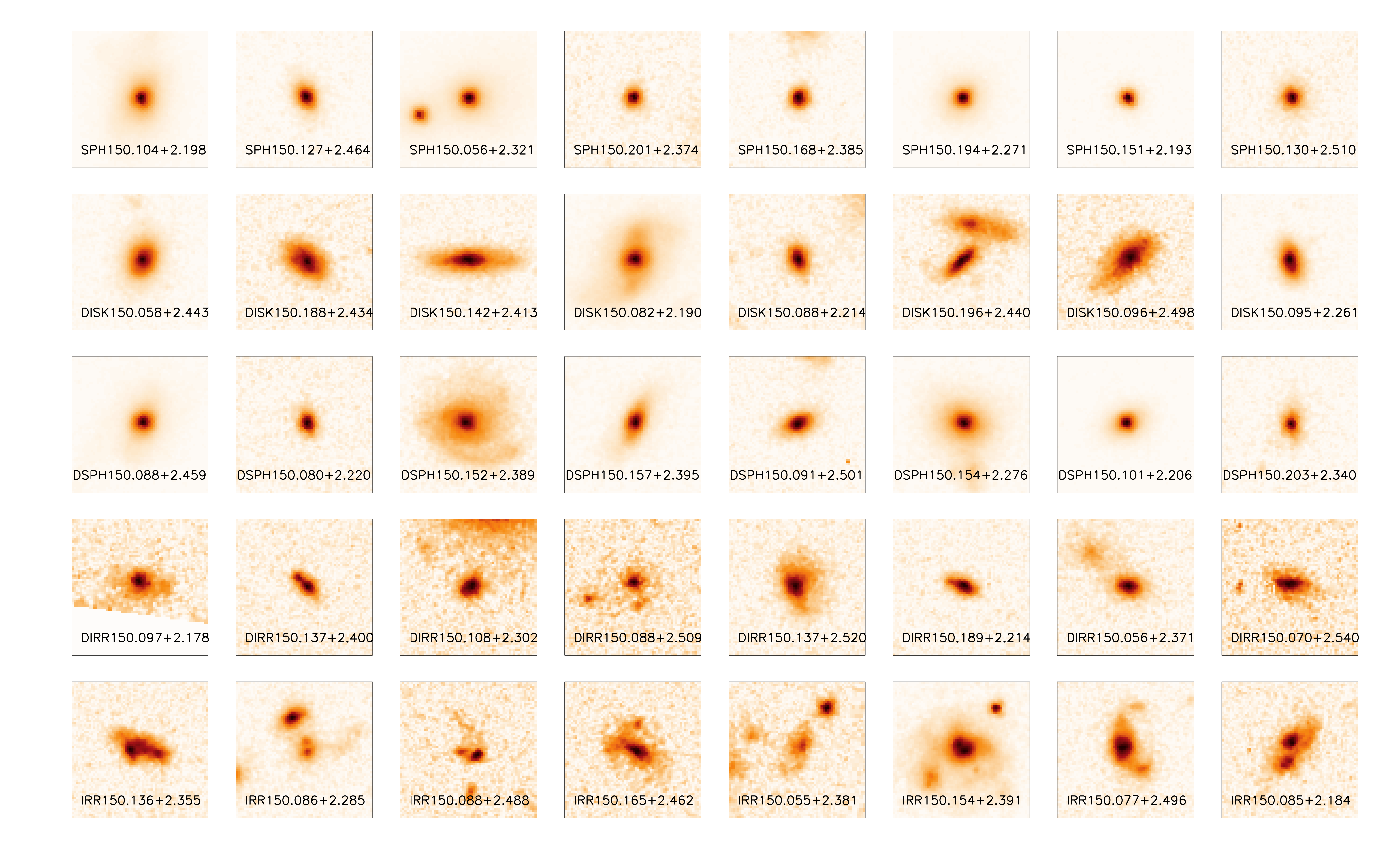}
\caption{Examples stamps of the 5 morphological classes defined for illustration in the COSMOS/CANDELS field. From top to bottom we show spheroids, disks, disk+spheroids. irregular disks and irregulars. The selection of these galaxies is done fully randomly. Recall that COSMOS galaxies have not been used for training the algorithm, therefore they are completely new for the best model. The size of the stamps is $3.8^{"}\times3.8^{"}$. }
\label{fig:stamps_pure}
\end{center}
\end{figure*}

\begin{figure*}
\begin{center}
\includegraphics[width=0.99\textwidth]{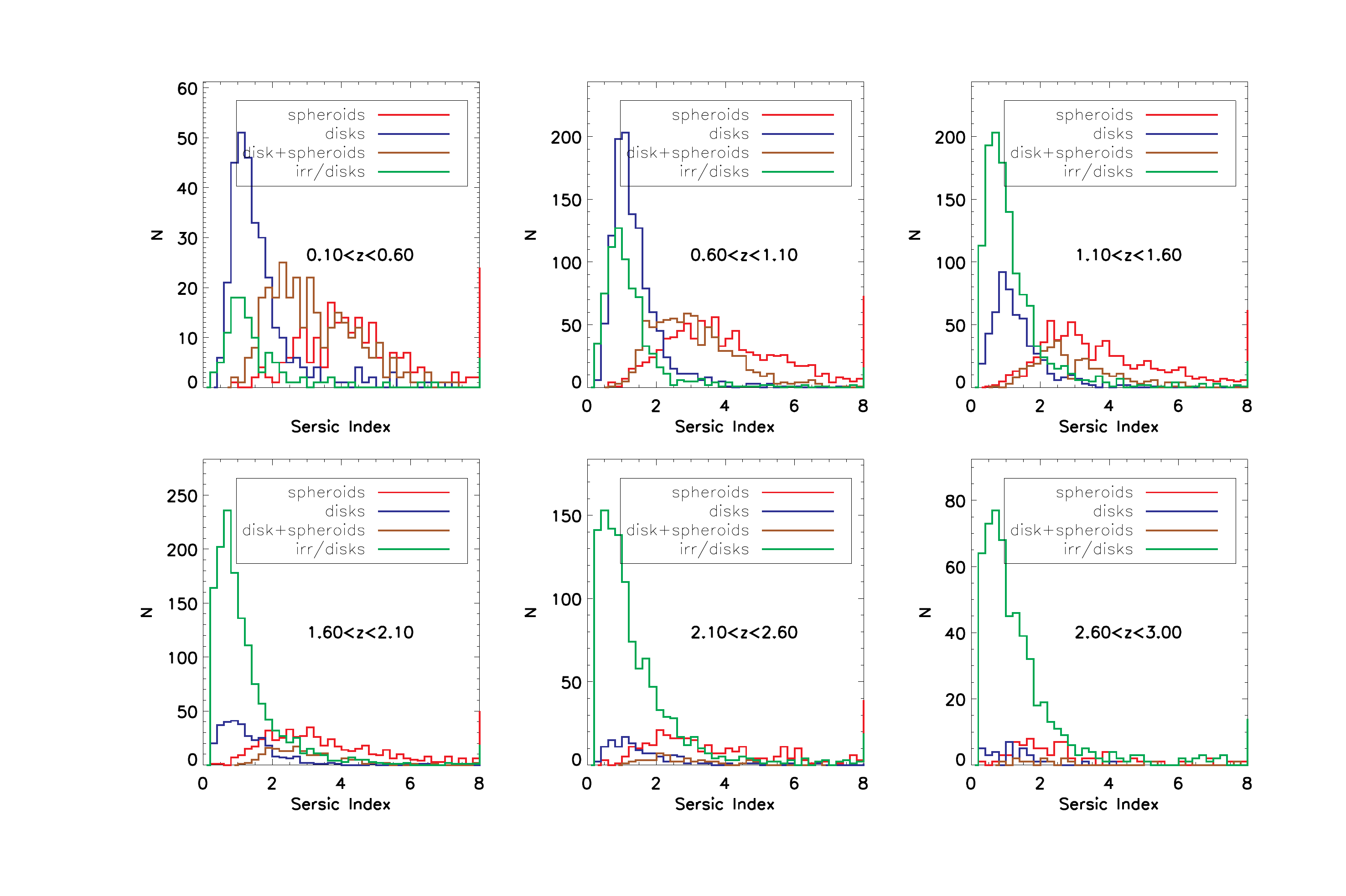}
\caption{S\'ersic index distribution for different morphological types  as labeled. We show galaxies with $M_*/M_\odot>10^{10}$. Each panel shows a different redshift bin. The expected trends are observed, i.e. bulge dominated systems tend to have high S\'ersic indices while more disky galaxies peak at lower values.  }
\label{fig:sersic}
\end{center}
\end{figure*}

\begin{figure*}
\begin{center}
\includegraphics[width=0.99\textwidth]{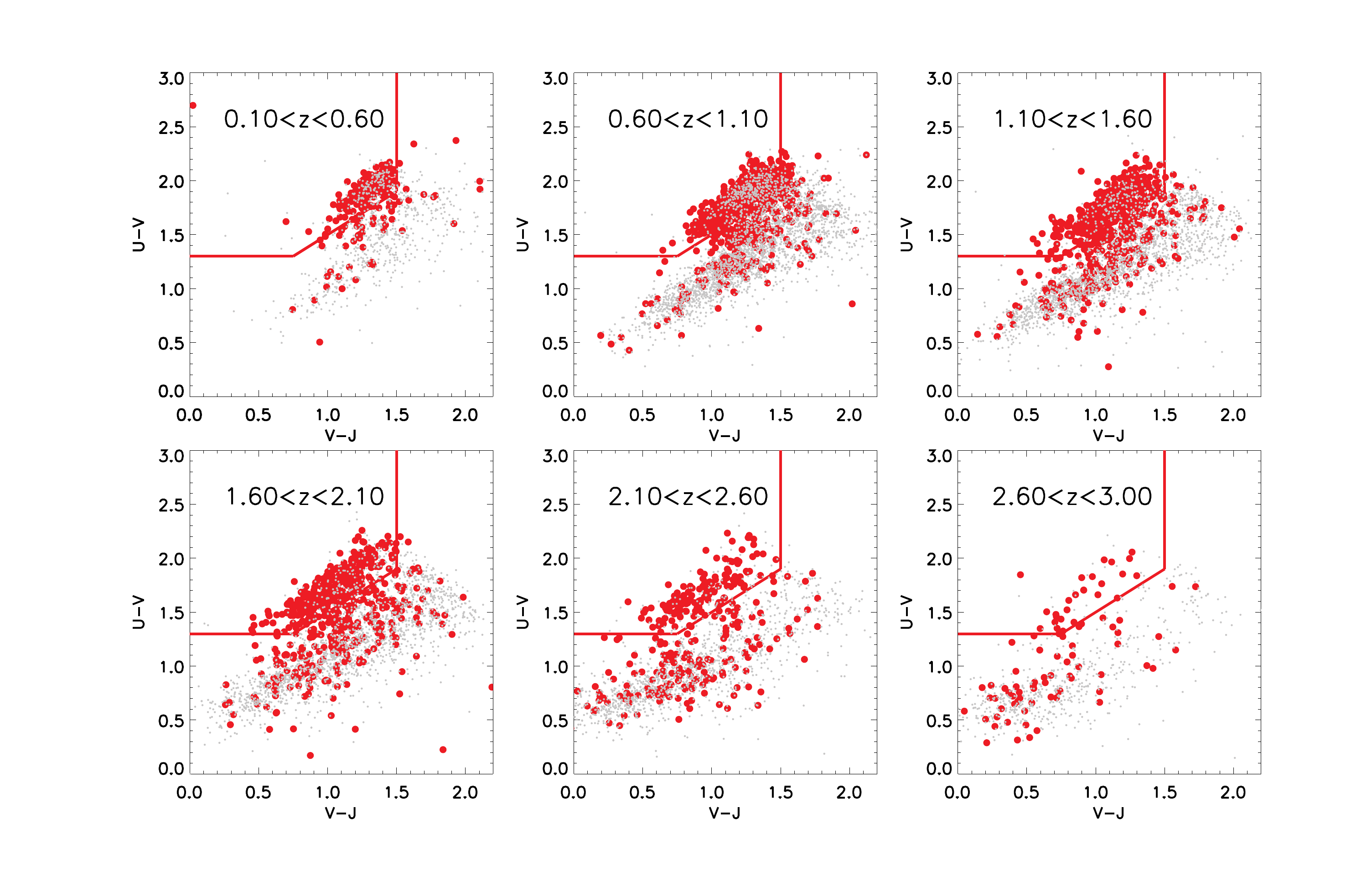}
\caption{ UVJ plane for $M_*/M_\odot>10^{10}$ galaxies in different redshift bins as labeled. Red dots show spheroids and gray points show all other galaxies. The red lines show the location of passive galaxies according to \protect\cite{2012ApJ...754L..29W} }
\label{fig:uvplanesph}
\end{center}
\end{figure*}

\begin{figure*}
\begin{center}
\includegraphics[width=0.99\textwidth]{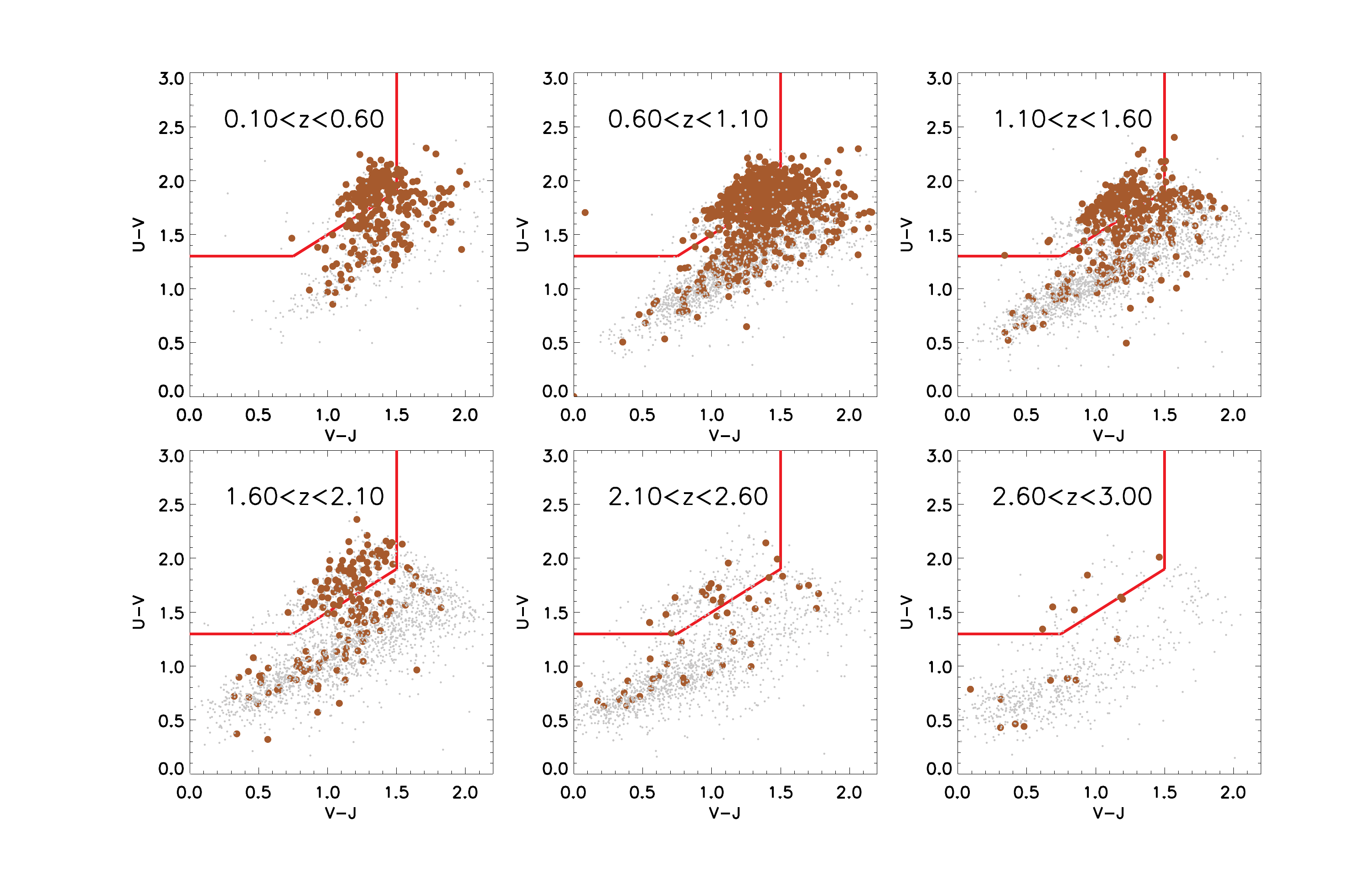}
\caption{ UVJ plane for $M_*/M_\odot>10^{10}$ galaxies in different redshift bins as labeled. Brown dots show disk+spheroids systems and gray points show all other galaxies. The red lines show the location of passive galaxies according to \protect\cite{2012ApJ...754L..29W}  }
\label{fig:uvplanedisksph}
\end{center}
\end{figure*}

\begin{figure*}
\begin{center}
\includegraphics[width=0.99\textwidth]{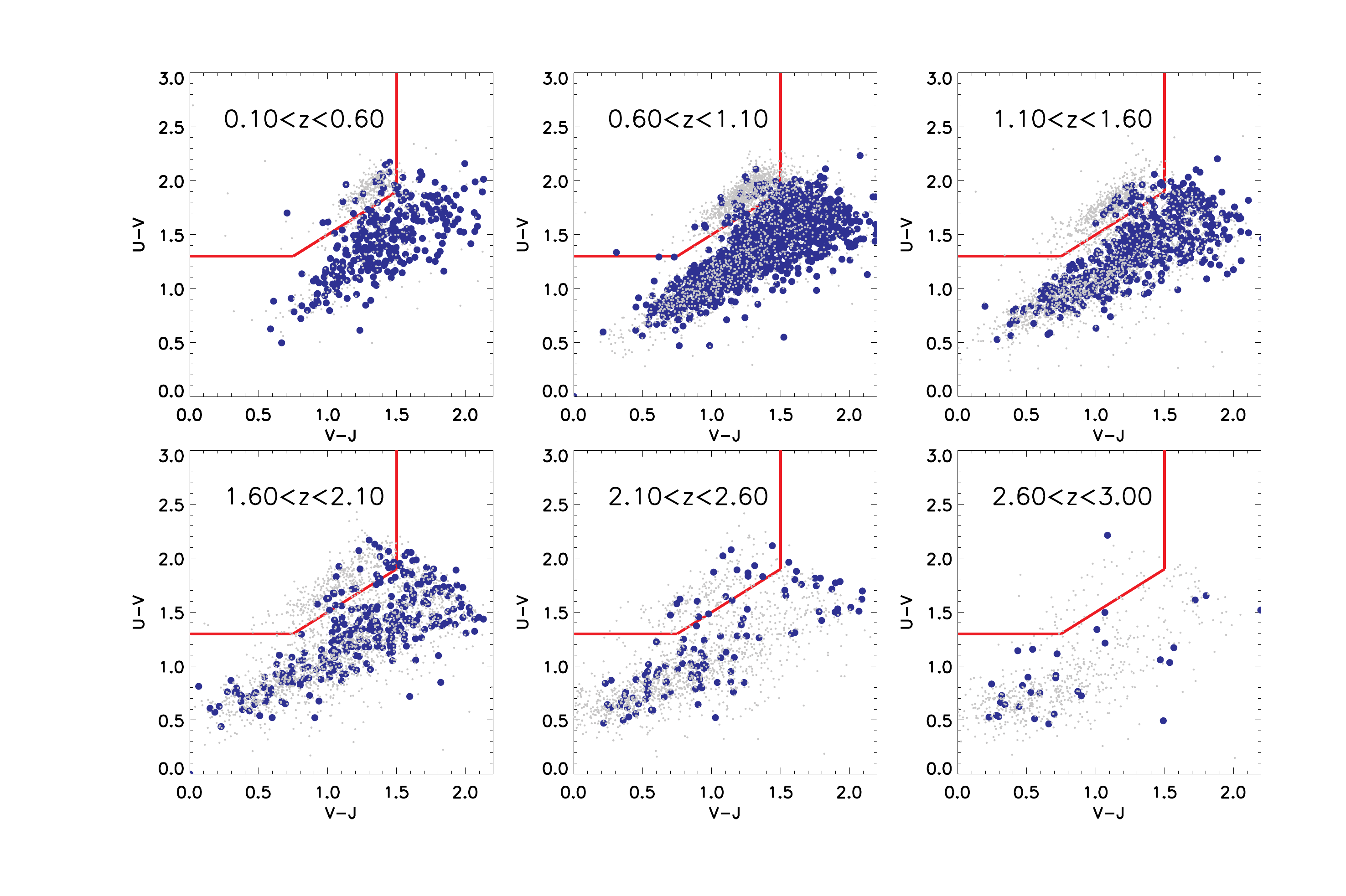}
\caption{ UVJ plane for $M_*/M_\odot>10^{10}$ galaxies in different redshift bins as labeled. Blue dots show disks and gray points show all other galaxies. The red lines show the location of passive galaxies according to \protect\cite{2012ApJ...754L..29W} }
\label{fig:uvplanedisk}
\end{center}
\end{figure*}

\begin{figure*}
\begin{center}
\includegraphics[width=0.99\textwidth]{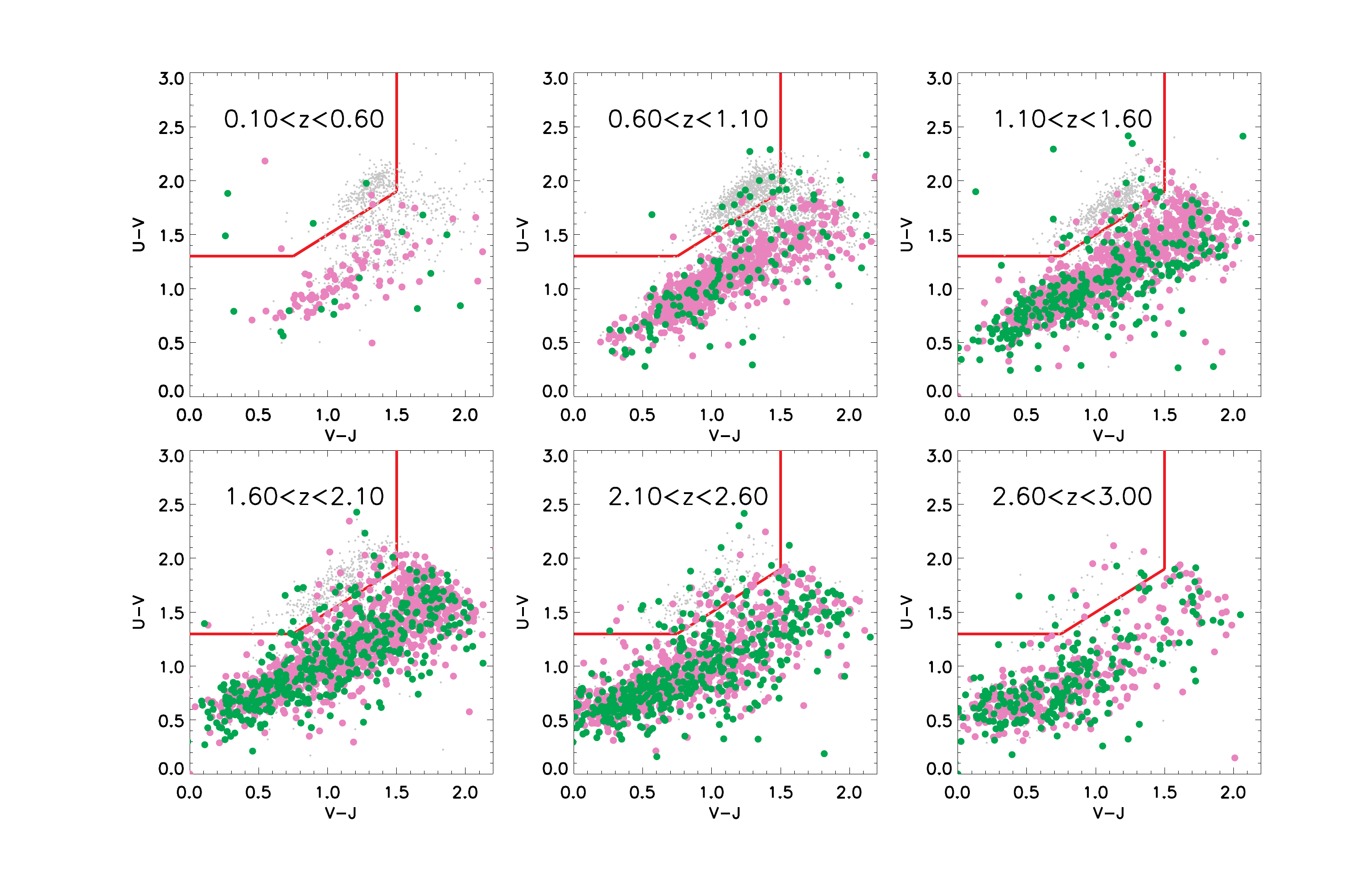}
\caption{ UVJ plane for $M_*/M_\odot>10^{10}$ galaxies in different redshift bins as labeled. Green and violet dots show irregular and disk/irregular galaxies respectively and gray points show all other galaxies. The red lines show the location of passive galaxies according to \protect\cite{2012ApJ...754L..29W} }
\label{fig:uvplaneirr}
\end{center}
\end{figure*}

\begin{figure*}
\begin{center}
\includegraphics[width=0.99\textwidth]{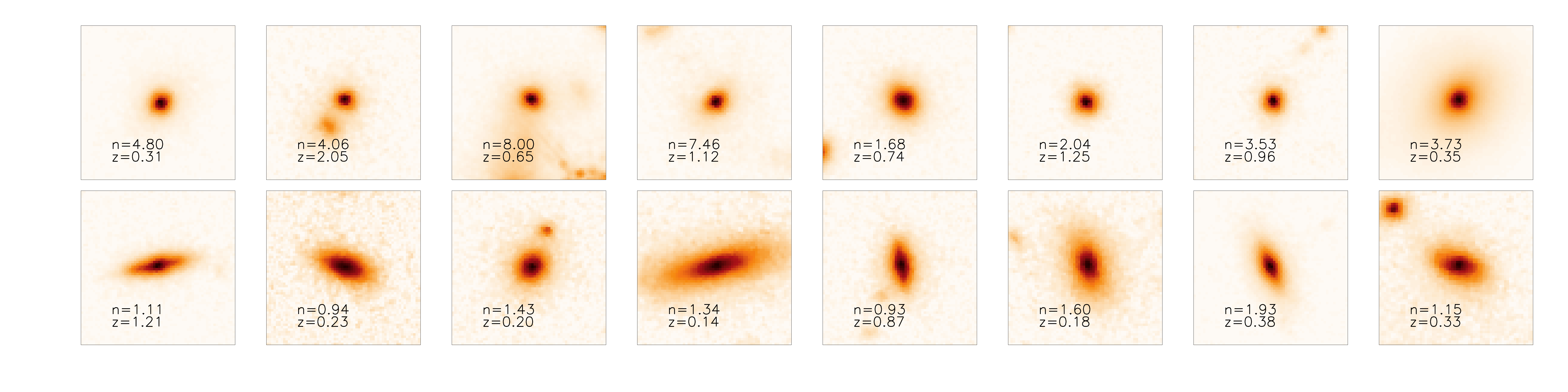}
\caption{ Example stamps of star-forming spheroids (top row) and passive disks (bottom row). For each galaxy we show the sersic index and the redshift.}
\label{fig:strange}
\end{center}
\end{figure*}

\section{Summary and conclusions}
\label{sec:summary}
This work presents a \emph{visual-like} morphological classification of $\sim50.000$ galaxies ($H<24.5$) in 5 CANDELS fields (GOODS-S, GOODS-N, UDS, COSMOS and EGS) in the H band, which probes optical rest-frame morphologies in the redshift range $1<z<3$. The sample is $\sim80\%$ complete down to $log(M_*/M_\odot)\sim10$. \\
Morphologies are estimated with a 5-layer Convolutional Neural Network (ConvNet) followed by 2 layers of fully connected perceptrons trained to reproduce the visual morphologies of $\sim8000$ galaxies in GOODS-S published by the CANDELS collaboration \citep{2014arXiv1401.2455K}. ConvNets are a particular family of neural networks that take advantage of the image stationarity to mimic the way the human brain cells behave to recognize specific patterns. \\

Following the approach in CANDELS, we associate to each galaxy 5 real numbers, $f_{spheroid}$, $f_{disk}$, $f_{irr}$, $f_{PS}$ and $f_{Unc}$, corresponding respectively to the frequency at which \emph{expert classifiers} flagged a galaxy as having a bulge, having a disk, presenting an irregularity, being compact or point-source and being unclassifiable. Galaxy images are interpolated to a fixed size, rotated and randomly perturbed before feeding the network to (i) avoid over-fitting and (ii) reach a comparable ratio of background vs. galaxy pixels in all images. \\

 ConvNets are able to predict the \emph{votes} of expert classifiers with a $<10\%$ bias and a $\sim10\%$ scatter. This makes the classification almost equivalent to a visual based one. The training took 10 days on a GPU and the classification is performed at a rate of 1000 galaxies/hour. As opposed to generalized CAS methods (i.e. galSVM), ConvNets are able to identify without ambiguity ($<1\%$ miss-classifications) objects that are not galaxies (high $f_{Unc}$ values), distinguish irregulars from disks at all redshifts and spheroids from disks. \\

The catalog of $\sim50.000$ galaxies is released with the present paper through the Rainbow database: \url{http://rainbowx.fis.ucm.es/Rainbow\_navigator\_public/}. The catalog actually increases by a factor of 5 the existing (public) morphologies in the CANDELS fields and is intended to be used for many diverse scientific applications (i.e. evolution of merger rates, morphological evolution from $z\sim3$, morphology-density/environment relation, morphology-AGN connection etc...). \\

Future efforts will be focused on optimizing deep-learning based approaches like the one presented here for EUCLID/WFIRST/LSST like data, analyzing deeper data such as the Hubble Frontier Fields as well as providing more detailed morphological descriptors in CANDELS (i.e tidal features etc...).

{\bf Acknowledgements:~} We thank the two anonymous referees for contributing to significantly improve this work. M.H.C acknowledges D.~Gratadour for kindly giving us access to the GPU cluster at LESIA. G.C.V gratefully acknowledges financial support from CONICYT-Chile through its doctoral scholarship and grant DPI20140090. S.M. acknowledges financial support from the Institut Universitaire de France (IUF), of which she is senior member. GB,DCK, and SMF acknowledge support from NSF grant AST-08-08133 and NASA grant HST-GO-12060.10A.

\clearpage

\end{document}